\documentclass{article}

\usepackage{PRIMEarxiv}

\usepackage[most]{tcolorbox}
\usepackage{varwidth}
\newtcolorbox{resboxblack}[2][]{enhanced,
before skip=2mm,after skip=2mm, colback=black!2,colframe=black!50,boxrule=0.2mm,
attach boxed title to top left={xshift=0.4cm,yshift*=0.7mm-\tcboxedtitleheight}, varwidth boxed title*=-3cm,
boxed title style={frame code={
            \path[fill=tcbcolback!30!black]
              ([yshift=-1mm,xshift=-1mm]frame.north west)
                arc[start angle=0,end angle=180,radius=1mm]
              ([yshift=-1mm,xshift=1mm]frame.north east)
                arc[start angle=180,end angle=0,radius=1mm];
            \path[left color=tcbcolback!60!black,right color=tcbcolback!60!black,
              middle color=tcbcolback!80!black]
              ([xshift=-2mm]frame.north west) -- ([xshift=2mm]frame.north east)
              [rounded corners=1mm]-- ([xshift=1mm,yshift=-1mm]frame.north east)
              -- (frame.south east) -- (frame.south west)
              -- ([xshift=-1mm,yshift=-1mm]frame.north west)
              [sharp corners]-- cycle;
            },interior engine=empty,
          }, fonttitle=\bfseries, title={#2},#1}

\newcommand{\specialitem}[3][white]{%
  \item[
    \colorbox{#2}{\textcolor{#1}{\makebox(1.8em, 0.8em){#3}}}
  ]
}

\newcommand\numberbox[2][]{\tikz[overlay]\node[fill=black, draw=white, inner sep=2pt, anchor=text, rectangle, rounded corners=1mm,#1] {#2};\phantom{#2}}

\usepackage{hyperref}
\usepackage{url}  
\usepackage{array}
\newcolumntype{N}{>{\centering\arraybackslash}m{0.41in}}
\newcolumntype{G}{>{\centering\arraybackslash}m{0.40in}}
\usepackage{multirow}
\usepackage{makecell}
\usepackage{subfigure}
\usepackage{cleveref}
\usepackage{booktabs, colortbl}

\usepackage{amsfonts}
\usepackage{fancyhdr}
\usepackage{nicefrac}       
\usepackage{microtype}

\usepackage{soul}
\makeatletter
\def\SOUL@hlpreamble{%
    \setul{}{3.5ex}% increased by 1ex
    \let\SOUL@stcolor\SOUL@hlcolor
    \dimen@\SOUL@ulthickness
    \dimen@i=-.75ex % increased by -0.25ex
    \advance\dimen@i-.5\dimen@
    \edef\SOUL@uldepth{\the\dimen@i}%
    \let\SOUL@ulcolor\SOUL@stcolor
    \SOUL@ulpreamble
}
\makeatother

\newcommand{\hlc}[2][yellow]{{%
    \colorlet{foo}{#1}%
    \sethlcolor{foo}\hl{#2}}%
}

\title{MeDeT: Medical Device Digital Twins Creation with Few-shot Meta-learning
}

\author{
  Hassan Sartaj, Shaukat Ali \\
  Simula Research Laboratory \\
  Oslo, Norway\\
  \texttt{\{hassan, shaukat\}@simula.no} \\
   \And
  Julie Marie Gjøby \\
  Welfare Technologies Section, Oslo Kommune Helseetaten \\
  Oslo, Norway\\
  \texttt{julie-marie.gjoby@hel.oslo.kommune.no} \\
}

\begin{document}
\maketitle

%% ---  approach name ---
\newcommand{\approach}{{\fontfamily{qhv}\selectfont MeDeT}}

% devices names
\newcommand{\Karie}{{\mbox{Karie}}}
\newcommand{\Medido}{{\mbox{Medido}}}
\newcommand{\Pilly}{{\mbox{Pilly}}}
\newcommand{\BPMeter}{{\mbox{BPMeter}}}
\newcommand{\POximeter}{{\mbox{POximeter}}}

% devices iDTs and aDTs
\newcommand{\KarieIdt}{{\fontfamily{cmtt}\selectfont \mbox{Karie-iDT}}}
\newcommand{\MedidoIdt}{{\fontfamily{cmtt}\selectfont \mbox{Medido-iDT}}}
\newcommand{\PillyIdt}{{\fontfamily{cmtt}\selectfont \mbox{Pilly-iDT}}}
\newcommand{\BPMeterIdt}{{\fontfamily{cmtt}\selectfont \mbox{BPMeter-iDT}}}
\newcommand{\POximeterIdt}{{\fontfamily{cmtt}\selectfont \mbox{POximeter-iDT}}}

\newcommand{\KarieIdtVi}{{\fontfamily{cmtt}\selectfont \mbox{Karie-v1-iDT}}}
\newcommand{\MedidoIdtVi}{{\fontfamily{cmtt}\selectfont \mbox{Medido-v1-iDT}}}
\newcommand{\PillyIdtVi}{{\fontfamily{cmtt}\selectfont \mbox{Pilly-v1-iDT}}}
\newcommand{\BPMeterIdtVi}{{\fontfamily{cmtt}\selectfont \mbox{BPMeter-v1-iDT}}}
\newcommand{\POximeterIdtVi}{{\fontfamily{cmtt}\selectfont \mbox{POximeter-v1-iDT}}}

\newcommand{\KarieAdtM}{{\fontfamily{cmtt}\selectfont \mbox{Karie-aDT-M}}}
\newcommand{\KarieAdtP}{{\fontfamily{cmtt}\selectfont \mbox{Karie-aDT-P}}}
\newcommand{\MedidoAdtK}{{\fontfamily{cmtt}\selectfont \mbox{Medido-aDT-K}}}
\newcommand{\MedidoAdtP}{{\fontfamily{cmtt}\selectfont \mbox{Medido-aDT-P}}}
\newcommand{\PillyAdtK}{{\fontfamily{cmtt}\selectfont \mbox{Pilly-aDT-K}}}
\newcommand{\PillyAdtM}{{\fontfamily{cmtt}\selectfont \mbox{Pilly-aDT-M}}}
\newcommand{\BPMeterAdtO}{{\fontfamily{cmtt}\selectfont \mbox{BPMeter-aDT-O}}}
\newcommand{\POximeterAdtB}{{\fontfamily{cmtt}\selectfont \mbox{POximeter-aDT-B}}}

\newcommand{\KarieAdtVii}{{\fontfamily{cmtt}\selectfont \mbox{Karie-v2-aDT}}}
\newcommand{\MedidoAdtVii}{{\fontfamily{cmtt}\selectfont \mbox{Medido-v2-aDT}}}
\newcommand{\PillyAdtVii}{{\fontfamily{cmtt}\selectfont \mbox{Pilly-v2-aDT}}}
\newcommand{\BPMeterAdtVii}{{\fontfamily{cmtt}\selectfont \mbox{BPMeter-v2-aDT}}}
\newcommand{\POximeterAdtVii}{{\fontfamily{cmtt}\selectfont \mbox{POximeter-v2-aDT}}}

\newcommand{\KarieAdtViii}{{\fontfamily{cmtt}\selectfont \mbox{Karie-v3-aDT}}}
\newcommand{\MedidoAdtViii}{{\fontfamily{cmtt}\selectfont \mbox{Medido-v3-aDT}}}
\newcommand{\PillyAdtViii}{{\fontfamily{cmtt}\selectfont \mbox{Pilly-v3-aDT}}}
\newcommand{\BPMeterAdtViii}{{\fontfamily{cmtt}\selectfont \mbox{BPMeter-v3-aDT}}}
\newcommand{\POximeterAdtViii}{{\fontfamily{cmtt}\selectfont \mbox{POximeter-v3-aDT}}}

\newcommand{\KarieAdtViv}{{\fontfamily{cmtt}\selectfont \mbox{Karie-v4-aDT}}}
\newcommand{\MedidoAdtViv}{{\fontfamily{cmtt}\selectfont \mbox{Medido-v4-aDT}}}
\newcommand{\PillyAdtViv}{{\fontfamily{cmtt}\selectfont \mbox{Pilly-v4-aDT}}}
\newcommand{\BPMeterAdtViv}{{\fontfamily{cmtt}\selectfont \mbox{BPMeter-v4-aDT}}}
\newcommand{\POximeterAdtViv}{{\fontfamily{cmtt}\selectfont \mbox{POximeter-v4-aDT}}}

\newcommand{\PD}{{\fontfamily{cmtt}\selectfont \mbox{PD}}}
\newcommand{\iDt}{{\fontfamily{cmtt}\selectfont \mbox{iDT}}}
\newcommand{\aDt}{{\fontfamily{cmtt}\selectfont \mbox{aDT}}}
\newcommand{\iDts}{{\fontfamily{cmtt}\selectfont \mbox{iDTs}}}
\newcommand{\aDts}{{\fontfamily{cmtt}\selectfont \mbox{aDTs}}}
\newcommand{\iDtVi}{{\fontfamily{cmtt}\selectfont \mbox{v1-iDT}}}
\newcommand{\aDtVii}{{\fontfamily{cmtt}\selectfont \mbox{v2-aDT}}}
\newcommand{\aDtViii}{{\fontfamily{cmtt}\selectfont \mbox{v3-aDT}}}
\newcommand{\aDtViv}{{\fontfamily{cmtt}\selectfont \mbox{v4-aDT}}}

\newcommand{\aDtK}{{\fontfamily{cmtt}\selectfont \mbox{aDT-K}}}
\newcommand{\aDtM}{{\fontfamily{cmtt}\selectfont \mbox{aDT-M}}}
\newcommand{\aDtP}{{\fontfamily{cmtt}\selectfont \mbox{aDT-P}}}
\newcommand{\aDtO}{{\fontfamily{cmtt}\selectfont \mbox{aDT-O}}}
\newcommand{\aDtB}{{\fontfamily{cmtt}\selectfont \mbox{aDT-B}}}
\newcommand{\aDtsK}{{\fontfamily{cmtt}\selectfont \mbox{aDTs-K}}}
\newcommand{\aDtsM}{{\fontfamily{cmtt}\selectfont \mbox{aDTs-M}}}
\newcommand{\aDtsP}{{\fontfamily{cmtt}\selectfont \mbox{aDTs-P}}}

\begin{abstract}
  \textcolor{black}{
  Testing healthcare Internet of Things (IoT) applications at system and integration levels necessitates integrating numerous medical devices of various types. Challenges of incorporating medical devices are: (i) their continuous evolution, making it infeasible to include all device variants, and (ii) rigorous testing at scale requires multiple devices and their variants, which is time-intensive, costly, and impractical. Our collaborator, Oslo City's health department, faced these challenges in developing automated test infrastructure, which our research aims to address. In this context, we propose a meta-learning-based approach (\approach{}) to generate digital twins (DTs) of medical devices and adapt DTs to evolving devices. 
  }
  We evaluate \approach{} in Oslo City's context using five widely-used medical devices integrated with a real-world healthcare IoT application. 
  Our evaluation assesses \approach{}'s ability to generate and adapt DTs across various devices and versions using different few-shot methods, the fidelity of these DTs, the scalability of operating 1000 DTs concurrently, and the associated time costs. 
  Results show that \approach{} can generate DTs with over 96\% fidelity, adapt DTs to different devices and newer versions with reduced time cost (around one minute), and operate 1000 DTs in a scalable manner while maintaining the fidelity level, thus serving in place of physical devices for testing.   
\end{abstract}


\section{Introduction}
Healthcare IoT applications are typically cloud-based, which allow central access from various stakeholders (e.g., patients, caregivers, and doctors) and efficiently deliver healthcare services such as remote monitoring and care~\cite{balasubramanian2021scalable}. Such applications must also incorporate healthcare services relying on various medical devices such as medicine dispensers, pulse measuring, and activity monitoring devices. 
Failure in the timely delivery of healthcare services may lead to severe consequences. 
Therefore, dependability assurance through rigorous system testing of such IoT applications is a primary consideration for healthcare authorities. 
System and integration level testing of healthcare IoT applications involves integrating medical devices of various types.
\textcolor{black}{
Medical devices constantly evolve due to the addition of new services and hardware/software upgrades, making it infeasible to incorporate all device variants for testing. 
Furthermore, rigorous testing at a large scale requires numerous evolving devices, which is time-intensive, costly, and impractical. 
Such testing could potentially lead to damaging these devices during test execution~\cite{sartaj2023testing}. 
}

\textcolor{black}{
This work is conducted in collaboration with Oslo City’s health department~\cite{wtsproject}. 
Oslo City aims to develop a cost-effective test infrastructure for the automated and rigorous testing of healthcare IoT applications at system and integration levels. 
During the development of this infrastructure, they encountered challenges related to integrating evolving medical devices and testing at a large scale. 
Our research is mainly devoted to devising a solution to address these challenges. 
Given the role of digital twins (DTs) as virtual replicas of physical entities~\cite{tao2019digital}, we presented the idea of using DTs of medical devices within the test infrastructure~\cite{sartaj2023hita}. 
Our research objective, in this context, is to propose an approach that facilitates the generation of DTs capable of adapting to device evolution and enables concurrently operating multiple DTs for large-scale testing. 
}

\textcolor{black}{
In this paper, we propose \approach{}, which employs meta-learning techniques~\cite{li2018learning} to generate DTs and adapt DTs to evolving devices, thereby enabling automated testing at scale in Oslo City. 
}
We chose to apply meta-learning since it is well-known for its domain generalization and adaptation capabilities~\cite{li2018learning,hospedales2021meta,nam2023stunt}. Such capabilities are needed since we aim to cost-effectively build and adapt DTs of different medical devices while considering their natural evolution (e.g., continuously introducing new or enhanced features), as opposed to constructing DTs for each new device from scratch or retraining an already-generated DT in response to any update in its corresponding physical medical device.

We evaluate \approach{} in Oslo City's context with five widely-used medical devices comprising three medicine dispensers (i.e., \Karie{}, \Medido{}, and \Pilly{}) and two measurement devices (i.e., \BPMeter{} and \POximeter{}) that are integrated with an industrial healthcare IoT application. 
These devices have been deployed in Oslo City and are currently being used in multiple municipalities of Norway. 
We conduct evaluation considering four perspectives: 1) \approach{}'s ability to generate (train) DTs and adapt (fine-tune) them across other types of medical devices and evolving versions, 2) the fidelity of generated and adapted DTs, 3) the scalability of concurrently simulating 1000 DTs of \Karie{}, \Medido{}, \Pilly{}, \BPMeter{}, and \POximeter{}, and 4) time cost involved in generating and adapting DTs. 
We assess each perspective by comparing the performance of few-shot learning methods, specifically 1-, 2-, and 5-shot methods.  
Results show that DTs generated and adapted with different few-shot methods have an average precision, recall, and F1-score of approximately 96\%. 
Results also show that DTs generated with all few shot methods have high fidelity (exceeding 96\% overall) with respect to their physical counterparts. 
In the case of DT adaptation across devices and versions, the 1-shot method demonstrates a high fidelity for simpler devices such as \Pilly{} and \BPMeter{}. 
Whereas for feature-rich devices like \Karie{}, DT adaptations require high-shot methods, such as the 2-shot or 5-shot methods. 
Results of operating 1000 DTs in different batches (i.e., 100, 200, 400, 600, 800, and 1000) reveal that \approach{} is scalable in operating multiple DTs while keeping high fidelity intact.  
Moreover, time cost analysis indicates that adapting a DT to another device or a newer version takes approximately one minute using various few-shot methods. 
This is a considerable reduction compared to the 16-25 minutes required to generate a DT from scratch with training, especially for IoT-based applications with continuously evolving numerous medical devices.
Based on the results, we also provide valuable insights and lessons learned for researchers and practitioners.

The paper's organization is as follows. 
\textcolor{black}{Related works are discussed in ~\Cref{relatedwork}. }
Background and industrial context are presented in \Cref{background}. 
Our approach is illustrated in \Cref{approach}. The empirical evaluation of our approach is described in \Cref{experiment}. 
Insights and lessons are presented in \Cref{insights}.  
The paper concludes in \Cref{conclusion}.

\section{Related Work}\label{relatedwork}

In this section, we present a review of the literature relevant to our research work. 
We focus on four main aspects, including the generation of DTs, industrial DTs, testing IoT applications, and IoT-based healthcare applications. 

\subsection{Digital Twins Generation}
Several approaches exist in the literature regarding the generation and application of DTs across different domains. 
In the following, we discuss existing approaches and relate them to our work. 
\subsubsection{Model-based Approaches} 
In the literature, some researchers have focused on conceptual works related to digital twins. For example, Yue et al.~\cite{yue2021understanding} proposed a conceptual model for developing DTs in the cyber-physical systems (CPS) domain. Along the same lines, Rivera et al.~\cite{rivera2020engineering} presented a reference architecture to support modeling and designing DTs of IoT devices. Such conceptual works can be implemented with different model-based approaches.
In the context of DTs generation, Kirchhof et al.~\cite{kirchhof2020model} presented a model-driven approach to create DTs of CPS and integrate DTs to synchronize CPS development. Moreover, Sleuters et al.~\cite{sleuters2019digital} proposed an approach for generating DTs to support the behavior analysis of IoT-enabled systems. 
In another work related to model-based DT, Christofi and Pucel~\cite{christofi2022novel} proposed operational models for creating DTs that can support operators in locating malfunctions. 
For the DT architectural work, Dalibor et al.~\cite{dalibor2020towards} presented a model-based architecture to integrate, operate, and monitor DTs using an interactive interface. 
Building on this, Dalibor et al.~\cite{dalibor2022generating} introduced a model-based platform that facilitates designing DTs of CPS using modeling languages and enables the generation and operation of these DTs. 
In the work on DT modeling notations, Corradini et al.~\cite{corradini2023dtmn} presented modeling language to model DTs for diverse domains such as CPS and IoT. 
All the works outlined above employ model-driven methods to generate DTs, which require considerable manual effort, particularly when devices are constantly evolving. 
In comparison, our work utilizes meta-learning to generate DTs and adapt to evolving medical devices with few-shot methods.

\subsubsection{ML-based Techniques} 
In this context, Robles et al.~\cite{robles2023opentwins} introduced an open-source framework designed to create and visualize DTs, along with the ability to analyze data and make predictions. 
Picone et al.~\cite{picone2021wldt} developed a generic framework to support engineers in designing, creating, and operating DTs within the IoT domain. 
Given that both above-mentioned frameworks are generic, they need to be customized for a specific type of device and each device's evolution. 
In contrast, our research is focused on generating DTs for IoT medical devices capable of adapting to the device's upgrades. 
Sciullo et al.~\cite{sciullo2024relativistic} proposed a framework dedicated to generating and calibrating generic DTs of various IoT entities. 
While this work facilitates the generation of general-purpose DTs, the creation of domain-specific DTs necessitates considerable customization. 
Unlike this, our work is focused on generating DTs for IoT medical devices and adapting DTs to accommodate evolving requirements. 
Xu et al.~\cite{xu2024pretrain} presented an approach to evolve CPS DTs by transferring knowledge from an older variant to a newer one. 
This approach utilizes transfer learning to evolve DTs of CPS, which require a significant amount of data to retrain. 
In comparison, our approach employs meta-learning to generate DTs for IoT medical devices, which enables rapid adaptation to new device variants with minimal data and few-shot learning. 
Zhou et al.~\cite{zhou2024toward} presented an approach for building DTs of human wearable motion-tracking devices used within the CPS environment. 
Compared to this, our work targets generating DTs of IoT medical devices. 
In addition, our work supports adapting DTs to new and upgraded devices.

\subsubsection{DTs in Healthcare} 
In the healthcare domain, Shoukat et al.~\cite{shoukat2024smart} proposed an architecture to model DTs of home devices, aiming to enhance healthcare services for the elderly. 
Although this work introduces a DT modeling architecture, it necessitates a specialized approach for generating DTs, which is the main focus of our work. 
Pirbhulal et al.~\cite{pirbhulal2024cognitive} presented a framework for creating cognitive DTs capable of addressing potential cyber security risks in healthcare applications. 
Compared to this, our objective is to generate DTs that can facilitate automated testing on a large scale. 
Bersani et al.~\cite{bersani2022engineering} introduced a preliminary concept of medical devices DTs that can handle security threats and environmental uncertainties. 
While this work focuses on security and uncertainty aspects, our work proposes an approach to generate and adapt DTs to enable testing automation. 
Elayan et al.~\cite{elayan2021digital} proposed a machine learning-based approach for creating DTs of electrocardiograms (ECG) in IoT-based healthcare systems, facilitating the prediction of heart diseases. 
While this work targets DTs of ECG, our work focuses on generating medical devices DTs capable of adapting to device evolution. 
In our previous work~\cite{sartaj2024modelbased}, we proposed a model-based approach to generate DTs of medicine dispensers, requiring substantial modeling effort for new/upgraded devices. 
In contrast to previous work, this paper addresses the challenge of continuously evolving medical devices by proposing a meta-learning-based approach for generating DTs adaptable to new/upgraded devices.

\subsubsection{Applications of DTs} 
Several techniques for utilizing DTs across various tasks are available in the existing literature.  
In this context, Kaul et al.~\cite{kaul2023role} used DTs for cancer care, David et al.~\cite{david2021inference} used DTs to infer simulation models, Kirchhoff et al.~\cite{kirchhof2021understanding} and Nguyen et al.~\cite{nguyen2022digital} used DTs for IoT cloud simulations, Lehner et al.~\cite{lehner2021aml4dt} used to maintain DTs in IoT, and Croatti et al.~\cite{croatti2020integration} used DTs as agents to manage severe health conditions. 
Moreover, DTs also played a key role in streamlining smart manufacturing~\cite{somers2023digital}. 
In this regard, Damjanovic and Behrendt~\cite{damjanovic2019open} utilized a DT as a demonstrator for smart CPS, Dobaj et al.~\cite{dobaj2022towards} presented a DT-based self-adaptive model for CPS to facilitate DevOps, Wang et al.~\cite{wang2019digital} used DTs to diagnose faults in rotating machines, and Xia et al.~\cite{xia2021digital} used DTs to automate smart control in industrial manufacturing. 
While all the works mentioned above utilize DTs in different engineering tasks, our research, in contrast, focuses on the generation of DTs specifically designed to handle the evolving needs of IoT medical devices.

\subsection{Industrial Digital Twins}
Several commercial and open-source tools are available for creating DTs, e.g., Amazon's AWS IoT TwinMaker~\cite{iottwinmaker}, Microsoft's Azure IoT Hub~\cite{iothub} with Azure Digital Twins~\cite{azuredt} service, and Eclipse's Vorto~\cite{vorto}, Hono~\cite{hono} and Ditto~\cite{ditto}. 
Some commercial tools like AWS IoT TwinMaker limit their use to healthcare due to safety concerns. 
Open-source tools use public data storage~\cite{eclipseterms}, raising privacy issues. 
Testing-wise, these tools limit the number of requests on DTs' web service, posing a bottleneck for rigorous and automated testing. 
Moreover, they are overly generic~\cite{dalibor2022cross}, requiring manual development of application/domain-specific features~\cite{lehner2022digital}.
The manual effort required increases significantly as medical devices evolve with hardware and upgrades.  
In comparison, our work presents an approach for generating DTs that can adapt to device evolution, support automated testing, and address safety and privacy concerns.

\subsection{Testing IoT Applications}
Various approaches for testing IoT applications exist, such as model-based testing of IoT edges and clouds~\cite{li2022domain} and search-based mutation testing
for IoT events~\cite{gutierrez2019evolutionary}. 
A few approaches use DTs to detect anomalies in various IoT applications like healthcare IoT~\cite{gupta2021hierarchical} and industrial IoT~\cite{de2023digital}. 
The works discussed above either propose testing techniques for IoT applications or employ DTs for various testing activities, such as anomaly detection. 
Compared to these works, our work does not present a testing technique. 
Rather, we propose an approach for generating DTs and using these DTs to facilitate test execution during system or integration-level testing of healthcare IoT applications. 
Moreover, our approach can be integrated with the existing testing techniques to enable automated testing with DTs.

\subsection{IoT-based Healthcare}

In the works related to IoT applications in the healthcare domain, Korzun~\cite{korzun2017internet} presented several use cases of IoT-based mobile health (mHealth) systems and proposed a conceptual framework for enhancing health services through IoT-enabled mHealth. 
For improving IoT-enabled healthcare, Demetriou et al.~\cite{demetriou2023internet} presented the concept of a new society with an efficient healthcare system featuring IoT. 
To analyze IoT's role in healthcare, Kelly et al.~\cite{kelly2020internet} provided a comprehensive review of the role played by IoT and IoT-enabled devices in delivering efficient healthcare services. 
Similarly, Rayan et al.~\cite{rayan2022internet} discussed the concept, applications, and challenges associated with IoT medical devices employed to deliver various healthcare services. 
To enhance the safety of IoT-enabled applications, Alsaig and Alagar~\cite{alsaig2023dependable} presented a service and role-based architecture for healthcare IoT systems to facilitate security and privacy. 
While the works mentioned above introduced various concepts for IoT and medical devices in healthcare, our work targets a different objective, i.e., generating DTs of IoT medical devices to support testing healthcare IoT applications.

\textcolor{black}{\section{Background and Research Proposal}\label{background}}
\textcolor{black}{
In this section, we initially introduce the background of meta-learning. 
This is followed by a discussion of our real-world application context and its associated challenges, and we then present our research proposal. 
}

\subsection{Meta-Learning}\label{metalearning}
Meta-learning is typically referred to as \emph{learning to learn}~\cite{thrun1998learning, hospedales2021meta, schmidhuber1987evolutionary}, which refines multiple learning experiences over several interrelated tasks and utilizes such experiences to improve performance in learning new tasks. 
This differs from a typical machine learning approach in which a model needs to be trained from scratch for each new task~\cite{hospedales2021meta}. 
Meta-learning has been used for efficient domain generalization and adaptation without requiring a large amount of data for training in several areas, such as object detection, image classification, robotics, and healthcare~\cite{li2018learning,hospedales2021meta,nam2023stunt}. 

A typical meta-learning process has a \emph{base/inner learning algorithm} to solve a particular task, such as binary classification, and a \emph{meta/outer learning algorithm} to guide the base algorithm in achieving a meta objective, e.g., few-shot learning~\cite{finn2017model,nashid2023retrieval}. 
Meta-learning problems are commonly formulated by considering three aspects: \emph{meta-representation}, \emph{meta-optimizer}, and \emph{meta-objective}~\cite{hospedales2021meta}. 
\emph{Meta-representation} entails creating a meta-taskset, specifying neural network architecture, and selecting hyperparameters and a loss function. 
\emph{Meta-optimizer} involves defining optimization strategies like gradient descent or reinforcement learning to optimize hyperparameters for quick adaptation to tasks in a meta-taskset. 
\emph{Meta-objective} requires choosing learning alternatives like few/many-shot, single/multi-task, or offline/online learning. 
Training with such formulation mainly focuses on rapid learning and adaptation to new tasks, distinguishing it from traditional machine learning.

\subsection{Industrial Context and Challenges}\label{industrycontext}
\begin{figure}[htbp]
\centerline{\includegraphics[width=8.0cm, height=4.0cm, keepaspectratio]{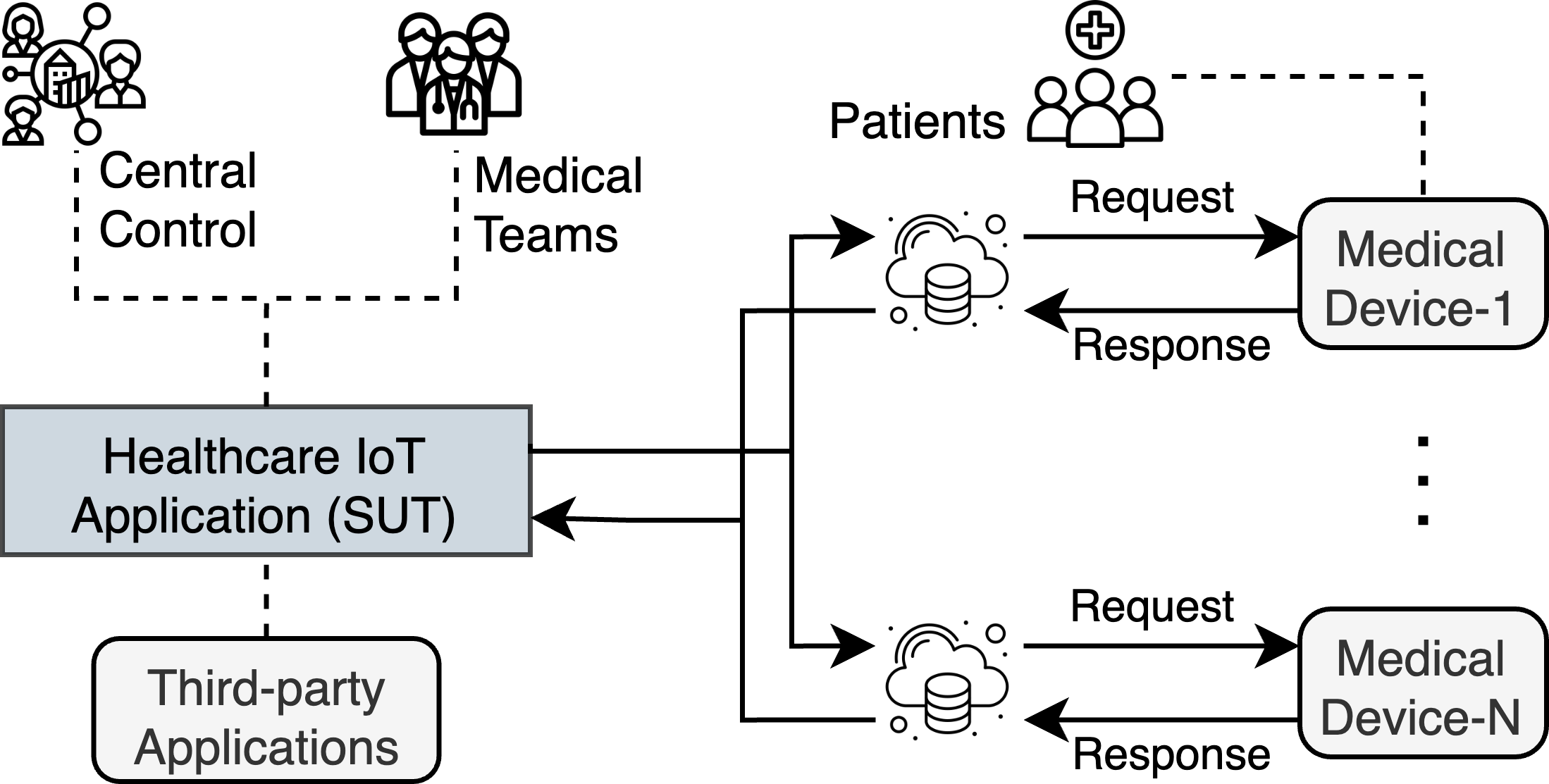}}
\caption{Overview of an IoT-based healthcare system}
\label{fig:iot}
\end{figure}

Under the national welfare technology program~\cite{wtsoslo}, Oslo City's healthcare department~\cite{oslocity} aims at delivering timely and high-quality healthcare services to its residents. 
For this purpose, Oslo City works together with industrial healthcare service providers to develop IoT-based healthcare applications. One such is illustrated in \Cref{fig:iot}: an industrial-scale application (the system under test (SUT)) is connected to various types of medical devices, third-party applications, and central control authorities. 
Each type of medical device has a dedicated web server and Application Programming Interfaces  (APIs) that allow integration and communication with the healthcare IoT application. 
Medical devices are assigned to patients based on their needs. For instance, medicine dispensers are assigned to patients for timely medication, and pulse oximeters are used for monitoring patients' health. These devices inform relevant stakeholders (e.g., caregivers) about the patient's health conditions, especially in emergencies. 

\textcolor{black}{
Healthcare IoT applications are mainly used for \emph{digital home care and monitoring}, \emph{facilitating home hospitals}, and \emph{managing medical equipment}. 
In \emph{digital home care and monitoring}, various medical devices such as medicine dispensers, heart rate monitors, and fitness trackers are used to monitor patients' health, deliver healthcare to patients, and continuously report patients' health conditions to healthcare professionals. 
Similarly, \emph{home hospitals}---also referred to as \emph{telehealth}---aim to facilitate virtual appointments and follow-up treatments with healthcare professionals, thereby reducing the need for in-person visits. 
This is especially beneficial in remote areas with limited access to healthcare services. 
Furthermore, in the context of \emph{managing medical equipment}, hospitals can utilize healthcare IoT applications to monitor and manage medical equipment, enabling them to detect malfunctions or damages. 
This allows for early precautions, ensures safe usage of medical devices, and ultimately improves patient care. 
}

\textcolor{black}{
Testing of healthcare IoT applications is performed at different levels, such as unit, integration, and system. 
Within our industrial context, unit-level testing is largely automated. However, testing at integration and system levels involves considerable manual effort. 
Integration and system-level testing require creating a test infrastructure incorporating numerous medical devices of various types, all connected to a healthcare IoT application. 
This testing infrastructure is essential for enabling automated and rigorous testing, ultimately ensuring the dependability of healthcare IoT applications~\cite{sartaj2023hita}. 
For such a test infrastructure, physically integrating numerous medical devices of different types is time-consuming, costly, and impractical. 
Therefore, the current practice is to designate one device of each type for testing purposes, manually integrate these devices with healthcare IoT applications, and run tests. 
However, this practice restricts the scope of test scenarios, does not facilitate large-scale testing with numerous devices, and could potentially compromise the dependability of healthcare IoT applications. 
In addition, the challenges associated with integrating multiple types of medical devices into a test infrastructure are outlined below. 
}

\textcolor{black}{
\hlc[violet!7!white]{\textbf{Challenge 1 (Evolution):}}
Medical devices constantly undergo different evolution stages with the addition of new features, hardware/software upgrades, and new services. 
This leads to multiple versions of devices, each with varying hardware/software characteristics. 
Incorporating these numerous versions of different types of devices into test infrastructure is infeasible. 
}

\textcolor{black}{
\hlc[violet!7!white]{\textbf{Challenge 2 (Large scale Testing):}}
System and integration testing at a large scale requires integrating thousands of evolving medical devices with a healthcare IoT application. 
This process is not only labor-intensive but also financially expensive. 
Furthermore, executing tests with numerous medical devices may result in excessive requests to the device server, potentially leading to service blockage or even damage to the devices~\cite{sartaj2023testing}.  
}

\textcolor{black}{\subsection{Research Proposal}\label{proposal}}
\textcolor{black}{
Oslo City's healthcare department experienced the above-mentioned challenges while creating a test infrastructure. 
Considering this, we presented the idea of using DTs of medical devices as a part of the test infrastructure~\cite{sartaj2023hita} to replace physical devices with their digital twins to support large-scale testing of evolving healthcare IoT applications, i.e., addressing the above two mentioned challenges. 
In our industrial context, creating test infrastructure is a \emph{development} task. 
Addressing the challenges associated with creating this test infrastructure, such as constructing DTs that can adapt to evolving requirements and support scalable and automated testing, is a \emph{research} activity. 
Therefore, in this paper, our research focuses on building DTs of constantly evolving medical devices with meta-learning and operating these DTs as a part of the test infrastructure. 
Given the lack of existing approaches to solve challenges outlined in \Cref{industrycontext}, our research proposal of utilizing meta-learning to generate DTs that are adaptable to evolving devices signifies a novel contribution. 
In the subsequent section, we present our proposed approach. 
}

% make two figures appear together
\begin{figure*}[!t]
\centering
\centerline{\includegraphics[width=9.2cm, height=4.6cm, keepaspectratio]{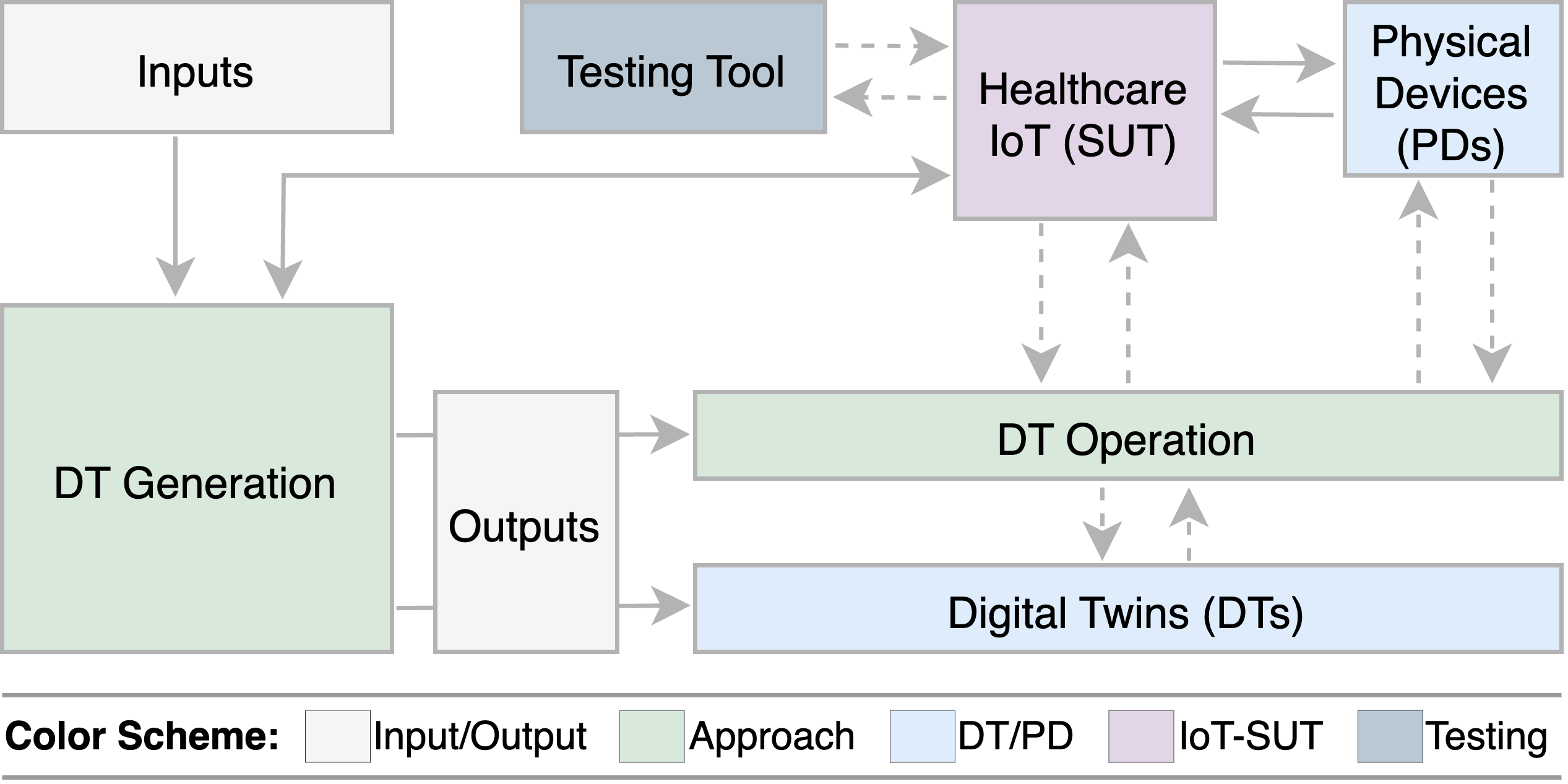}}
\caption{\textcolor{black}{High-level overview of the approach, showing main components and their interactions. The detailed workflow is shown in \Cref{fig:app}}}
\label{fig:app-hl}
\centerline{\includegraphics[width=\textwidth, keepaspectratio]{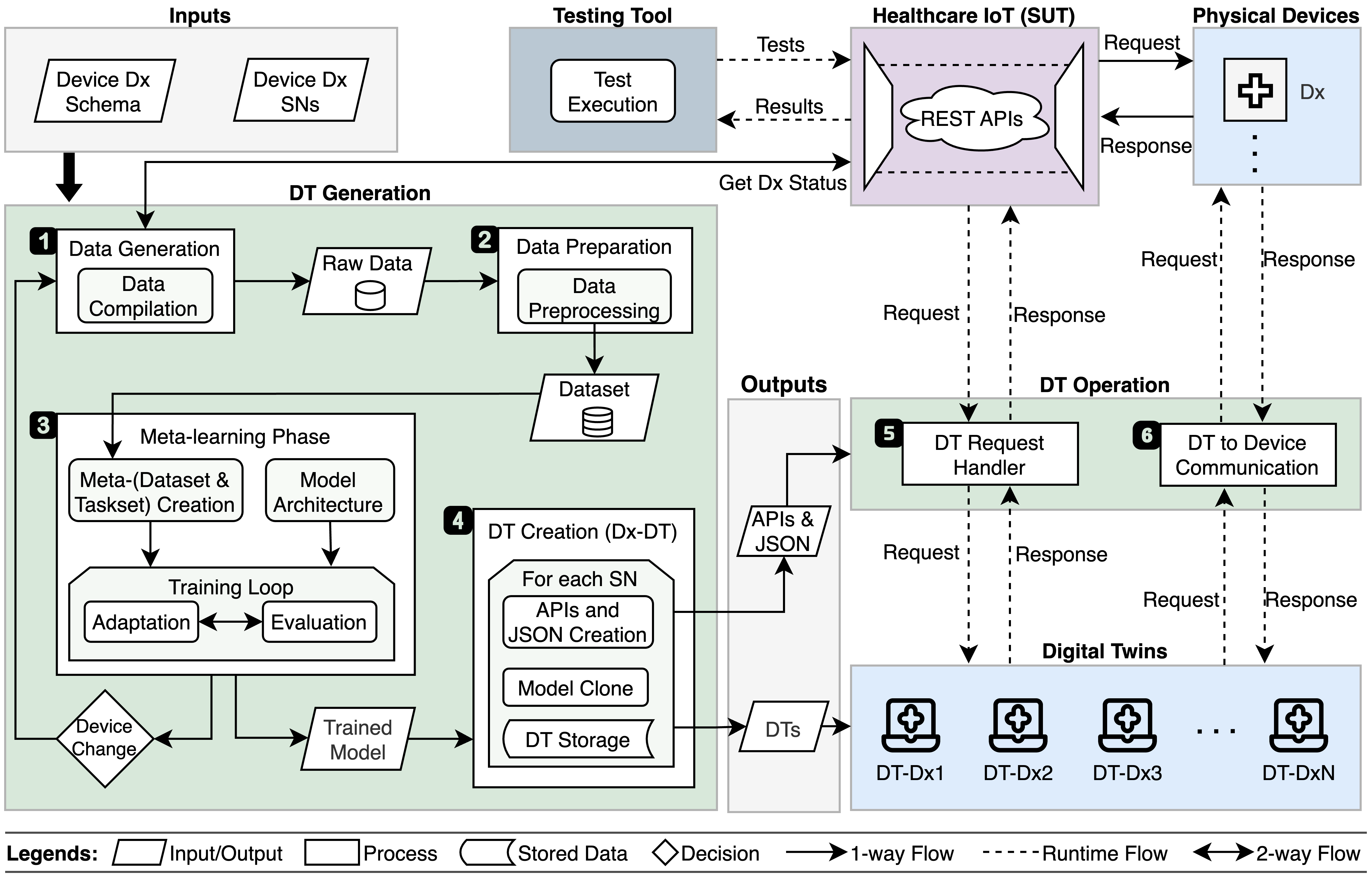}}
\caption{\textcolor{black}{Detailed workflow of \approach{}, including its six phases for generating a device's DTs and operating these DTs during the testing process of a healthcare IoT application.}}
\label{fig:app}
\end{figure*}

\section{Approach}\label{approach}

This section introduces our approach (\approach{}) for \textcolor{black}{
generating DTs of medical devices and operating these DTs when they are integrated with a healthcare IoT application for testing purposes. 
}
\approach{} utilizes meta-learning to build DTs for constantly evolving medical devices. 
\textcolor{black}{
\Cref{fig:app-hl} shows a high-level overview of the \approach{}, illustrating inputs, outputs, DT generation and operation components of \approach{}, and their interactions with the healthcare IoT application and physical devices. 
Following this overview, a more detailed workflow of \approach{} is presented in \Cref{fig:app}, which consists of six phases within the DT generation and operation components. 
}

Phase \hspace{1pt}\textbf{\textcolor{white}{\numberbox{1}}}\hspace{1pt} is \emph{Data Generation} which has a data compilation process. 
In this phase, we use a medical device (\emph{Dx}) schema \textcolor{black}{
in the form of an OpenAPI specification containing a device's API documentation. 
Using the device \emph{Dx} schema, we create and send API requests to a healthcare IoT application, which communicates with the device and returns responses. 
}
We compile raw data from responses containing information about device configurations, request processing time, and response status codes. 
These raw data are used in the next \textcolor{black}{phase}. 
Phase \hspace{1pt}\textbf{\textcolor{white}{\numberbox{2}}}\hspace{1pt} is \emph{Data Preparation} which involves data preprocessing for meta-learning. 
In this phase, we prepare data for training by applying data preprocessing techniques such as handling missing values, outliers, and various data types. 
Phase \hspace{1pt}\textbf{\textcolor{white}{\numberbox{3}}}\hspace{1pt} is \emph{Meta-Learning} which comprises three sub-phases: meta-dataset \& meta-taskset creation, determining model architecture, and training loop with continuous evaluation and adaptation steps. First, we use the dataset from the previous phase to create a meta-dataset and meta-taskset for the meta-learning algorithm. Next, we determine a neural network model architecture for training. Third, with the meta-dataset, meta-taskset, and neural network model, the meta-learning algorithm trains a model during the training loop involving continuous evaluation and adaptation.  
Phase \hspace{1pt}\textbf{\textcolor{white}{\numberbox{4}}}\hspace{1pt} is \emph{Build Digital Twins (Dx-DT)}. 
In this phase, we use the device (\emph{Dx}) schema, device serial numbers \textcolor{black}{(\emph{Dx SNs} as shown in \Cref{fig:app})}, and the trained model from the previous phase to build DTs, which involves creating DT APIs and JSON objects, cloning the trained model, and creating DTs storage. 
Phase \hspace{1pt}\textbf{\textcolor{white}{\numberbox{5}}}\hspace{1pt} is \emph{DT Request Handler} which allows integration with the healthcare IoT application and processes requests during testing via DTs' APIs and JSON objects. 
Phase \hspace{1pt}\textbf{\textcolor{white}{\numberbox{6}}}\hspace{1pt} is \emph{DT to Device Communication}, which establishes DTs communication with their corresponding physical devices via device APIs for sending requests and getting responses.

\subsection{Data Generation}\label{sec:datagen}

In the operating phase of a healthcare IoT application, medical devices are assigned to patients, and their usages produce data required for training machine learning models. %Different patients likely produce different data. For example, a patient's medication plan in a medicine dispenser differs from another patient's. 
Such data are, however, inaccessible due to privacy concerns. Thus, in this phase, we utilize medical (test) devices to generate data sufficient for building DTs that can serve testing purposes. 

\approach{} takes input a medical device (\emph{Dx}) schema in the form of OpenAPI specification format, 
%Such schema are commonly available as a part of API specification documents. 
which is used to determine device properties along with associated types, value ranges, device APIs, and JSON object format. 
We randomly generate within-range and out-of-range values for each device property, aiming to create a balanced dataset representative enough for potential success and failure responses. 
Presumably, the values generated within the range for a given feature may resemble actual patient data. 
Due to data protection and privacy regulations, we lack access to real patients' datasets. 
Given the unavailability of actual patient data, assessing the resemblance between generated and actual patient data is impossible. 
Furthermore, despite the availability of REST API testing tools (e.g., EvoMaster~\cite{arcuri2019restful}), we choose not to use any such tool. 
This is because these tools, without being tailored for specific domains, tend to generate data representing a high number of failure responses~\cite{isaku2023cost,sartaj2024restapi}, which consequently leads to imbalanced datasets.

Values generated for each property are then used to create a JSON object and send a POST request containing the JSON object in the request's body to the healthcare IoT application. During request processing, the healthcare IoT application communicates with a corresponding medical device and returns a response received from the device. The response is then used to assemble device information containing device configurations, request processing time, and response status codes, which are stored in the tabular format as raw data. The process continues until the specified maximum limit is reached, which is configurable based on either the time budget or the maximum number of requests, often defined by a medical device service provider. 

\textcolor{black}{
The raw data ($D_{raw}$), produced during this phase, is represented by~\Cref{eq-rawdt}. 
In this equation, $x_i \in [\mathbb{R}, \mathbb{Z}, \mathcal{S}, \mathbb{B}]$ symbolizes a feature vector and $c_i$ denotes a target class. 
Here $i=1, 2, ..., n$ and symbols $\mathbb{R}$, $\mathbb{Z}$, $\mathcal{S}$, and $\mathbb{B}$ represent Real, Integer, String, and Boolean values, respectively. 
}

\textcolor{black}{
\begin{equation}\label{eq-rawdt}
D_{raw} = \{(x_1, c_1), (x_2, c_2), ..., (x_n, c_n)\}
\end{equation}
}

\subsection{Data Preparation}\label{sec:dataprep}

Since raw data generated in the Data Generation phase contains mixed types of values, % corresponding to different data types, such as integers, strings, boolean, and enumerations. 
null/empty values, special characters (like \%), and outliers, such data must be preprocessed before being utilized for training machine learning models~\cite{kotsiantis2006data}. 
We employ data preprocessing techniques, including data encoding, handling null/empty values, feature engineering, handling outliers, and label encoding. 
Specifically, 1) we use integer data encoding to make different data types' values consistent; 2) we fill null and empty values with zeros; 3) we use a variance-based feature selection technique to remove features with low and high variance to avoid underfitting or overfitting~\cite{guyon2003introduction}; 4) we remove special characters (e.g., \%, *, and \_) and transform outlier values with the maximum possible value from the range; and 5) since the labels in our case correspond to HTTP response status codes (i.e., 2XX for success responses and 4XX/5XX for failure responses~\cite{rfc9110}), we assign each target label with a value between 0 and the total number of unique status codes. 

\textcolor{black}{
The preprocessed data ($D_{pro}$) is represented by~\Cref{eq-prodt}. 
In this equation, $f$ symbolizes the processing function that is applied to each data point in $D_{raw}$, i.e., $\{f(x_1, c_1), f(x_2, c_2), ..., f(x_n, c_n)\}$ to obtain the $D_{pro}$. 
Here, $X_i$ and $C_i$ denote the transformed feature vector and target class respectively. 
Both $X_i$ and $C_i$ belong to the set of real numbers $\mathbb{R}$, ensuring a uniform data type across all values. 
}

\textcolor{black}{
\begin{equation}\label{eq-prodt}
D_{pro} = f(D_{raw}) = \{(X_1, C_1), (X_2, C_2), ..., (X_n, C_n)\}
\end{equation}
}

\begin{figure*}[!t]
\centerline{\includegraphics[width=14.0cm, height=3.4cm, keepaspectratio]{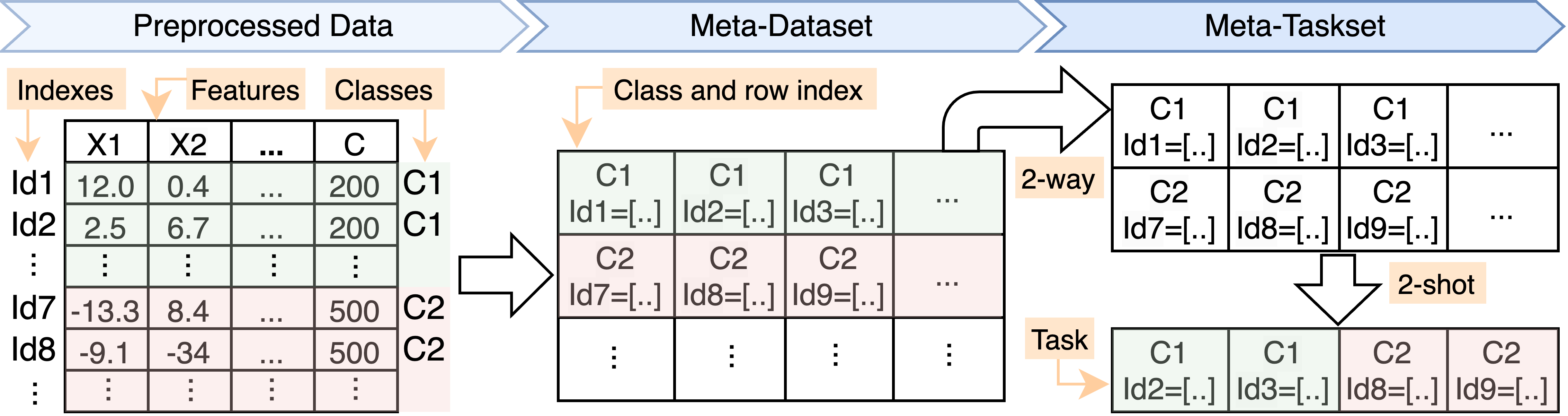}}
\caption{An example of meta-dataset and meta-taskset creation using preprocessed data. \textcolor{black}{The data corresponding to class C1 is highlighted in green, while the data for class C2 is indicated in red. }}
\label{fig:example}
\end{figure*}

\subsection{Meta-Learning Phase} \label{subsec:metalearningpphase}
In this phase, we present \emph{meta-representation} in Subsections: \Cref{subsec:dts}, which discusses the creation of meta-dataset and meta-taskset, and \Cref{subsec:model}, which focuses on determining the model architecture. 
Next, we describe the \emph{meta-optimizer} and \emph{meta-objective} suitable for our problem in \Cref{subsec:training}.

\subsubsection{Meta-Dataset \& Meta-Taskset Creation}\label{subsec:dts}
The preprocessed dataset (the output of the \emph{Data Preparation} phase) is loaded as a meta-dataset, which is then used to create a meta-taskset. 
This requires defining transformations to create a meta-taskset for training, based on configurations of \emph{N}-ways, \emph{K}-shots, \emph{M} tasks, remapping labels, and handling consecutive labels. Creating \emph{N}-way and \emph{K}-shot tasks involves selecting \emph{N} new distinct classes and \emph{K} different data points corresponding to each of these \emph{N} classes. Remapping labels and handling consecutive labels are for creating diverse meta-tasksets; concretely, each label in a meta-taskset is remapped to a new label between 0 and the maximum number of unique labels present in the meta-taskset, and task labels are reordered consecutively according to classes. 

\textcolor{black}{
Formally, the meta-dataset ($D_{meta}$) is expressed by a set of meta-data points (up to a total of $g$), as defined in~\Cref{eq-metadt1}. 
Within this set, a particular meta-data point ($D_{z} \in D_{pro}$) is denoted by~\Cref{eq-metadt2}. 
In this equation, $C_i$ corresponds to a target class and the $[X_o, ..., X_p]$ represents the associated data rows. 
The taskset, as defined in~\Cref{eq-taskset}, constitutes a collection of $D_{meta}$, \emph{N}-ways, \emph{K}-shots, and \emph{M} tasks. 
In the following, we explain the process of creating the meta-dataset and meta-taskset with an example. 
}

\textcolor{black}{
\begin{equation}\label{eq-metadt1}
D_{meta} = \{D_1, D_2, ..., D_g\}
\end{equation}
}

\textcolor{black}{
\begin{equation}\label{eq-metadt2}
D_{z} = \{(C_1, [X_o, ..., X_p]), (C_2, [X_o, ..., X_p]), ..., (C_n, [X_o, ..., X_p])\}
\end{equation}
}

\textcolor{black}{
\begin{equation}\label{eq-taskset}
T = \{D_{meta}, N_{ways}, K_{shots}, M_{tasks}\}
\end{equation}
}

As illustrated in \Cref{fig:example}, the preprocessed data obtained from the previous step (\Cref{sec:dataprep}) has data rows with unique indexes (\emph{Id1}, \emph{Id2}, ...) and classes/labels (e.g., \emph{C1} and \emph{C2}). For demonstration, we remap each class with labels. For meta-dataset creation, each class label with all associated data indexes is stored in a table. Suppose we need to create tasks containing two data samples according to 2-way and 2-shot methods. From the meta-dataset, first, we randomly pick two classes (2-way) with two data samples (the top-right side of \Cref{fig:example}). Next, for the 2-shot, we randomly select two data samples for each class (\emph{C1} and \emph{C2}). This results in creating one task, as shown on the bottom-right side of \Cref{fig:example}. The same process is utilized to create \emph{M} tasks in meta-taskset for \emph{N}-way and \emph{K}-shot methods.

\subsubsection{Determining Model Architecture}\label{subsec:model}
We define a neural network model architecture with linear layers, an input layer, one hidden layer, and an output layer. 
The size of the input layer depends on the number of features in the dataset because every medical device has a different set of features. 
We identify the number of features from the dataset during the meta-dataset and meta-taskset creation phase. 
The hidden layer dimension is 128, which is determined based on our pilot experiments for hyperparameters search. 
The output layer dimension corresponds to status codes. 
We use the Sigmoid activation function in our model architecture, which is suitable for linear layered neural networks and classification problems~\cite{daqi2005classification}. 

For each new medical device, the architecture of its model needs to be determined initially before training, based on the features of the specific device. 
In the case of a device upgrade, the trained model corresponding to the earlier device version is loaded for fine-tuning. 
If the device upgrade involves the addition/removal of features, the model architecture can incorporate such changes because the input layer is determined based on the number of features. 
\textcolor{black}{
To formalize the model, we represent the input layer as $A = [A_1, A_2, ..., A_j]^T$, where $a_i \in X_i$ and $||A|| = ||X_i||$. 
The output layer can be denoted as $Y = [Y_1, Y_2, ..., Y_k]^T$, where each $Y_i \in C_i$ and $||Y|| = ||C_i||$.  
Furthermore, the hidden layer can be denoted as $H = [H_1, H_2, ..., H_{128}]^T$, indicating that the hidden layer has a dimension of 128, i.e., $dim(H) = 128$. 
Using these notations, \Cref{eq-model} represents a neural network model. 
Within this model, $f$ symbolizes the activation function (e.g., Sigmoid), $W_{1}$ and $W_{2}$ denote the weight matrices associated with the respective layers, and the bias terms for the hidden and output layers are denoted by $b_1$ and $b_2$ respectively. 
}

\textcolor{black}{
\begin{equation}\label{eq-model}
Y = f(W_{2}f(W_{1} A + b_1) + b_2)
\end{equation}
}

\subsubsection{Training with Meta-Learning Algorithm}\label{subsec:training}

We use a gradient descent-based \emph{meta-optimizer}, named Model-Agnostic Meta-Learning (MAML)~\cite{finn2017model}, which is very efficient to learn a new task, as it only requires a small set of training data and a few fine-tuning steps to learn new tasks~\cite{hospedales2021meta}. 
The \emph{meta-objective} utilized in MAML is few-shot learning in which the objective is to learn from a small amount of training data and train a model that can be easily fine-tuned with a few training steps. 
We leverage the Adam optimizer, which is suitable for gradient-based algorithms~\cite{finn2017model,hospedales2021meta}, to minimize training loss iteratively. 

\textcolor{black}{
The hyperparameters necessary for training must be determined based on a pilot study. 
While several tools such as Optuna~\cite{akiba2019optuna} are available for automated hyperparameter search, to the best of our knowledge, these tools do not support meta-learning algorithms (like MAML) and integration with meta-learning libraries such as \emph{learn2learn}~\cite{arnold2020learn2learn}. 
Therefore, following the common practice for hyperparameter search~\cite{akiba2019optuna}, we set up a pilot study tailored to our specific context. 
% Hence, we conducted a pilot study to determine the model architecture and hyperparameters without these tools. 
During the pilot study, we explored a variety of initial configurations for determining model architecture and hyperparameter values. 
When defining the model architecture, we specified hidden layers in various dimensions, including 16, 32, 64, 128, 256, and 512. 
Following this, we configured different activation functions into our model, such as Sigmoid, Tanh, Softmax, ReLU, and Leaky ReLU. 
When determining hyperparameter values, we employed widely-used loss functions such as Mean Squared Error (MSE), Cross-Entropy, and Kullback-Leibler (KL) Divergence. 
Subsequently, we configured diverse values for \emph{N}-ways, learning rates, time steps for training and adaptation, and the number of iterations. 
Each configuration was set up with different K-shot methods, such as 1-shot, 2-shot, and 5-shot. 
For each run corresponding to a particular configuration, we examined the accuracy. 
Using these accuracy results, we selected model architecture (as discussed in \Cref{subsec:model}) and hyperparameter values that achieved the highest accuracy for our work. 
The pilot study outcomes led us to select the \emph{Cross-Entropy} loss function for MAML. 
The remaining hyperparameter values are provided in \Cref{subsec:expsetup}. 
}

The training process starts with the meta-taskset \textcolor{black}{($T$)}, model \textcolor{black}{($Y$)}, and MAML algorithm for a specified number of iterations. 
In each iteration time step, data from the meta-taskset is sampled randomly. 
This data is divided into \emph{adaptation} and \emph{evaluation} sets, in which adaptation set size is calculated as \emph{N}-ways$\times$\emph{K}-shots and the evaluation set size is the meta-taskset size - adaptation set size. 
The MAML algorithm learns from the adaptation set for a defined number of adaptation steps, which is determined based on the pilot experiment results.
After the adaptation steps are completed, the evaluation set is used to estimate the prediction accuracy of the MAML learner using the loss function. 
This process continues until the maximum number of iterations is reached. The maximum number of iterations can be determined from model accuracy and loss function values. 
If there are minor improvements in model accuracy and loss function values are close to zero repeatedly, this indicates the maximum limit is reached. 
At the end of the training process, we save the trained model, which is used as a part of a DT corresponding to the device (\emph{Dx}). 
In the case of another device, we repeat the \emph{Data Generation}, \emph{Data Preparation}, and \emph{Meta-Learning} phases to train a model for a new device. 
Training a model from scratch is only initially needed for a new medical device. 
If the medical device upgrades, the trained model is used for fine-tuning with few-shot methods.

\subsection{Building Digital Twins}\label{sec:builddt}
Each medical device has a unique serial number (SN) that is used to integrate with the healthcare IoT application and storage to preserve the device configurations. 
The default configurations are loaded from storage at startup to initialize a device. 
Moreover, each device uses APIs for request handling and JSON format for data exchange during requests. 

In this phase, we build DTs while considering medical devices' integration and operating mechanisms.  
%First, we utilize device (\emph{Dx}) schema, \emph{Dx} SNs, and a trained model to build an integratable and operatable DT. 
To do this, we first use the \emph{Dx} schema to identify APIs and JSON format for DTs to process requests from the healthcare IoT application, assign a physical device's SN to a DT, and create DT APIs with the same SN. 
Next, we load the trained model from the meta-learning phase to operate a DT. 
At the beginning of the DT operation, we use default configurations from \emph{Dx} schema to create the initial state of a DT. 
The DT's state changes during request handling and communicating with the physical device. 

For testing at a large scale, it is desired to integrate multiple DTs representing multiple medical devices with the healthcare IoT application. 
We use a list of \emph{Dx} SNs to create multiple DTs representing physical devices of \emph{Dx} type. 
For each SN, we clone the trained model, create storage to hold the state of \emph{Dx}, create APIs for integration with a healthcare IoT application, and define JSON format for data exchange during request handling. 
\textcolor{black}{
It is important to note that the space and memory requirements for cloning the model depend on multiple factors, including model complexity (number of layers, size of each layer, and data types of model parameters), training data, machine specifications, operating system, and the specific deep learning library utilized. 
Similarly, the space and memory requirements for the DT storage rely on the amount of a specific device's data (i.e., device configurations) that needs to be saved, along with the JSON library being used. 
In our work, the model architecture (\Cref{subsec:model}) was not complex, containing three linear layers, the size of hidden layer 128, and float64 data type for model parameters. 
Moreover, the devices utilized do not possess a large amount of configuration data, thus eliminating the need for extensive storage space. 
During our experiments, we observed that the memory consumed for model cloning was up to 13.8 KBs, while the memory used for DT storage was up to 17.1 KBs. 
Furthermore, we noted that the space utilized by the cloned model ranged from 16 to 27 KBs, and the space used for the DT storage ranged from 2 to 749 Bytes. 
}

The DTs start operating with default configurations after creating a specified number of clones of the trained model. 
Making clones of a trained model is beneficial in the case of only one device with specific SN upgrades. 
For this, we need to fine-tune the trained model of only a particular DT by repeating phases 1, 2, and 3 with few-shot learning~\cite{finn2017model}. 
While repeating these phases for fine-tuning, we use the updated device \emph{Dx} schema to generate a dataset and set few-shot parameters of MAML. 
Fine-tuning with MAML works with a small amount of dataset~\cite{finn2017model}; thus, the data generation phase can be configured to collect sufficient data. 
This way, the model is fine-tuned for an upgraded device with a small amount of data and a few shots for learning.

\subsection{DT Request Handler}\label{sec:dtrequest}
During testing with a physical device in place, the IoT application makes API requests (via HTTP) to obtain and update the device configurations using HTTP methods GET and POST, respectively. 
Similar to the communication mechanism supported by physical devices, this phase creates a DT request handler using APIs and JSON format produced as outputs of the previous phase.  
Specifically, we utilize HTTP methods GET and POST to handle API requests for a DT and use JSON data interchange format to create and send responses. 
Each response consists of a JSON object with DT configurations and an HTTP status code representing success or failure.

A GET request obtains current device configurations, e.g., medication plans or settings. 
To handle GET requests for a particular DT API, first, we validate the request to check if the required configurations are available for the device. 
If not, return a response with an error message and an HTTP error code. 
If the required configurations are available, we obtain these configurations from storage and return a response with a JSON object and an HTTP success code. 
To handle POST requests for a specific DT API, we retrieve a JSON object from the request body. 
We parse the JSON object to accumulate the values for each configuration parameter, which are supplied to a DT's trained model to get the output. 
In the case of a success code as output, the device configurations are updated with these values, and a response is returned with DT's current configurations in JSON and an HTTP success code. 
In the case of a failure code as output, a response with an error message and an HTTP error code is returned without updating DT's configurations.

\subsection{DT to Device Communication}\label{sec:dt2pt}
DTs are desired to operate with healthcare IoT applications in place of physical devices to enable rigorous testing. In this context, DTs may be required to communicate with their corresponding physical devices in cases such as calibrating the DT model. 
Therefore, in this phase, we facilitate establishing a communication channel between a DT and a physical device. 

% DT APIs are also used to establish communication with physical devices.
To allow a DT to communicate with its corresponding physical device, we use device APIs to send requests and JSON format for data exchange. 
For each GET request received from a healthcare IoT application, we send a GET request to the physical device. 
Based on the response received from the physical device, we update the DT's configurations and return a response to the healthcare IoT application.
To handle a POST request, we process it using the DT model and get its output. 
We also send a POST request to the physical device with the same JSON object (received from the healthcare IoT application). 
We compare the response received from the physical device with the output of DT. 
Regarding output differences, we update DT's configuration using physical device configurations, store JSON configurations for fine-tuning the DT model, and return a response to the healthcare IoT application.

\section{Empirical Evaluation}\label{experiment}
\textcolor{black}{
The aim of this evaluation is to assess the effectiveness of our proposed approach (\approach{}) in addressing the key challenges (specifically, \emph{evolution} and \emph{large-scale testing}) outlined in \Cref{industrycontext}. 
Considering this, 
}
we evaluate \approach{} from four aspects: (i) its ability to generate DTs and adapt them to different types of medical devices, (ii) the fidelity of DTs generated from scratch with training and adapted from other DTs with fine-tuning, (iii) the scalability in simulating DTs of multiple devices, and (iv) time cost involved in generating and adapting DTs. 
We assess each aspect by comparing the performance of few-shot learning methods, i.e., 1-, 2-, and 5-shot utilized for generating, adapting, and operating DTs. 
Based on these aspects, we formulate the following \textcolor{black}{four} research questions (RQs). 
\textcolor{black}{
Among these research questions, RQ1, RQ2, and RQ4 are associated with the \emph{evolution} challenge, while RQ4 is related to the \emph{large-scale testing} challenge. 
}

\begin{itemize}
    \specialitem{darkgray!90}{RQ1} \hlc[gray!7!white]{What is the performance of \approach{} in generating DTs and adapting them to different types of medical devices?}\\
    To address this RQ, we apply \approach{} to generate DTs from scratch with training (\iDts{}) and adapt DTs (\aDts{}) through fine-tuning. We then observe \approach{}'s performance in terms of the number of shots required to precisely learn the behavior of different medical devices during the creation of \iDts{} and \aDts{}. 
    % We apply \approach{} to generate DTs (\iDts{}) and observe its effectiveness in terms of the number of shots required to adapt DTs (\aDts{}) for different medical devices.
    \specialitem{darkgray!90}{RQ2} \hlc[gray!7!white]{What is the fidelity of the DTs generated and adapted by \approach{} for a variety of medical devices?}\\
    In this RQ, we assess DTs' fidelity regarding the behavioral similarity among \iDts{}, \aDts{}, and their corresponding physical devices (PDs). The aim is to analyze if DTs are interchangeable with their respective medical devices. 
    \specialitem{darkgray!90}{RQ3} \hlc[gray!7!white]{How does the fidelity of multiple DTs vary while simulating various medical devices?} \\
    This RQ aims to evaluate \approach{}'s scalability in generating and operating multiple \iDts{} and \aDts{} in different batch sizes. We also analyze whether the fidelity is consistent when operating multiple DTs simultaneously. 
    \specialitem{darkgray!90}{RQ4} \hlc[gray!7!white]{What is the time cost associated with generating \iDts{} and adapting \aDts{}?} \\
    The objective of this RQ is to examine the time cost involved in generating DTs from scratch through training, as well as adapting these DTs to different devices and their subsequent versions using fine-tuning methods. 
\end{itemize}

\subsection{Implementation}
We implemented \approach{} in Python, relying on Scikit-learn~\cite{scikit-learn} for data preparation, and used the PyTorch framework and the meta-learning library \emph{learn2learn}~\cite{arnold2020learn2learn}. 
We developed \emph{DT Request Handler} and \emph{DT to Device Communication} using the Flask framework~\cite{flask}. 
For DT APIs, we utilized Flask-RESTful (compatible with Flask) and JSON to realize the data exchange among API requests. 
We provided an open-source generic framework implementation \textcolor{black}{on GitHub~\cite{Sartaj_MeDeT_2023}}, excluding the association with real devices and healthcare IoT applications due to non-disclosure agreements.  

\subsection{Medical Devices}
We employed five medical devices including three medicine dispensers (i.e., \emph{\Karie{}}~\cite{karie}, \emph{\Medido{}}~\cite{medido}, and \emph{\Pilly{}}~\cite{pilly}) and two measurement devices (i.e., \emph{Blood Pressure Meter (\BPMeter{})} and \emph{Pulse Oximeter (\POximeter{})}), provided by Oslo City's health department. 
These medical devices have been extensively used, representing low- to high-end and basic to advanced featured devices. A brief description of each device is given as follows.

%\textbf{\Karie{}.} 
\textbf{\Karie{}} is an advanced automatic medicine dispenser that allows users to customize settings with various options, e.g., alarm tune, volume, and time zone, through a graphical interface (for patients) or using a healthcare IoT application (for caregivers). 
It automatically loads a medication plan from a healthcare IoT application and follows 
it to generate an alarm for the patient and dispense medicine. 
It also notifies the IoT application about the device status and missed and intake doses, which is important for caregivers. 

\textbf{\Medido{}} is an automatic medicine dispenser that allows caregivers/medical professionals to change settings using an IoT application. In addition, it supports alarm and medication settings and three languages. Its basic function is to load medication plans from the healthcare IoT application, deliver medicine doses according to the plan, and report to the healthcare IoT application about the medicine delivery issue, intake dose, and forgotten medicine. 

\textbf{\Pilly{}} is a simple, low-end medicine dispenser with basic medication functionalities such as alarm duration, volume, and intake modes. 
It allows a fixed medication plan with a specified number of doses. 
These settings can only be updated from a healthcare IoT application, which informs the healthcare IoT application about medicine intake and forgotten medicine. 

\textbf{Blood Pressure Meter (\BPMeter{}) } is an advanced user-friendly device designed to provide precise and reliable blood pressure readings. 
It features a clear display of systolic and diastolic blood pressure measurements and pulse rate. 
The device can store up to 200 past readings, enabling users to track their blood pressure tendencies over time. 
The device operates on battery power, making it portable for travel. 

\textbf{Pulse Oximeter (\POximeter{})} is a health monitoring device designed to provide accurate and instantaneous measurements of oxygen saturation levels in the blood. 
It is compact, lightweight, and easy to use, making it ideal for home and travel use.
It also features a low power consumption design, an automatic power-off function, and can provide up to 30 hours of continuous use, ensuring high uptime.

\begin{table*}[!t]
	\centering
	\noindent
	\caption{Digital twin adaptation scenarios for two types of adaptations: (i) device-to-device adaptations, illustrating adaptations across similar nature of devices like medicine dispensers; and (ii) adaptations from one version of a device to its subsequent versions, reflecting upgrades or updates in the device.}
	\begin{tabular}{p{.12\textwidth}p{.35\textwidth}p{.40\textwidth}}\toprule
		\multicolumn{1}{l }{\textbf{}} & \multicolumn{2}{c }{\textbf{DT adaptation types}}  \\ 
		\cmidrule(lr){2-3}
		\multicolumn{1}{ l }{\textbf{Device}} & \textbf{Adaptation across devices} & \textbf{Adaptation across versions}\\ 
		\cmidrule(lr){1-1}
		\cmidrule(lr){2-2}\cmidrule(lr){3-3}
		\multicolumn{1}{ l }{\Karie{} (K)}&\makecell[l]{\KarieIdtVi{} $\rightarrow$ \MedidoAdtK{},\\ \KarieIdtVi{} $\rightarrow$ \PillyAdtK{}}&\makecell[l]{\KarieIdtVi{} $\rightarrow$ \KarieAdtVii{},\\ \KarieAdtVii{} $\rightarrow$ \KarieAdtViii{},\\ \KarieAdtViii{} $\rightarrow$ \KarieAdtViv{}}\\
        \arrayrulecolor{black!20}\cmidrule[0.05pt]{2-3}
		\multicolumn{1}{ l }{\Medido{} (M)}&\makecell[l]{\MedidoIdtVi{} $\rightarrow$ \KarieAdtM{},\\ \MedidoIdtVi{} $\rightarrow$ \PillyAdtM{}}&\makecell[l]{\MedidoIdtVi{} $\rightarrow$ \MedidoAdtVii{},\\ \MedidoAdtVii{} $\rightarrow$ \MedidoAdtViii{},\\ \MedidoAdtViii{} $\rightarrow$ \MedidoAdtViv{}}\\
        \arrayrulecolor{black!20}\cmidrule[0.05pt]{2-3}
        \multicolumn{1}{ l }{\Pilly{} (P)}&\makecell[l]{\PillyIdtVi{} $\rightarrow$ \KarieAdtP{},\\ \PillyIdtVi{} $\rightarrow$ \MedidoAdtP{}}&\makecell[l]{\PillyIdtVi{} $\rightarrow$ \PillyAdtVii{},\\ \PillyAdtVii{} $\rightarrow$ \PillyAdtViii{},\\ \PillyAdtViii{} $\rightarrow$ \PillyAdtViv{}}\\
		\arrayrulecolor{black!20}\cmidrule[0.05pt]{2-3}
        \multicolumn{1}{ l }{\BPMeter{} (B)}&\makecell[l]{\BPMeterIdtVi{} $\rightarrow$ \POximeterAdtB{}}&\makecell[l]{\BPMeterIdtVi{} $\rightarrow$ \BPMeterAdtVii{},\\ \BPMeterAdtVii{} $\rightarrow$ \BPMeterAdtViii{},\\ \BPMeterAdtViii{} $\rightarrow$ \BPMeterAdtViv{}}\\
        \arrayrulecolor{black!20}\cmidrule[0.05pt]{2-3}
        \multicolumn{1}{ l }{\POximeter{} (O)}&\makecell[l]{\POximeterIdtVi{} $\rightarrow$ \BPMeterAdtO{}}&\makecell[l]{\POximeterIdtVi{} $\rightarrow$ \POximeterAdtVii{},\\ \POximeterAdtVii{} $\rightarrow$ \POximeterAdtViii{},\\ \POximeterAdtViii{} $\rightarrow$ \POximeterAdtViv{}}\\
		\arrayrulecolor{black}\bottomrule
	\end{tabular}
	\label{tab:adts}
\end{table*}

\begin{table}[!t]
	\centering
    \small
	\noindent
	\caption{Comparisons and analyses for each RQ}
	% \begin{tabular}{l p{4.2cm} p{5.7cm}}
    \begin{tabular}{p{.1\textwidth}p{.28\textwidth}p{.37\textwidth}}
        \toprule
		\multicolumn{1}{ l }{\textbf{RQ}} & \textbf{Comparison} & \textbf{Metrics and Statistical Analyses} \\%& \textbf{Statistical Test} \\ 
		\cmidrule(lr){1-1}\cmidrule(lr){2-2}\cmidrule(ll){3-3} %\cmidrule(ll){4-4}
		\multicolumn{1}{l}{1} &\iDts{} and \aDts{} independently& Precision, Recall, F1-score, Mean, and STD\\
		\multicolumn{1}{l}{2} &\iDts{} with \aDts{} and both with PDs & \% Similarity, Wilcoxon test, and Cliff's Delta ($\delta$)\\
        \multicolumn{1}{l}{3} &\iDts{} with \aDts{} and both with PDs& \% Similarity, Wilcoxon test, and Cliff's Delta ($\delta$) \\
        \multicolumn{1}{l}{4} &\iDts{} and \aDts{}& Time cost\\
		\bottomrule
	\end{tabular}
	\label{tab:rqmapping}
\end{table}

\subsection{Evaluation Setup, Execution, and Metrics}
\subsubsection{Setup.}\label{subsec:expsetup}
%We created an experiment setup considering the constraints prescribed by Oslo City's technical team.
We used five medical devices in our experiment, \Karie{}, \Medido{}, \Pilly{}, \BPMeter{}, and \POximeter{}, integrated with a real-world healthcare IoT application and built by different vendors. 
Additionally, we selected four evolving software versions for these devices. 
Given that each device is equipped with dedicated APIs for integration and communication, we relied on the API schema and documentation provided by Oslo City's healthcare department to facilitate communication with these medical devices. 
Using devices' API information, we configured \approach{}'s \emph{Data Generation} phase to compile datasets for each device. 
Each medical device's service provider permits a specific number of API requests, and exceeding the limit may block further requests or damage a medical device. 
Therefore, this phase was set to run for a specified duration with a delay between two consecutive API requests. 
In particular, we executed the \emph{Data Generation} phase for eight hours for training with the base version and four hours for adaptation with each evolving version for every device, incorporating a three-second delay between two successive requests. 
\textcolor{black}{
Note that the medical devices with a stable software version installed are typically not flaky. 
This means that sending the same request multiple times typically results in the same response for a stable version of a device, as observed in our experiments. 
However, in the case of a new version (i.e., with a hardware or software upgrade), responses may vary compared to the older version. 
Despite this variation in responses across different releases, it does not impact the data collected for a particular device version. 
}

\begin{figure}[htbp]
\centerline{\includegraphics[width=8.2cm, height=4.7cm, keepaspectratio]{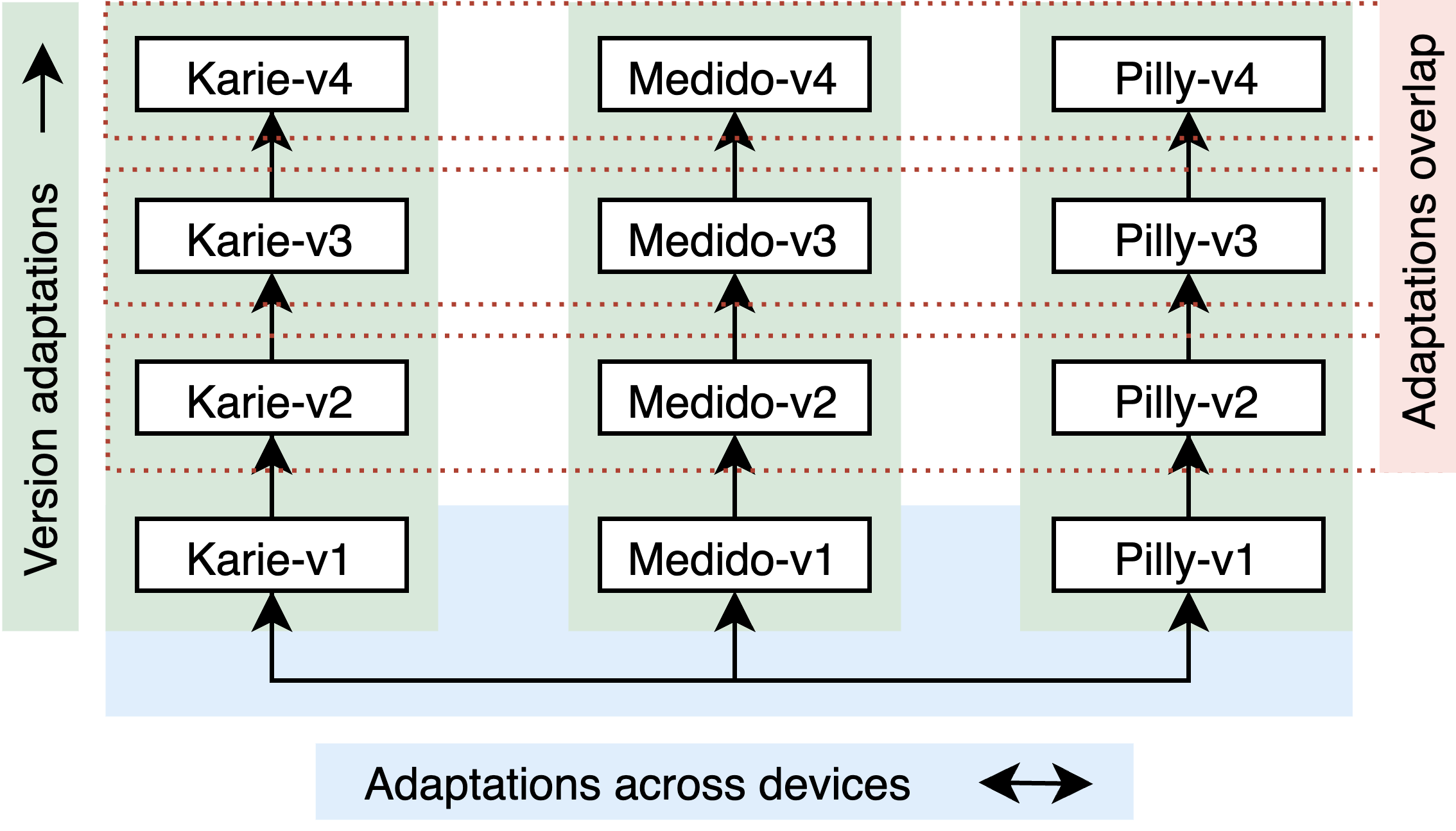}}
\caption{
\textcolor{black}{DT adaptations employed in our experiment: it demonstrates adaptations across devices (\Karie{}, \Medido{}, and \Pilly{}) using their based version (\emph{v1}), and depicts the progression of version adaptations for each device ranging from version \emph{v1} to \emph{v4}. 
}}
\label{fig:adaptations}
\end{figure}

Initially, we used each device's dataset for training to generate an \iDt{} for every device, i.e., \KarieIdtVi{}, \MedidoIdtVi{}, \PillyIdtVi{}, \BPMeterIdtVi{}, and \POximeterIdtVi{}. 
We then used each device \iDt{} to create \aDts{} considering two types of adaptations: (i) device-to-device adaptations, representing adaptations across devices of similar nature like medicine dispensers; and (ii) adaptations from one version of a device to its subsequent versions, reflecting upgrades or updates in the device software. 
\Cref{tab:adts} outlines scenarios for both types of DT adaptations. 
For the device-to-device adaptation, we created separate adaptation scenarios for medicine dispensers and measurement devices, due to the lack of common features between these device types. % because both device types have no common features. 
In addition, adapting between two distinct device types is impractical. 
In the case of the medicine dispenser, we utilized \KarieIdtVi{} to adapt to \Medido{} and \Pilly{}, resulting in the creation of \MedidoAdtK{} and \PillyAdtK{} respectively. 
Similarly, we utilized \MedidoIdtVi{} to create \KarieAdtM{} and \PillyAdtM{}, as well as \PillyIdtVi{} to generate \KarieAdtP{} and \MedidoAdtP{}. 
In the case of measurement devices, we used \BPMeterIdtVi{} to adapt to \POximeterAdtB{}, and vice versa. 
For adaptations across versions, we employed the base version (\emph{v1}) to adapt to the successive version  (\emph{v2}) of the same device. 
Using \emph{v2}, we created an adaptation for the next version, i.e., \emph{v3}, and continued this process for subsequent versions. 
\textcolor{black}{
As illustrated in \Cref{fig:adaptations}, for \Karie{} version adaptations, we used \KarieIdtVi{} to adapt to \KarieAdtVii{}, used \KarieAdtVii{} to adapt to \KarieAdtViii{}, and used \KarieAdtViii{} to adapt to \KarieAdtViv{}. 
Similar version adaptations were conducted for \Medido{}, \Pilly{}, \BPMeter{}, and \POximeter{}. 
For the adaptations across devices, we only used the base version (\emph{v1}) of both types of devices when adapting from one device to another. 
The primary reason was to analyze the individual impact of each type of adaptation, i.e., version and device adaptation. 
As depicted in \Cref{fig:adaptations}, while it is possible to adapt from one device to another using \emph{v2}, this would result in a device adaptation that has already undergone version adaptation (i.e., from \emph{v1} to \emph{v2}). 
This type of device adaptation overlaps with the version's adaptation. 
Consequently, it becomes challenging to distinguish whether any performance improvement or degradation is due to version or device adaptation. 
In our experiment, we opted for the base version (\emph{v1}) when performing adaptations across devices. 
}

We selected hyperparameters adhering to the common practice with meta-learning experiments~\cite{nam2023stunt}. 
In particular, we performed a pilot study for hyperparameters search with a partial dataset (generated for one hour) with different combinations of parameters including \emph{N}-shot, \emph{K}-way, learning rate, optimizer, loss function, adaptation steps, time steps for training and adaptation, and the number of iterations. 
To analyze the performance of few-shot methods in generating and adapting DTs with \approach{}, we opted for {\fontfamily{qcr}\selectfont n\_shots=1,2,5} inspired by existing works~\cite{finn2017model,nam2023stunt}. 
The remaining \textcolor{black}{hyperparameter values determined based on the pilot study (see Section \ref{subsec:metalearningpphase})} are {\fontfamily{qcr}\selectfont meta-learning rate=0.001}, 
{\fontfamily{qcr}\selectfont MAML learning rate=0.05},  
{\fontfamily{qcr}\selectfont time steps=256(train)\&64(adapt)}, and {\fontfamily{qcr}\selectfont adaptation\_steps=1}.

To analyze the performance of each \iDt{} and \aDt{} for RQ1, we generated a testing dataset by executing \emph{Data Generation} phase for four hours.  
The collection of this dataset was carefully timed to ensure data variety and prevent overlap with the training and adaptation datasets. 
Mainly, we initiated the data collection for the testing dataset a week after we had finished gathering data for the training and adaptation stages. 
The objective was to collect an independent testing dataset to enable unbiased performance analysis of the \iDt{} and \aDt{}.
\textcolor{black}{
For the training phase, we employed five datasets, each corresponding to a different medical device. 
Furthermore, for adaptations across versions, we utilized four additional datasets per device, each corresponding to one of four versions (i.e., v1--v4). 
The datasets used in the experiments were systematically generated using \approach{}'s \emph{Data Generation} phase. 
Although random data could potentially be used for comparison, it is important to note that machine learning models typically perform poorly when trained with such data~\cite{goodfellow2016deep,bishop2006pattern}. 
This is primarily because machine learning models try to find patterns in the training data~\cite{bishop2006pattern}. 
If the data is random and there are no patterns to find, the model cannot learn to make accurate predictions. 
This scenario was also observed during our pilot experiment. 
Consequently, we could not incorporate random datasets for comparison as it would not result in a fair comparison. 
}

For RQ2, we integrated the \iDts{} and \aDts{} of \Karie{}, \Medido{}, \Pilly{}, \BPMeter{}, and \POximeter{} with the healthcare IoT application. 
We used APIs and JSON objects of \iDt{}, \aDt{}, and physical devices to create API requests with randomly generated within-range and out-of-range values (same as for \textit{Data Generation} in \Cref{sec:datagen}). 
Next, we sent the same API request to each \iDt{}, \aDt{}, and their corresponding \Karie{}-PD, \Medido{}-PD, \Pilly{}-PD, \BPMeter{}-PD, and \POximeter{}-PD. 
Each \iDt{}, \aDt{}, and PD processed the API requests independently and returned responses in JSON, which are used to analyze the fidelity of \iDts{} and \aDts{}. 

For RQ3, first, we generated 1000 \iDts{} and 1000 \aDts{} for each of \Karie{}-PD, \Medido{}-PD, \Pilly{}-PD, \BPMeter{}-PD, and \POximeter{}-PD. % according to the design in \Cref{tab:adts}. 
Next, we ran 1000 DTs (i.e., \iDts{} and \aDts{}) of each device concurrently and in different size batches, i.e., 100, 200, 400, 600, 800, and 1000 DTs. 
Finally, we analyzed the fidelity of all \iDts{} and \aDts{} by comparing them with their corresponding PDs using the returned responses.  

To record time for RQ4, we captured the initial timestamp before starting the training and adaptation processes. 
Upon completion, we noted the final timestamp for each execution. 
Using these timestamps, we calculated the time spent during training to create \iDts{} and during adaptations to generate \aDts{}.

\subsubsection{Execution.}
We executed experiments using a machine with an 8-core CPU, 24 GB of RAM, and macOS as an operating system. 
For the stopping criteria, we set the total number of iterations to 5000 for training and 1000 for adaptation.  
We terminated the training and adaptation process when no significant improvement in the model was observed after a certain number of iterations \textcolor{black}{(100 for training and 20 for adaptation)}.

\subsubsection{Metrics and Statistical Tests.}
\Cref{tab:rqmapping} presents comparisons, metrics, and statistical tests utilized for addressing each RQ. 
For RQ1, we calculated precision, recall, and F1-score for all \iDts{} and \aDts{}. 
We also reported descriptive statistics using mean and standard deviation (STD). 
\textcolor{black}{
To compute precision, recall, and F1-score, we consider all classes represented by success and failure status codes (specifically, 2XX, 4XX, and 5XX). 
These codes are obtained from the responses returned by \iDts{} and \aDts{}. 
Among \emph{n} total classes, precision ($P_i$), recall ($R_i$), and F1-score ($F_i$) for a particular class are calculated using \Cref{eq1,eq2,eq3}. 
In these equations, $TP_i$ represents \textit{true positives}, $FP_i$ denotes \textit{false positives}, and $FN_i$ signifies \textit{false negatives} for a given class \emph{i}. 
Upon calculating individual $P_i$, $R_i$, and $F_i$ for each class, the overall precision, recall, and F1-score are computed by taking an average of values corresponding to all classes (a total of $n$). 
The formula for precision is given in \Cref{eq4}, for recall in \Cref{eq5}, and for the F1-score in \Cref{eq6}. 
}

\textcolor{black}{
\begin{equation}\label{eq1}
P_i = \frac{TP_i}{\left(TP_i + FP_i\right)}
\end{equation}
}

\textcolor{black}{
\begin{equation}\label{eq2}
R_i = \frac{TP_i}{\left(TP_i + FN_i\right)}
\end{equation}
}

\textcolor{black}{
\begin{equation}\label{eq3}
F_i  = \frac{2 \times (R_i \times P_i)}{R_i + P_i}
\end{equation}
}

\textcolor{black}{
\begin{equation}\label{eq4}
Precision=\frac{\sum^n_{i=1}{P_i}}{n}
\end{equation}
}

\textcolor{black}{
\begin{equation}\label{eq5}
Recall=\frac{\sum^n_{i=1}{R_i}}{n}
\end{equation}
}

\textcolor{black}{
\begin{equation}\label{eq6}
F1-score=\frac{\sum^n_{i=1}{F_i}}{n}
\end{equation}
}

To analyze the similarity between a DT (\iDt{}/\aDt{}) and a PD (for RQ2 and RQ3), we employed the Hamming distance to calculate precise differences while comparing the response strings of a DT and a PD. 
Using the Hamming distance fits our scenario well, as it effectively compares fixed-length strings while considering minor variations.
\textcolor{black}{
We calculate the similarity between two response strings using \Cref{eq7} and \Cref{eq8}. 
Assume that \emph{x} represents the response string of a DT and \emph{y} represents the response string of a PD, each having a length of \emph{m}. 
The Hamming distance between \emph{x} and \emph{y} is calculated using \Cref{eq7}. 
In this equation, the expression $|x_i-y_i|$ yields a value of 1 when $x_i$ is not equal to $y_i$, and 0 otherwise. 
The overall Hamming distance value approaching 1 signifies a high degree of dissimilarity between the two strings. 
As the Hamming distance measures the dissimilarity between two strings, it returns a larger value when the strings are more dissimilar.  
Therefore, to calculate the percentage similarity between two responses, we employ \Cref{eq8}. 
In this equation, we subtract the Hamming distance from 1 to get a similarity score and then multiply by 100 to obtain the percentage similarity. 
}

\textcolor{black}{
\begin{equation}\label{eq7}
H_d(x, y) = \sum_{i=1}^{m} |x_i-y_i|
\end{equation}
}

\textcolor{black}{
\begin{equation}\label{eq8}
\%~Similarity = \left(1 - H_d(x, y)\right) * 100
\end{equation}
}

For RQ2 and RQ3, we also statistically analyzed the similarities between each DT and its corresponding medical device to test the hypothesis that the DT's behavior is similar to the medical device. 
We employed non-parametric tests for hypothesis testing, following guidelines by Arcuri and Briand~\cite{arcuri2011practical}. 
Specifically, we used the Wilcoxon signed-rank test with a significance level ($\alpha$) 0.05 and Cliff's Delta ($\delta$) as an effect size measure. 
For RQ4, we computed the total time duration for creating \iDts{} and \aDts{}. 

%% Results tables

% RQ1 table
\begin{table*}[!t]
	\centering
    \small
	\noindent
	\caption{RQ1 results for \iDts{} and \aDts{} of each device. An \iDt{} refers to a DT generated from scratch; an \aDt{} refers to an adapted DT with \approach{}. A DT adaption scenario is encoded with the convention of DeviceName-aDT-M/P/K, where M, P, and K refer to \Medido{}, \Pilly{}, and \Karie{}, respectively. For example, \PillyAdtM{} means a DT of \Pilly{} adapted from \Medido{}'s \iDt{} with \approach{}. \textcolor{black}{Precision, recall, and F1-score values ranging from 90-95\% are highlighted with a green color, while values falling below 90\% are marked with a red color.}}
	\begin{tabular}{l N N N N N N N N N}
    \toprule
		\multicolumn{1}{l }{\textbf{}} & \multicolumn{3}{c}{\textbf{1-shot}} & \multicolumn{3}{c }{\textbf{2-shot}} & \multicolumn{3}{c }{\textbf{5-shot}} \\ 
		\cmidrule(lr){2-4}
		\cmidrule(ll){5-7}
            \cmidrule(ll){8-10}
		\multicolumn{1}{ l }{\textbf{DT}} & \textbf{Precision} & \textbf{Recall} & \textbf{F1} & \textbf{Precision} & \textbf{Recall}& \textbf{F1}& \textbf{Precision} & \textbf{Recall}& \textbf{F1}\\ 
		\cmidrule(lr){1-1}
		\cmidrule(lr){2-4}
		\cmidrule(ll){5-7}\cmidrule(ll){8-10}
		\multicolumn{1}{ l }{\textbf{\KarieIdtVi{}}} &100.0\%&99.87\%&99.93\%&\cellcolor{red!5}87.27\%&\cellcolor{red!5}78.33\%&\cellcolor{red!5}82.56\%&99.89\%&99.91\%&99.90\%\\
		\multicolumn{1}{ l }{\textbf{\MedidoIdtVi{}}}  &\cellcolor{green!5}93.54\%&\cellcolor{green!5}92.81\%&\cellcolor{green!5}93.18\%&99.48\%&99.03\%&99.26\%&95.98\%&\cellcolor{red!5}87.57\%&\cellcolor{green!5}91.58\%\\
		\multicolumn{1}{ l }{\textbf{\PillyIdtVi{}}} & 100.0\%&99.89\%&99.95\%&\cellcolor{green!5}94.03\%&\cellcolor{red!5}88.09\%&\cellcolor{green!5}90.97\%&100.0\%&98.40\%&99.19\%\\
        \multicolumn{1}{ l }{\textbf{\BPMeterIdtVi{}}} &100.0\%&99.99\%&99.99\%&100.0\%&99.98\%&99.99\%&100.0\%&100.0\%&100.0\%\\
        \multicolumn{1}{ l }{\textbf{\POximeterIdtVi{}}} &100.0\%&100.0\%&100.0\%&100.0\%&100.0\%&100.0\%&100.0\%&100.0\%&100.0\%\\
        
        \arrayrulecolor{black!20}\cmidrule[0.05pt]{1-10}
        
        \multicolumn{1}{ l }{\textbf{\KarieAdtM{}}} &97.60\%&97.60\%&97.60\%&97.72\%&\cellcolor{green!5}95.0\%&96.34\%&98.86\%&98.75\%&98.80\%\\
         \multicolumn{1}{ l }{\textbf{\KarieAdtP{}}} &\cellcolor{green!5}90.85\%&\cellcolor{green!5}90.85\%&\cellcolor{green!5}90.85\%&96.97\%&96.66\%&96.82\%&\cellcolor{red!5}79.62\%&\cellcolor{red!5}70.21\%&\cellcolor{red!5}74.62\%\\
        \multicolumn{1}{ l }{\textbf{\MedidoAdtK{}}}  &99.94\%&99.93\%&99.93\%&99.60\%&99.77\%&99.68\%&97.38\%&97.38\%&97.38\%\\
        \multicolumn{1}{ l }{\textbf{\MedidoAdtP{}}}  &\cellcolor{green!5}90.56\%&\cellcolor{red!5}77.01\%&\cellcolor{red!5}83.23\%&96.36\%&\cellcolor{red!5}57.21\%&\cellcolor{red!5}71.79\%&\cellcolor{green!7}94.19\%&\cellcolor{green!7}93.77\%&\cellcolor{green!7}93.98\%\\
		\multicolumn{1}{ l }{\textbf{\PillyAdtK{}}}  &\cellcolor{red!5}64.24\%&\cellcolor{red!5}64.24\%&\cellcolor{red!5}64.24\%&99.94\%&99.88\%&99.91\%&\cellcolor{red!5}81.38\%&\cellcolor{red!5}81.38\%&\cellcolor{red!5}81.38\%\\
		\multicolumn{1}{ l }{\textbf{\PillyAdtM{}}}  &\cellcolor{green!5}90.31\%&\cellcolor{green!5}90.31\%&\cellcolor{green!5}90.31\%&\cellcolor{red!5}63.88\%&\cellcolor{red!5}41.67\%&\cellcolor{red!5}50.44\%&\cellcolor{green!7}94.88\%&\cellcolor{green!7}93.62\%&\cellcolor{green!7}94.24\%\\
        \multicolumn{1}{ l }{\textbf{\BPMeterAdtO{}}} &100.0\%&99.99\%&99.99\%&100.0\%&99.98\%&99.99\%&100.0\%&\cellcolor{red!5}50.0\%&\cellcolor{red!5}66.67\%\\
        \multicolumn{1}{ l }{\textbf{\POximeterAdtB{}}} &100.0\%&100.0\%&100.0\%&100.0\%&100.0\%&100.0\%&100.0\%&\cellcolor{red!5}50.0\%&\cellcolor{red!5}66.67\%\\
        
        \arrayrulecolor{black!20}\cmidrule[0.05pt]{1-10}
        
        \multicolumn{1}{ l }{\textbf{\KarieAdtVii{}}} &97.59\%&97.59\%&97.59\%&\cellcolor{green!5}93.60\%&\cellcolor{green!5}93.71\%&\cellcolor{green!5}93.66\%&98.07\%&98.07\%&98.07\%\\
        \multicolumn{1}{ l }{\textbf{\KarieAdtViii{}}} &\cellcolor{red!5}83.18\%&\cellcolor{red!5}83.18\%&\cellcolor{red!5}83.18\%&97.98\%&96.67\%&97.32\%&97.73\%&\cellcolor{green!5}94.34\%&96.0\%\\
        \multicolumn{1}{ l }{\textbf{\KarieAdtViv{}}} &\cellcolor{red!5}74.86\%&\cellcolor{red!5}74.86\%&\cellcolor{red!5}74.86\%&100.0\%&\cellcolor{red!5}45.0\%&\cellcolor{red!5}62.07\%&98.82\%&99.44\%&99.13\%\\
        \multicolumn{1}{ l }{\textbf{\MedidoAdtVii{}}} &99.70\%&99.58\%&99.64\%&99.25\%&99.25\%&99.25\%&\cellcolor{green!5}91.48\%&\cellcolor{green!5}93.09\%&\cellcolor{green!5}92.28\%\\
        \multicolumn{1}{ l }{\textbf{\MedidoAdtViii{}}} &98.74\%&98.74\%&98.74\%&99.05\%&97.55\%&98.29\%&97.92\%&98.63\%&98.28\%\\
        \multicolumn{1}{ l }{\textbf{\MedidoAdtViv{}}} &99.88\%&99.76\%&99.82\%&99.79\%&99.78\%&99.79\%&\cellcolor{red!5}87.32\%&\cellcolor{red!5}82.65\%&\cellcolor{red!5}84.92\%\\
        \multicolumn{1}{ l }{\textbf{\PillyAdtVii{}}} &99.62\%&99.41\%&99.52\%&99.52\%&99.05\%&99.28\%&\cellcolor{red!5}72.87\%&\cellcolor{red!5}72.87\%&\cellcolor{red!5}72.87\%\\
        \multicolumn{1}{ l }{\textbf{\PillyAdtViii{}}} &100.0\%&99.89\%&99.95\%&\cellcolor{red!5}80.59\%&\cellcolor{red!5}80.59\%&\cellcolor{red!5}80.59\%&97.87\%&97.25\%&97.56\%\\
        \multicolumn{1}{ l }{\textbf{\PillyAdtViv{}}} &95.45\%&95.45\%&95.45\%&96.89\%&96.55\%&96.72\%&96.81\%&96.81\%&96.81\%\\
        \multicolumn{1}{ l }{\textbf{\BPMeterAdtVii{}}} &100.0\%&99.99\%&99.99\%&100.0\%&99.98\%&99.99\%&100.0\%&100.0\%&100.0\%\\
        \multicolumn{1}{ l }{\textbf{\BPMeterAdtViii{}}} &100.0\%&99.99\%&99.99\%&100.0\%&99.98\%&99.99\%&100.0\%&100.0\%&100.0\%\\
        \multicolumn{1}{ l }{\textbf{\BPMeterAdtViv{}}} &100.0\%&99.99\%&99.99\%&100.0\%&99.98\%&99.99\%&100.0\%&100.0\%&100.0\%\\
        \multicolumn{1}{ l }{\textbf{\POximeterAdtVii{}}} &100.0\%&100.0\%&100.0\%&100.0\%&100.0\%&100.0\%&100.0\%&100.0\%&100.0\%\\
        \multicolumn{1}{ l }{\textbf{\POximeterAdtViii{}}} &100.0\%&100.0\%&100.0\%&100.0\%&100.0\%&100.0\%&100.0\%&100.0\%&100.0\%\\
        \multicolumn{1}{ l }{\textbf{\POximeterAdtViv{}}} &100.0\%&100.0\%&100.0\%&100.0\%&100.0\%&100.0\%&100.0\%&100.0\%&100.0\%\\
        \arrayrulecolor{black}\cmidrule(ll){1-10}
		\multicolumn{1}{ l }{\textbf{Mean}} &95.57\%&95.03\%&95.28\%&96.49\%&\cellcolor{green!5}91.56\%&\cellcolor{green!5}93.38\%&95.75\%&\cellcolor{green!5}91.22\%&\cellcolor{green!5}92.87\%\\
        \multicolumn{1}{ l }{\textbf{STD}} &$\pm$8.42&$\pm$9.05&$\pm$8.68&$\pm$7.61&$\pm$16.21&$\pm$12.41&$\pm$6.92&$\pm$13.96&$\pm$10.35\\ 
		\bottomrule
	\end{tabular}
	\label{tab:rq1results}
\end{table*}

% RQ2 table
\begin{table*}[!t]
	\centering
    \small
	\noindent
	\caption{RQ2 results for fidelity among \iDt{}, \aDt{}, and \PD{}. \iDt{} and \aDt{} represent a generated and adapted DT with \approach{}, whereas \PD{} refers to a physical device. \aDt{}-K/M/P/B/O denote an \aDt{} adapted from \Karie{}, \Medido{}, \Pilly{}, \BPMeter{} (BPM), or \POximeter{} (POxM), respectively. \textcolor{black}{Similarity values ranging from 90-95\% are highlighted with a green color.}}
	\begin{tabular}{p{0.2cm} p{2.4cm} G G G G G G G G G}\toprule
        \multicolumn{2}{l }{\textbf{}} & \multicolumn{3}{c}{\textbf{1-shot}} & \multicolumn{3}{c }{\textbf{2-shot}} & \multicolumn{3}{c }{\textbf{5-shot}} \\ 
		\cmidrule(lr){3-5}
		\cmidrule(ll){6-8}
        \cmidrule(ll){9-11}
		\multicolumn{2}{ l }{\textbf{Comparison}} & \textbf{Sim. \%} & \textbf{\textit{p}-value} & \textbf{Cliff $\delta$}& \textbf{Sim. \%} & \textbf{\textit{p}-value} & \textbf{Cliff $\delta$}& \textbf{Sim. \%} & \textbf{\textit{p}-value} & \textbf{Cliff $\delta$} \\
		\cmidrule(lr){1-2}
		\cmidrule(lr){3-5}
            \cmidrule(ll){6-8}
            \cmidrule(ll){9-11}
            \multirow{6}{*}{\textbf{\rotatebox[origin=c]{90}{\makecell{\Karie{}}}}}
            &\textbf{\iDtVi{} \& \PD{}}&99.63\%&0.45&0.002&95.46\%&0.99&-0.32&99.77\%&0.45&0.003
            \\&\textbf{\iDtVi{} \& \aDtM{}}&97.97\%&0.91&-0.02&95.60\%&0.99&0.34&98.98\%&0.73&-0.01
            \\&\textbf{\iDtVi{} \& \aDtP{}}&\cellcolor{green!5}91.03\%&0.31&0.01&95.31\%&0.98&0.33&99.88\%&1.0&0.16
            \\&\textbf{\aDtM{} \& \aDtP{}}&99.10\%&0.95&-0.029&95.03\%&0.41&-0.001&99.86\%&1.0&-0.17
            \\&\textbf{\aDtM{} \& \PD{}}&97.59\%&0.88&-0.018&95.10\%&0.40&0.032&98.75\%&0.68&-0.008
            \\&\textbf{\aDtP{} \& \PD{}}&\cellcolor{green!5}90.85\%&0.31&0.011&96.67\%&0.30&0.033&99.85\%&1.0&0.16\\
		\cmidrule(ll){1-11}
            \multirow{6}{*}{\textbf{\rotatebox[origin=c]{90}{\makecell{\Medido{}}}}}
            &\textbf{\iDtVi{} \& \PD{}}&\cellcolor{green!5}92.81\%&1.0&0.07&99.03\%&0.72&-0.009&99.34\%&1.0&0.34
            \\&\textbf{\iDtVi{} \& \aDtK{}}&\cellcolor{green!5}92.75\%&1.0&-0.072&98.60\%&0.20&0.014&99.34\%&1.0&-0.35
            \\&\textbf{\iDtVi{} \& \aDtP{}}&\cellcolor{green!5}93.09\%&1.0&0.05&99.40\%&1.0&0.81&99.02\%&1.0&-0.33
            \\&\textbf{\aDtK{} \& \aDtP{}}&95.43\%&1.0&-0.10&99.39\%&1.0&-0.82&\cellcolor{green!5}93.44\%&0.74&-0.02
            \\&\textbf{\aDtK{} \& \PD{}}&99.82\%&0.56&-0.002&99.57\%&0.39&0.004&97.38\%&0.59&-0.006
            \\&\textbf{\aDtP{} \& \PD{}}&95.50\%&1.0&0.10&99.43\%&1.0&0.81&\cellcolor{green!5}93.77\%&0.29&0.014\\
		\cmidrule(ll){1-11}
            \multirow{6}{*}{\textbf{\rotatebox[origin=c]{90}{\makecell{\Pilly{}}}}}
            &\textbf{\iDtVi{} \& \PD{}}&99.79\%&0.42&0.002&98.55\%&1.0&-0.12&98.40\%&0.33&0.007
            \\&\textbf{\iDtVi{} \& \aDtK{}}&96.01\%&1.0&-0.35&98.52\%&1.0&0.12&99.28\%&0.99&-0.16
            \\&\textbf{\iDtVi{} \& \aDtM{}}&\cellcolor{green!5}90.43\%&1.0&-0.10&\cellcolor{green!5}90.75\%&1.0&-0.38&\cellcolor{green!5}94.68\%&0.61&-0.01
            \\&\textbf{\aDtK{} \& \aDtM{}}&\cellcolor{green!5}93.87\%&1.0&-0.26&\cellcolor{green!5}94.12\%&1.0&0.49&99.17\%&0.99&-0.14
            \\&\textbf{\aDtK{} \& \PD{}}&95.98\%&1.0&-0.35&99.88\%&0.47&0.002&99.22\%&0.99&-0.13
            \\&\textbf{\aDtM{} \& \PD{}}&\cellcolor{green!5}90.31\%&1.0&-0.09&\cellcolor{green!5}94.05\%&1.0&-0.49&\cellcolor{green!5}93.62\%&0.49&-0.004\\
        \cmidrule(ll){1-11}
            \multirow{3}{*}{\textbf{\rotatebox[origin=c]{90}{\makecell{BPM}}}}
            &\textbf{\iDtVi{} \& \PD{}}&99.98\%&0.50&-0.0001&99.97\%&0.52&-0.0003&100.0\%&0.5&0.0
            \\&\textbf{\iDtVi{} \& \aDtO{}}&100.0\%&0.5&0.0&100.0\%&0.5&0.0&100.0\%&1.0&0.0
            \\&\textbf{\aDtO{} \& \PD{}}&99.99\%&0.51&-0.0001&99.96\%&0.51&-0.0003&100.0\%&1.0&0.0\\
        \cmidrule(ll){1-11}
            \multirow{3}{*}{\textbf{\rotatebox[origin=c]{90}{\makecell{POxM}}}}
            &\textbf{\iDtVi{} \& \PD{}}&100.0\%&0.5&0.0&100.0\%&0.5&0.0&100.0\%&0.5&0.0
            \\&\textbf{\iDtVi{} \& \aDtB{}}&100.0\%&0.5&0.0&100.0\%&0.5&0.0&100.0\%&1.0&0.0
            \\&\textbf{\aDtB{} \& \PD{}}&100.0\%&0.5&0.0&100.0\%&0.5&0.0&100.0\%&1.0&0.0\\
		% \cmidrule(ll){1-5}
		% \multicolumn{1}{ l }{\textbf{Mean}} &&&\\
  %           \multicolumn{1}{ l }{\textbf{STD}} &&&\\ 
		\bottomrule
	\end{tabular}
	\label{tab:rq2results}
\end{table*}

\begin{table*}[!t]
	\centering
    \small
	\noindent
	\caption{RQ2 results for fidelity among \iDt{}, \aDt{}, and \PD{} across versions. \iDt{} and \aDt{} represent a generated and adapted DT with \approach{}, whereas \PD{} refers to a physical device. \aDt{}-K/M/P/B/O denote an \aDt{} adapted from \Karie{}, \Medido{}, \Pilly{}, \BPMeter{}, or \POximeter{}, respectively. \textcolor{black}{Similarity values ranging from 90-95\% are highlighted with a green color.}}
	\begin{tabular}{p{0.2cm} p{2.4cm} G G G G G G G G G}\toprule
        \multicolumn{2}{l }{\textbf{}} & \multicolumn{3}{c}{\textbf{1-shot}} & \multicolumn{3}{c }{\textbf{2-shot}} & \multicolumn{3}{c }{\textbf{5-shot}} \\ 
		\cmidrule(lr){3-5}
		\cmidrule(ll){6-8}
        \cmidrule(ll){9-11}
		\multicolumn{2}{ l }{\textbf{Comparison}} & \textbf{Sim. \%} & \textbf{\textit{p}-value} & \textbf{Cliff $\delta$}& \textbf{Sim. \%} & \textbf{\textit{p}-value} & \textbf{Cliff $\delta$}& \textbf{Sim. \%} & \textbf{\textit{p}-value} & \textbf{Cliff $\delta$} \\ 
		\cmidrule(lr){1-2}
		\cmidrule(lr){3-5}
        \cmidrule(ll){6-8}
        \cmidrule(ll){9-11}
            \multirow{6}{*}{\textbf{\rotatebox[origin=c]{90}{\makecell{\Karie{}}}}}
            &\textbf{\iDtVi{} \& \aDtVii{}}&97.87\%&0.81&-0.014&95.13\%&0.99&-0.29&98.30\%&0.16&0.017
            \\&\textbf{\aDtVii{} \& \aDtViii{}}&98.89\%&0.98&-0.047&\cellcolor{green!5}93.33\%&0.83&-0.074&\cellcolor{green!5}92.86\%&0.99&-0.07
            \\&\textbf{\aDtViii{} \& \aDtViv{}}&96.44\%&1.0&0.16&96.44\%&0.17&-0.18&\cellcolor{green!5}93.20\%&0.97&0.04
            % \\&\textbf{\iDtVi{} \& \PD{}}&&&&&&&&&
            \\&\textbf{\aDtVii{} \& \PD{}}&97.60\%&0.17&0.016&\cellcolor{green!5}93.33\%&0.68&-0.033&98.07\%&0.80&-0.015
            \\&\textbf{\aDtViii{} \& \PD{}}&98.93\%&0.99&0.063&96.67\%&0.31&0.041&\cellcolor{green!5}94.34\%&0.99&0.056
            \\&\textbf{\aDtViv{} \& \PD{}}&97.38\%&1.0&-0.12&96.57\%&0.80&0.21&98.41\%&0.18&0.016\\
        \cmidrule(ll){1-11}
            \multirow{6}{*}{\textbf{\rotatebox[origin=c]{90}{\makecell{\Medido{}}}}}
            &\textbf{\iDtVi{} \& \aDtVii{}}&\cellcolor{green!5}92.51\%&1.0&0.071&98.28\%&0.60&-0.004&99.39\%&1.0&0.42
            \\&\textbf{\aDtVii{} \& \aDtViii{}}&98.80\%&0.50&-0.001&\cellcolor{green!5}94.62\%&0.99&-0.0538&99.77\%&0.99&-0.10
            \\&\textbf{\aDtViii{} \& \aDtViv{}}&98.92\%&0.49&0.001&95.27\%&0.99&0.045&99.31\%&1.0&0.43
            % \\&\textbf{\iDtVi{} \& \PD{}}&&&&&&&&&
            \\&\textbf{\aDtVii{} \& \PD{}}&99.58\%&0.52&-0.001&99.24\%&0.63&-0.005&\cellcolor{green!5}90.16\%&0.99&-0.08
            \\&\textbf{\aDtViii{} \& \PD{}}&98.74\%&0.51&0.0001&95.16\%&0.99&0.05&97.70\%&0.21&0.023
            \\&\textbf{\aDtViv{} \& \PD{}}&99.76\%&0.54&-0.001&99.78\%&0.45&0.003&99.32\%&1.0&-0.40\\
        \cmidrule(ll){1-11}
            \multirow{6}{*}{\textbf{\rotatebox[origin=c]{90}{\makecell{\Pilly{}}}}}
            &\textbf{\iDtVi{} \& \aDtVii{}}&99.62\%&0.64&-0.004&98.43\%&1.0&-0.11&95.61\%&1.0&0.29
            \\&\textbf{\aDtVii{} \& \aDtViii{}}&99.62\%&0.36&0.004&97.55\%&1.0&0.18&95.68\%&1.0&-0.34
            \\&\textbf{\aDtViii{} \& \aDtViv{}}&95.32\%&1.0&0.04&97.68\%&1.0&-0.22&\cellcolor{green!5}92.02\%&0.98&0.08
            % \\&\textbf{\iDtVi{} \& \PD{}}&&&&&&&&&
            \\&\textbf{\aDtVii{} \& \PD{}}&99.42\%&0.28&0.006&99.05\%&0.71&-0.01&95.38\%&1.0&-0.28
            \\&\textbf{\aDtViii{} \& \PD{}}&99.79\%&0.42&0.002&97.65\%&1.0&-0.19&\cellcolor{green!5}92.02\%&0.98&0.08
            \\&\textbf{\aDtViv{} \& \PD{}}&95.45\%&1.0&-0.042&96.55\%&0.97&0.037&96.81\%&0.5&0.0\\
        \cmidrule(ll){1-11}
            \multirow{6}{*}{\textbf{\rotatebox[origin=c]{90}{\makecell{\BPMeter{}}}}}
            &\textbf{\iDtVi{} \& \aDtVii{}}&100.0\%&0.5&0.0&100.0\%&0.5&0.0&100.0\%&0.5&0.0
            \\&\textbf{\aDtVii{} \& \aDtViii{}}&100.0\%&0.5&0.0&100.0\%&0.5&0.0&100.0\%&0.5&0.0
            \\&\textbf{\aDtViii{} \& \aDtViv{}}&100.0\%&0.5&0.0&100.0\%&0.5&0.0&100.0\%&0.5&0.0
            % \\&\textbf{\iDtVi{} \& \PD{}}&&&&&&&&&
            \\&\textbf{\aDtVii{} \& \PD{}}&99.98\%&0.51&-0.0001&99.97\%&0.52&-0.0003&100.0\%&0.5&0.0
            \\&\textbf{\aDtViii{} \& \PD{}}&99.99\%&0.51&-0.0001&99.98\%&0.52&-0.0003&100.0\%&0.5&0.0
            \\&\textbf{\aDtViv{} \& \PD{}}&99.98\%&0.50&-0.0001&99.97\%&0.51&-0.0003&100.0\%&0.5&0.0\\
        \cmidrule(ll){1-11}
            \multirow{6}{*}{\textbf{\rotatebox[origin=c]{90}{\makecell{\POximeter{}}}}}
            &\textbf{\iDtVi{} \& \aDtVii{}}&100.0\%&0.5&0.0&100.0\%&0.5&0.0&100.0\%&0.5&0.0
            \\&\textbf{\aDtVii{} \& \aDtViii{}}&100.0\%&0.5&0.0&100.0\%&0.5&0.0&100.0\%&0.5&0.0
            \\&\textbf{\aDtViii{} \& \aDtViv{}}&100.0\%&0.5&0.0&100.0\%&0.5&0.0&100.0\%&0.5&0.0
            % \\&\textbf{\iDtVi{} \& \PD{}}&&&&&&&&&
            \\&\textbf{\aDtVii{} \& \PD{}}&100.0\%&0.5&0.0&100.0\%&0.5&0.0&100.0\%&0.5&0.0
            \\&\textbf{\aDtViii{} \& \PD{}}&100.0\%&0.5&0.0&100.0\%&0.5&0.0&100.0\%&0.5&0.0
            \\&\textbf{\aDtViv{} \& \PD{}}&100.0\%&0.5&0.0&100.0\%&0.5&0.0&100.0\%&0.5&0.0\\
		\bottomrule
	\end{tabular}
	\label{tab:rq2resultsversions}
\end{table*}

\subsection{Results and Discussion}
In this section, we discuss the results for each RQ in detail, referring to \Cref{tab:rq1results,tab:rq2results,tab:rq2resultsversions,tab:timeresults} and \Cref{fig:rq3results}. 
% \Cref{tab:rq2results}, \Cref{tab:rq2resultsversions}, \Cref{fig:rq3results}, and \Cref{tab:timeresults}. 
\Cref{tab:rq1results} presents the results for RQ1, which includes precision, recall, and F1-score values for each device's \iDts{} and \aDts{} generated using the 1-, 2-, and 5-shot methods. 
The fidelity comparison results for RQ2, in the context of adaptations across devices and versions, are presented in \Cref{tab:rq2results} and \Cref{tab:rq2resultsversions}.
\Cref{fig:rq3results} illustrates the fidelity results for multiple DTs corresponding to RQ3.
\Cref{tab:timeresults} presents the time cost results relevant to RQ4. 
% Following, we discuss the results for each RQ in detail. 

\subsubsection{RQ1 Results}\label{RQ1}
\textcolor{black}{
As shown in \Cref{tab:rq1results}, for} training with the 1-shot method, precision, recall, and F1-score values for \iDts{} of \Karie{}, \Pilly{}, \BPMeter{}, and \POximeter{} are approximately 100\%. 
The only exception is \MedidoIdtVi{}, where the precision, recall, and F1-score are slightly lower, i.e., around 93\%.
In the 2-shot training method, precision, recall, and F1-score for \MedidoIdtVi{} increase to 99\%. 
For the \iDts{} of \BPMeter{} and \POximeter{}, the performance of the 2-shot method is comparable to that of the 1-shot method. 
In the case of \Karie{}, all values for precision, recall, and F1-score are below 90\%, indicating that the 2-shot method underperforms compared to the 1-shot method. 
In the case of \Pilly{}, precision decreased to 94\%, while recall dropped to 88\%, resulting in an approximate F1-score of 91\%.
For training with the 5-shot method, the results for \Karie{}, \Pilly{}, \BPMeter{}, and \POximeter{} are similar to the 1-shot method. 
Only for \MedidoIdtVi{}, precision is increased, while recall and F1-score are marginally lower compared to those from the 1-shot method.
The overall results for training with different methods show that 1-shot and 5-shot methods are suitable compared to the 2-shot method. 

In the case of device-to-device adaptations of DTs, \Karie{} adapted from \Medido{} (denoted as \KarieAdtM{}) shows consistent performance across all shot methods. 
A similar pattern is observed with \Medido{} adapted from \Karie{} (i.e., \MedidoAdtK{}). 
These findings indicate that each shot method performs comparably when adapting to and from between \Karie{} and \Medido{}. 
For \Karie{} adapted from \Pilly{} (i.e., \KarieAdtP{}) results show approximately 91\% and 97\% for 1-shot and 2-shot methods respectively. 
However, with the 5-shot method, the precision decreases to around 80\%, recall to 70\%, and F1-score to 75\%.
For \Medido{} adapted from \Pilly{} (i.e., \MedidoAdtP{}), precision is greater than 90\%, whereas recall and F1-score have lower values for 1-shot and 2-shot methods. 
In contrast, with the 5-shot method, performance improves, indicating precision, recall, and F1-score values around 94\%. 
For \Pilly{} adapted from \Karie{}, the 2-shot method exhibits superior performance. 
Conversely, when \Pilly{} is adapted from \Medido{}, the 1-shot and 5-shot methods outperform the 2-shot method. 
For both \BPMeterAdtO{} and \POximeterAdtB{}, the 1-shot and 2-shot methods demonstrate comparable performance, achieving 100\% in precision, recall, and F1-score. 
However, in the 5-shot method, while precision is maintained at 100\%, the recall and F1-score are around 50\% and 67\%. 
The overall results indicate that for adaptations from a low-featured device such as \Pilly{} to a high-featured device like \Karie{} and \Medido{} (or the reverse), the 2-shot or 5-shot method is required. 
For the adaptations between similar-featured devices like \Karie{} and \Medido{}, even the 1-shot method is sufficient. 
Moreover, simpler devices like \BPMeter{} and \POximeter{} can be easily adapted with the 1-shot or 2-shot method.

In the case of DT adaptations across versions, all shot methods perform well for \KarieAdtVii{}. 
However, for \KarieAdtViii{} and \KarieAdtViv{}, the 1-shot method underperforms compared to the 2-shot and 5-shot methods. 
Among \Karie{} adaptations from \emph{v1} to \emph{v4}, the 5-shot method outperforms, achieving approximately 96\% in precision, recall, and F1-score. 
For \Medido{} adaptations from \emph{v1} to \emph{v4}, both 1-shot and 2-shot surpass the 5-shot method, each indicating approximately 99\% performance. 
For \Pilly{} adaptations from \emph{v1} to \emph{v4}, the 1-shot method shows outstanding performance, especially when compared to the 2-shot method, which underperforms for \PillyAdtViii{}, and the 5-shot method, which underperforms for \PillyAdtVii{}. 
As for \BPMeter{} and \POximeter{} adaptations, all shot methods demonstrate comparable performance, each achieving around 100\% in precision, recall, and F1-score. 
The overall results for adaptations across versions suggest that for a high-featured device like \Karie{}, which undergoes major changes with each release, it is necessary to employ either the 2-shot or 5-shot method. 
However, for devices that undergo minor changes across evolving releases, the 1-shot method is sufficient to adapt to a newer version. 

The analysis of mean and STD suggests that the 1-shot method demonstrates consistent performance, reflected by a stable mean and a comparatively lower standard deviation than other methods. 
In comparison, the 2-shot and 5-shot methods, despite having high overall precision, show a low average recall and F1 score accompanied by a high standard deviation. 
These results indicate that the 1-shot method exhibits more consistent performance when compared to the 2-shot and 5-shot methods. 
\textcolor{black}{
The overall high performance of the 1-shot method can be attributed to its lower likelihood of overfitting compared to the 2-shot and 5-shot methods. 
Given that 1-shot learning utilizes fewer data points (specifically, a single example for each class), a model employing this method is less likely to overfit. 
Consequently, the results for the 1-shot method demonstrated higher precision, recall, and F1-score compared to the other methods. 
Furthermore, the results indicated some cases (such as \MedidoIdtVi{}, \PillyAdtK, and \KarieAdtViii) where the 1-shot method underperformed, whereas the 2/5-shot method outperformed. 
In these cases, we noted that the quantity of training data was lesser than in other cases. 
As a result, the 1-shot method, which uses a single example per class, could not learn sufficiently. 
The primary cause for the reduced data is device request processing and response time, which sometimes leads to a smaller quantity of data in a given time for data generation. 
Therefore, in such scenarios, the results suggest that the 2/5-shot method might be a more suitable option. 
}

\begin{center}
    \begin{resboxblack}[colbacktitle=gray,left=3.5pt,right=3.5pt,top=3.5pt,bottom=3.5pt]{RQ1 Outcome}
        \textcolor{black}{
        To generate \iDts{} with \approach{}, the 1-shot method with predominantly $\approx$99\% F1-score outperformed the 2-shot or 5-shot methods. 
        For DT adaptations across devices, 2-shot demonstrated higher performance with an F1-score of mostly above 98\%. This is mainly due to device feature variance (low/high number of features) during adaptation. Therefore, the 2-shot method is deemed suitable when adapting between devices with low and high features. 
        For DT adaptations across versions, 1-shot outperformed with an F1-score of $\approx$98\%. This indicates that 1-shot is appropriate for adapting DTs to evolving device variants.  
        In summary, these results demonstrate \approach{}'s capability in dealing with the \emph{evolution} challenge. 
        }
    \end{resboxblack}
\end{center}

\subsubsection{RQ2 Results}\label{RQ2}
\textcolor{black}{
\Cref{tab:rq2results} presents results for adaptations across different devices and \Cref{tab:rq2resultsversions} presents results for adaptations across versions. 
}
\textcolor{black}{
As shown in \Cref{tab:rq2results}, the} similarity of \KarieIdtVi{} with \PD{} is approximately 99\% for the 1-shot and 5-shot methods. 
However, for the 2-shot method, the similarity is a bit lower, at around 95\%.
For \Medido{}, the similarity of \MedidoIdtVi{} with \PD{} is approximately 99\% when using the 2-shot and 5-shot methods. 
However, with the 1-shot method, the similarity is slightly lower at around 92\%. 
The comparison of \PillyIdtVi{} with its corresponding \PD{} demonstrates a similarity exceeding 98\% for all shot methods. 
Similarly, when comparing \BPMeterIdtVi{} and \POximeterIdtVi{} with their respective \PD{}, the similarity is nearly 100\% across all shot methods.

For the device-to-device adaptations of DTs (\Cref{tab:rq2results}), \Karie{} \aDts{} demonstrate a high similarity for the 1-shot method, exceeding 97\% in most cases. 
Nonetheless, \KarieAdtP{} exhibits a slightly lower similarity, approximately 91\%, with \iDt{} and PD. 
With the 2-shot and 5-shot methods, the similarity of \Karie{} \aDts{} is approximately 95\% and 99\% respectively. 
When comparing \Medido{} \iDts{}, \aDts{}, and \PD{}, the 1-shot method shows a similarity of more than approximately 93\%. 
Notably, the 2-shot method demonstrates a higher similarity of around 99\% in all comparisons. 
As for the 5-shot method, the similarity exceeds 97\% in most cases, while in the two cases, it is around 93\%. 
In the case of \Pilly{}, the similarity of \aDt{} adapted from \Medido{} is comparatively lower (around 90\% for 1-shot and 93\% for 2-shot and 5-shot) compared to other \iDts{} and \aDts{}. 
In comparisons involving \iDts{} and \aDts{} of \BPMeter{} and \POximeter{}, a perfect similarity score of 100\% is observed for each shot method.

In the case of DT adaptations across versions (\Cref{tab:rq2resultsversions}), all \Karie{} DT's version adaptations using the 1-shot method show a similarity of 96\% or higher. 
In contrast, the 2-shot and 5-shot methods demonstrate a varied similarity, ranging approximately from 92\% to 98\%. 
The adaptation results for \Karie{} suggest that \aDts{} generated using the 1-shot method exhibit consistent similarity. 
For \Medido{} adaptations using the 1-shot method, the similarity is approximately 99\% in the majority of comparisons. 
However, there is an exception in the comparison between \iDtVi{} and \aDtVii{}, where the similarity is slightly lower at around 92\%. 
In \Medido{} adaptations using the 2-shot method, the similarity exceeds 94\% in all comparisons. 
With the 5-shot method for \Medido{} adaptations, every comparison exhibits a similarity greater than 97\%, except for \aDtVii{} versus \PD{}, where the similarity is slightly lower at around 90\%. 
In the case of \Pilly{} adaptations, both 1-shot and 2-shot methods demonstrate high similarity approximately around 99\%. 
However, when utilizing the 5-shot method for \Pilly{} adaptations, the similarity decreases slightly, varying from 92\% to 96\%, which is marginally lower than that of the 1-shot and 2-shot methods. 
The results for \BPMeter{} and \POximeter{} DT adaptations across versions consistently demonstrate a similarity of approximately 100\% for all shot methods.

The overall results for DTs' similarity comparisons show that \iDts{} generated using all methods exhibit high similarity with their corresponding PDs. 
Moreover, for the adaptions across devices and versions, \aDts{} generated using the 1-shot method show consistent similarity with their corresponding \PD{} and other \aDts{}.
For all DT comparisons, the Wilcoxon test results show $p-value>\alpha$, suggesting no significant difference between any two distributions. 
The effect size ($\delta$) values are nearly zero in all cases, highlighting that the magnitude difference is negligible. 
These observations imply that DTs (\iDts{} and \aDts{}) behaviors are nearly similar to their corresponding PDs. 

\textcolor{black}{
The overall results demonstrate that the 2-shot method outperforms other methods for DT adaptations across devices. 
The primary reason is that the 2-shot method utilizes two examples corresponding to each class, which is useful when adapting between devices with low and high features. 
The 1-shot method's best overall performance compared to the 2-shot and 5-shot methods can be due to its ability for quick adaptation using just a single example per class. 
Furthermore, using a few examples in the 1-shot method facilitates handling noisy data and overfitting, thereby enhancing its stability for adaptations. 
Thus, the 1-shot method is considered appropriate for DT adaptations across evolving versions. 
}

\begin{center}
    \begin{resboxblack}[colbacktitle=gray,left=3.5pt,right=3.5pt,top=3.5pt,bottom=3.5pt]{RQ2 Outcome}
        \textcolor{black}{
        The fidelity of DTs generated with training and adapted across devices exceeds 95\% in most cases when using the 2-shot method,  compared to other methods. 
        For adaptations across versions, the 1-shot method mostly demonstrated a fidelity of over 96\%, outperforming 2- and 5-shot methods. 
        Therefore, when applying \approach{} to generate or adapt DTs, 1-shot or 2-shot methods are deemed suitable. 
        Overall, these results indicate \approach{}'s ability to tackle the \emph{evolution} challenge. 
        }
    \end{resboxblack}
\end{center}

% RQ3: figures
\begin{figure*}
	\subfigure[100 DTs]{\includegraphics[width=8.5cm,height=5.2cm]{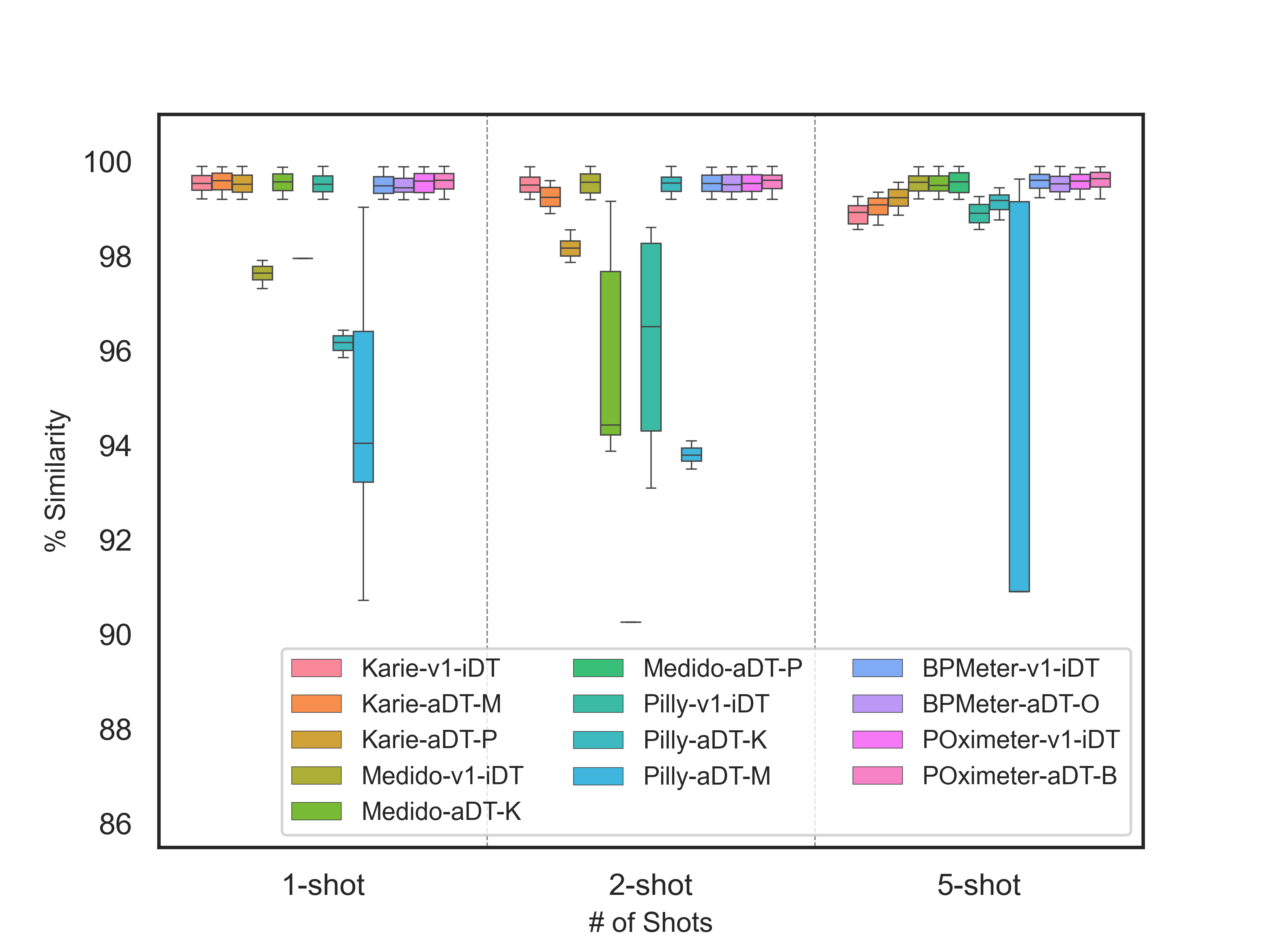}\label{fig:b1}}
	\subfigure[200 DTs]{\includegraphics[width=8.5cm,height=5.2cm]{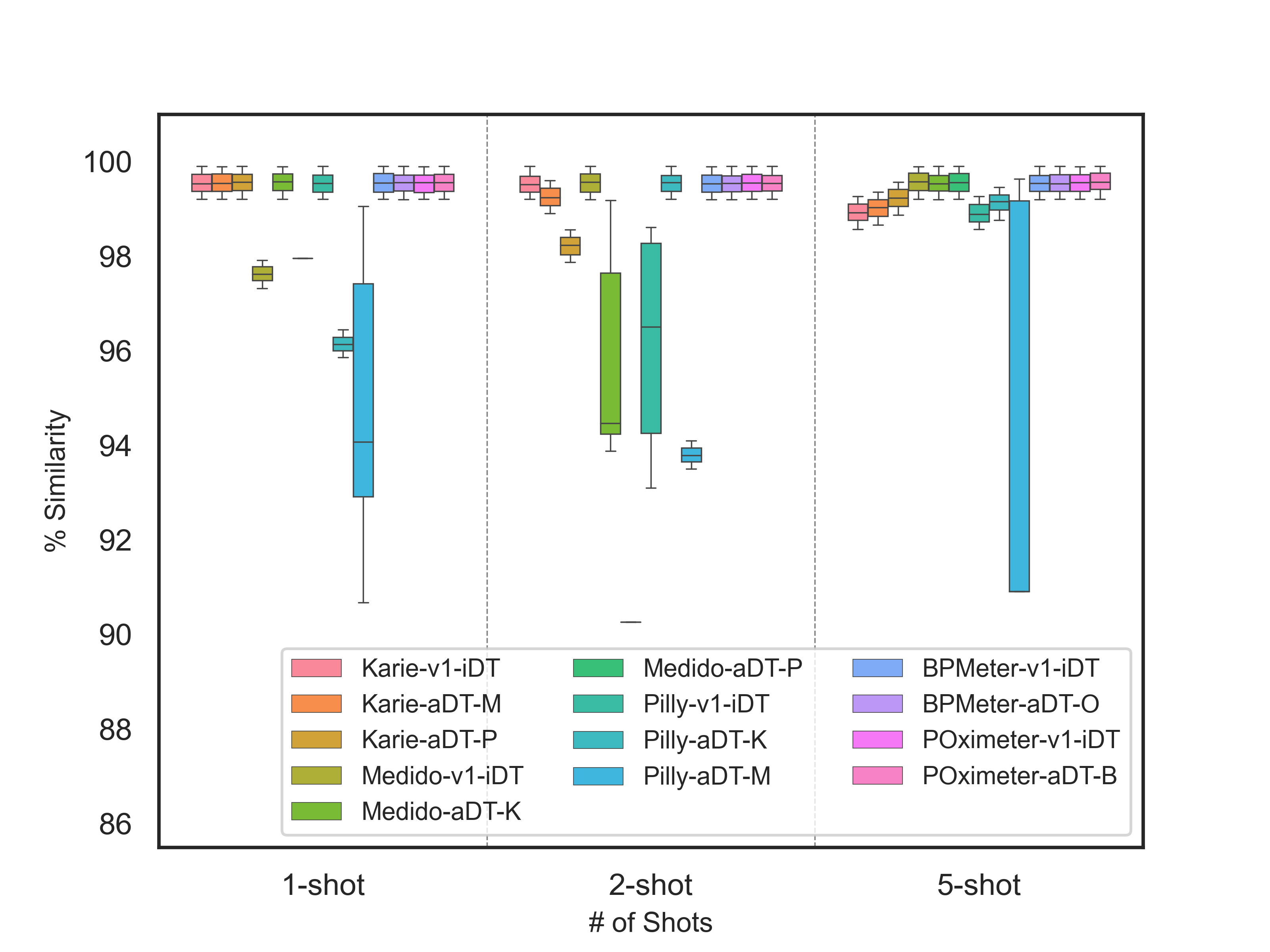}\label{fig:b2}}
	\subfigure[400 DTs]{\includegraphics[width=8.5cm,height=5.2cm]{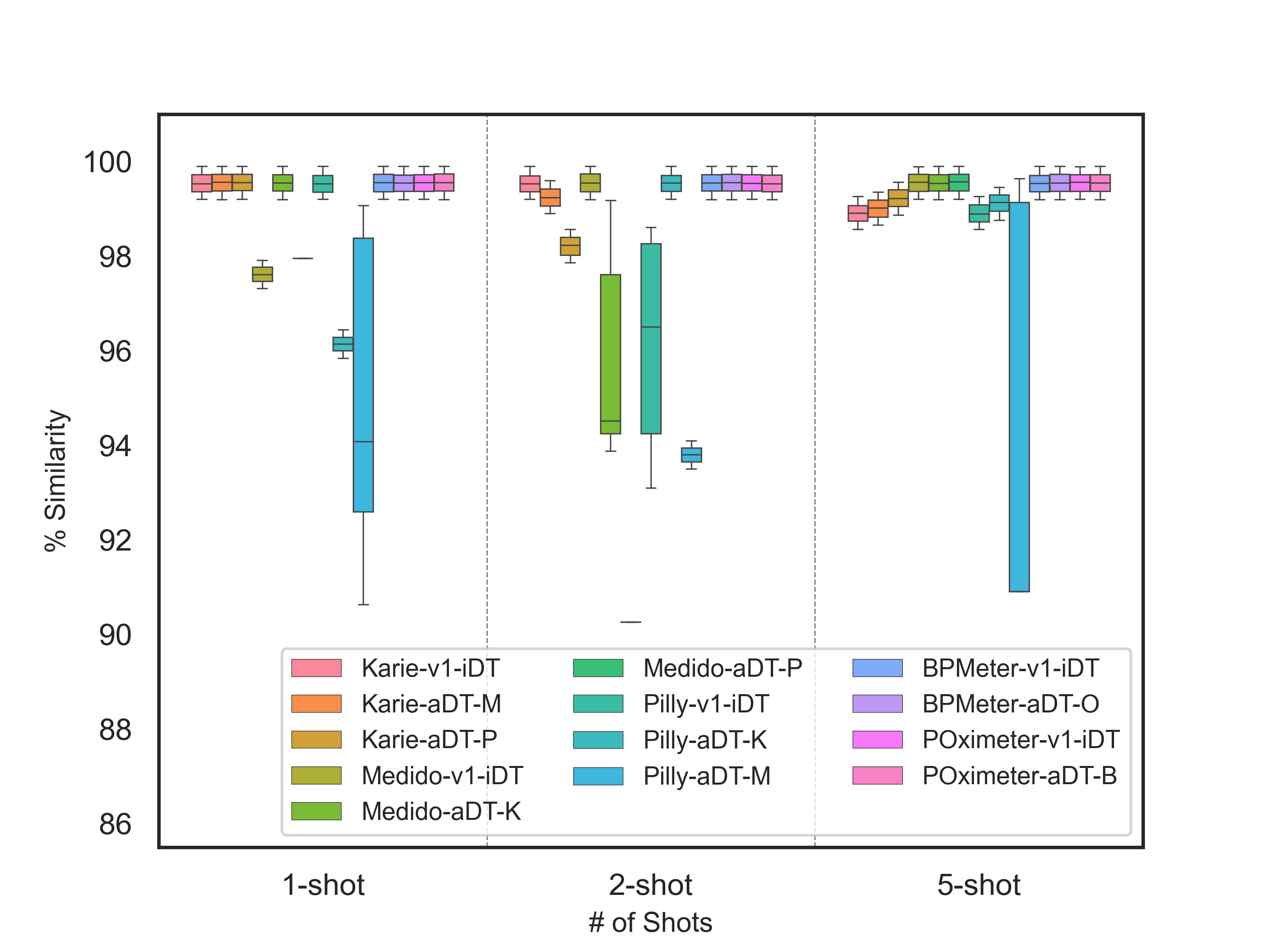}\label{fig:b3}}
	\subfigure[600 DTs]{\includegraphics[width=8.5cm,height=5.2cm]{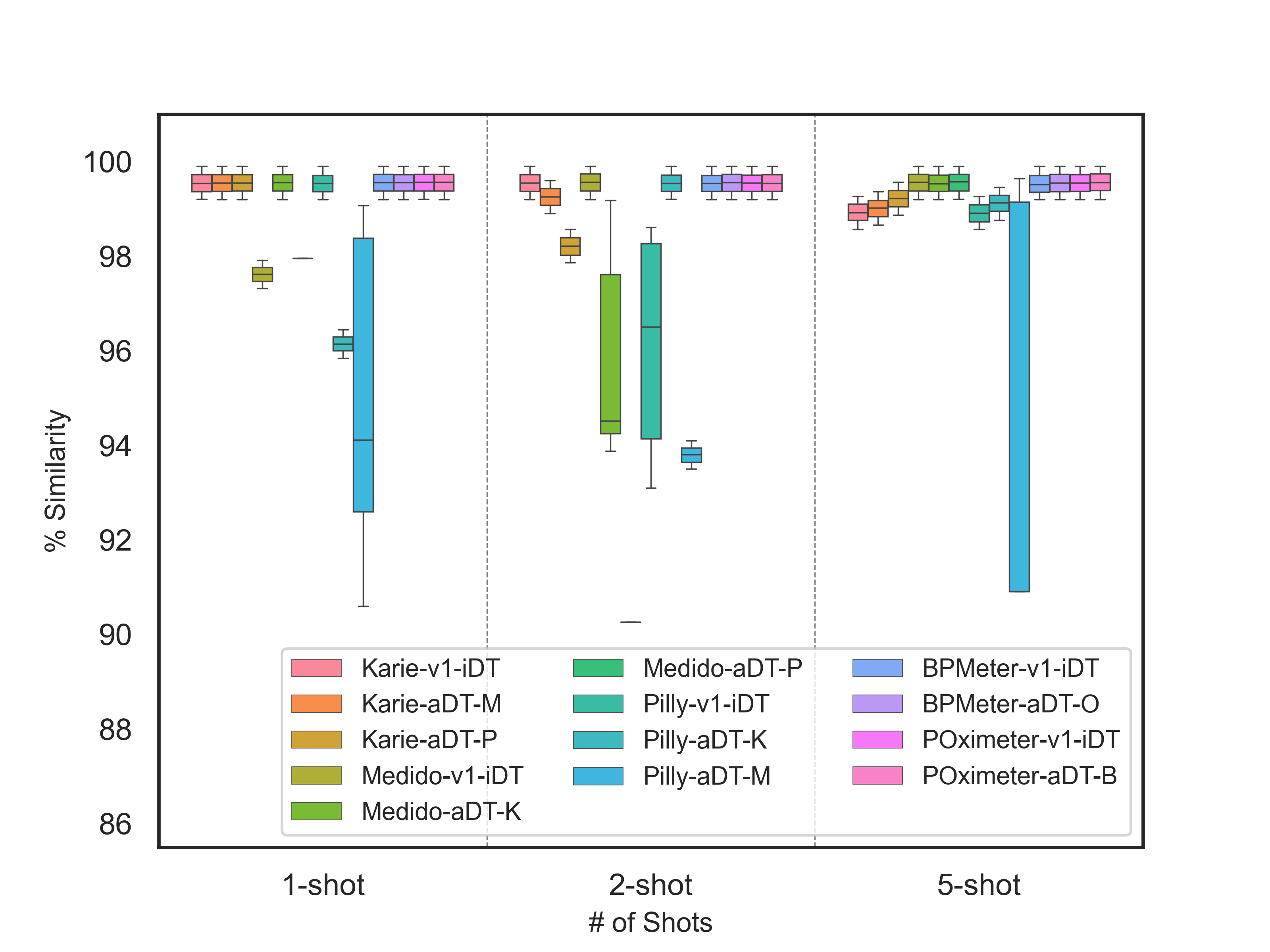}\label{fig:b4}}
	\subfigure[800 DTs]{\includegraphics[width=8.5cm,height=5.2cm]{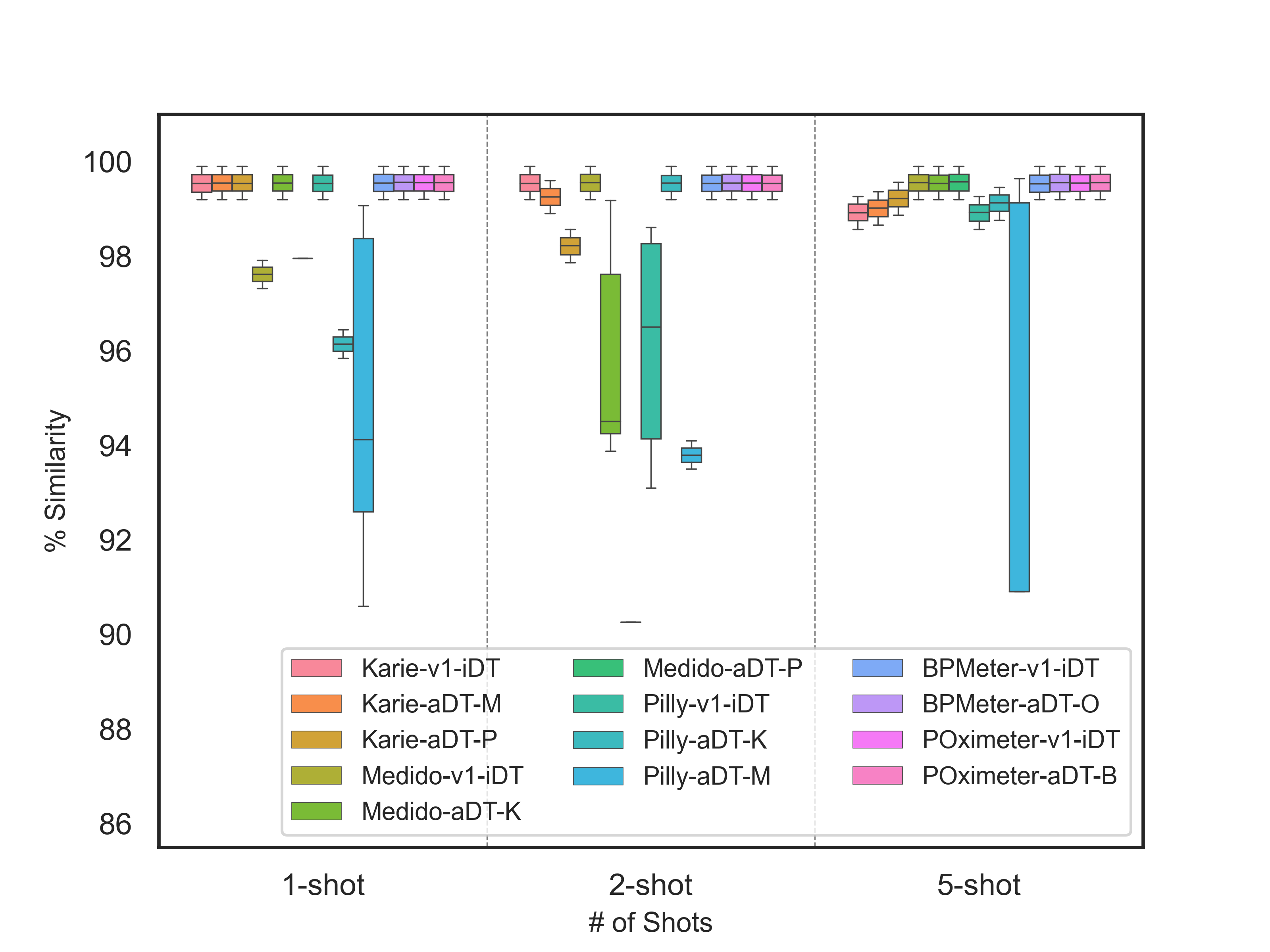}\label{fig:b5}}
	\subfigure[1000 DTs]{\includegraphics[width=8.5cm,height=5.2cm]{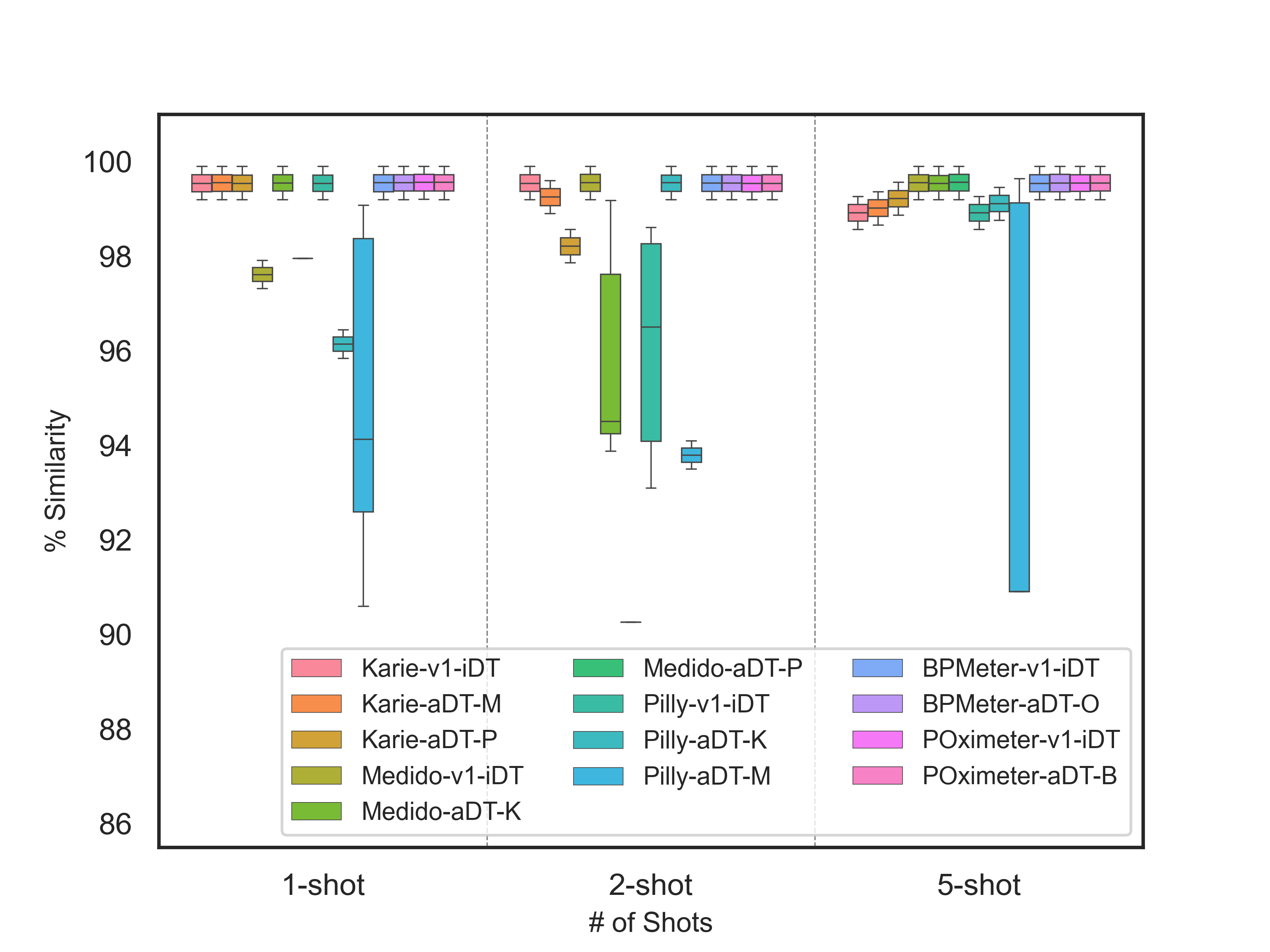}\label{fig:b6}}
	\caption{RQ3 results showing fidelity of 1000 DTs in different batch sizes}\label{fig:rq3results}
\end{figure*}

% RQ4 table - new one
\begin{table*}[!t]
	\centering
	\noindent
	\caption{RQ4 results for comparison between the time taken to generate DTs from scratch during training versus the time taken to adapt DTs across different devices and versions. \textcolor{black}{High total time values are highlighted with a red color, while low values are marked with a green color.}}
    \renewcommand{\arraystretch}{1.0}%0.5
	\begin{tabular}{p{0.3cm} p{5.6cm} G G G}\toprule
        \multicolumn{2}{l }{\textbf{}} & \multicolumn{3}{c}{\textbf{Time (m)}} \\ 
		\cmidrule(lr){3-5}
		\multicolumn{2}{ l }{\textbf{DT Generation}} & \textbf{1-shot} & \textbf{2-shot} & \textbf{5-shot} \\ 
		\cmidrule(lr){1-2}
		\cmidrule(lr){3-5}
            \multirow{5}{*}{\textbf{\rotatebox[origin=c]{90}{\makecell{Training}}}}
            &\textbf{\KarieIdtVi{}}&18.335&21.834&25.223
            \\&\textbf{\MedidoIdtVi{}}&17.525&19.672&20.685
            \\&\textbf{\PillyIdtVi{}}&17.57&19.71&21.476
            \\&\textbf{\BPMeterIdtVi{}}&16.416&17.841&19.785
            \\&\textbf{\POximeterIdtVi{}}&16.397&17.845&19.549
            \\
        \arrayrulecolor{black!20}\cmidrule[0.05pt]{2-5}
        \multirow{2}{*}{\textbf{\rotatebox[origin=c]{90}{\makecell{}}}}
            &\textbf{Total Time}&\cellcolor{red!5}86.243&\cellcolor{red!5}96.902&\cellcolor{red!5}106.718\\
        \arrayrulecolor{black}
		\cmidrule(ll){1-5}
            \multirow{8}{*}{\textbf{\rotatebox[origin=c]{90}{\makecell{Device Adaptation}}}}
            &\textbf{\KarieAdtM{}}&1.766&0.831&0.873
            \\&\textbf{\KarieAdtP{}}&0.671&0.843&0.875
            \\&\textbf{\MedidoAdtK{}}&0.644&0.808&0.803
            \\&\textbf{\MedidoAdtP{}}&0.637&0.748&0.798
            \\&\textbf{\PillyAdtK{}}&0.62&0.765&0.801
            \\&\textbf{\PillyAdtM{}}&0.625&0.738&0.789
            \\&\textbf{\BPMeterAdtO{}}&0.598&0.664&0.758
            \\&\textbf{\POximeterAdtB{}}&0.578&0.678&0.739
            \\
        \arrayrulecolor{black!20}\cmidrule[0.05pt]{2-5}
        \multirow{2}{*}{\textbf{\rotatebox[origin=c]{90}{\makecell{}}}}
            &\textbf{Total Time}&\cellcolor{green!5}6.138&\cellcolor{green!5}6.076&\cellcolor{green!5}6.435\\
        \arrayrulecolor{black}
        \cmidrule(ll){1-5}
            \multirow{15}{*}{\textbf{\rotatebox[origin=c]{90}{\makecell{Version Adaptation}}}}
            &\textbf{\KarieIdtVi{}$\rightarrow$\KarieAdtVii{}}&0.677&0.827&1.025
            \\&\textbf{\KarieAdtVii{}$\rightarrow$\KarieAdtViii{}}&0.681&0.824&1.006
            \\&\textbf{\KarieAdtViii{}$\rightarrow$\KarieAdtViv{}}&0.941&0.936&0.935
            \\&\textbf{\MedidoIdtVi{}$\rightarrow$\MedidoAdtVii{}}&0.88&0.842&0.858
            \\&\textbf{\MedidoAdtVii{}$\rightarrow$\MedidoAdtViii{}}&0.848&0.843&0.855
            \\&\textbf{\MedidoAdtViii{}$\rightarrow$\MedidoAdtViv{}}&0.875&0.843&0.875
            \\&\textbf{\PillyIdtVi{}$\rightarrow$\PillyAdtVii{}}&0.852&0.838&0.848
            \\&\textbf{\PillyAdtVii{}$\rightarrow$\PillyAdtViii{}}&0.876&0.841&0.865
            \\&\textbf{\PillyAdtViii{}$\rightarrow$\PillyAdtViv{}}&0.865&0.838&0.859
            \\&\textbf{\BPMeterIdtVi{}$\rightarrow$\BPMeterAdtVii{}}&0.816&0.804&0.837
            \\&\textbf{\BPMeterAdtVii{}$\rightarrow$\BPMeterAdtViii{}}&0.812&0.807&0.818
            \\&\textbf{\BPMeterAdtViii{}$\rightarrow$\BPMeterAdtViv{}}&0.807&0.806&0.81
            \\&\textbf{\POximeterIdtVi{}$\rightarrow$\POximeterAdtVii{}}&0.804&0.803&0.818
            \\&\textbf{\POximeterAdtVii{}$\rightarrow$\POximeterAdtViii{}}&0.805&0.834&0.813
            \\&\textbf{\POximeterAdtViii{}$\rightarrow$\POximeterAdtViv{}}&0.806&0.847&0.85
            \\
            \arrayrulecolor{black!20}\cmidrule[0.05pt]{2-5}
        \multirow{2}{*}{\textbf{\rotatebox[origin=c]{90}{\makecell{}}}}
            &\textbf{Total Time}&\cellcolor{green!5}12.346&\cellcolor{green!5}12.532&\cellcolor{green!5}13.071\\
        \arrayrulecolor{black}
		\bottomrule
	\end{tabular}
	\label{tab:timeresults}
\end{table*}

\subsubsection{RQ3 Results}\label{RQ3}
\Cref{fig:rq3results} shows the results of multiple DTs' fidelity when compared with their corresponding PDs in different batch sizes: 100, 200, 400, 600, 800, and 1000 in \Cref{fig:b1}, \Cref{fig:b2}, \Cref{fig:b3}, \Cref{fig:b4}, \Cref{fig:b5}, and \Cref{fig:b6}, respectively. 
For \iDts{} and \aDts{} of \Karie{}, we notice consistent similarity as the number of DTs increases from 100 to 1000, irrespective of the few-shot method applied. 
A similar pattern of constant similarity is observed with \iDts{} and \aDts{} of \BPMeter{} and \POximeter{} devices across all shot methods. 
These observations suggest a stable fidelity level when operating multiple \iDts{} and \aDts{} across varying batch sizes for the respective devices---\Karie{}, \BPMeter{}, and \POximeter{}.

In the case of \Medido{}, there is a minimal fluctuation in similarity for the 2-shot method, with the median similarity maintaining a proximate value of 94\%. 
Whereas, for both 1-shot and 5-shot methods, the similarity between \Medido{}'s \iDts{} and \aDts{} remains consistent across each DT batch. 
For \PillyAdtM{} when using the 1-shot method, a slight similarity variance is observed in the boxplots. 
This indicates a minor spread in the distribution of data points, ranging from 92\% to 98\%. 
Nonetheless, the median similarity remains consistent at around 94\% across all batches. 
This implies a stable central tendency despite the slight variations.
When utilizing the 2-shot method, \PillyAdtM{} operates consistently across each DTs batch. 
However, with the 5-shot method, \PillyAdtM{} shows a variance in similarity ranging from 91\% to 98\%. 

\textcolor{black}{
The stability of the 1-shot and 5-shot methods compared to the 2-shot method could be attributed to the quality of data examples used in the 2-shot method. 
Specifically, if the two examples selected in the 2-shot method are noisy or unrepresentative, this can negatively affect the learning process. 
In the 1-shot method, the learning is based on a single example which can have a less negative effect with the example being noisy. 
In the 5-shot method, using five examples increases the likelihood of having more representative examples which could make the learning more stable. 
Thus, the results suggest that DTs generated using the 1-shot and 5-shot methods demonstrate consistent similarity with their physical counterparts across different batches. 
}

\begin{center}
    \begin{resboxblack}[colbacktitle=gray,left=3.5pt,right=3.5pt,top=3.5pt,bottom=3.5pt]{RQ3 Outcome}
        \textcolor{black}{
        The overall similarity of all \iDts{} and \aDts{} generated with the 1-shot method is mostly close to 98\%, though it exhibits minor variations with increasing batch size. 
        A similar pattern is observed with the 5-shot method, while the 2-shot method, despite similarity lower than 98\% in some cases, remains relatively consistent across different batches. 
        Therefore, \approach{} with fewer-shot (1/2-shot) methods is scalable in operating multiple DTs of different medical devices while maintaining high fidelity. 
        In summary, these results demonstrate \approach{}'s capability in handling challenges related to \emph{large-scale testing}.
        }
    \end{resboxblack}
\end{center}

\subsubsection{RQ4 Results}\label{RQ4}
From the results in \Cref{tab:timeresults}, it can be observed that training with the 1-shot method to create \iDts{} took approximately 16-18 minutes, depending on the device features. 
\Karie{}, with a higher number of features, required the longest time. 
\Medido{} and \Pilly{} took around 17 minutes each, whereas \BPMeter{} and \POximeter{} required slightly less time, approximately 16 minutes each. 
As the number of shots increased, i.e., 2- and 5-shot methods, training time also increased correspondingly. 
It took approximately 17-21 minutes for the 2-shot method and further increased to 19-25 minutes for the 5-shot method. 
The notable increase in training time can be seen in the case of \Karie{}, suggesting that devices with many features require more time as the number of shots increases.

Analyzing the results for adaptations across various devices, it is noticeable that the time required for each adaptation normally ranges from approximately 0.5 minutes to 0.9 minutes for all shot methods, including 1-shot, 2-shot, and 5-shot methods. 
Only in the case of \KarieAdtM{}, the adaptation process required approximately 1.7 minutes. 
For the adaptations across versions, the time needed for adapting to newer versions ranges approximately from 0.6 minutes to 1.0 minutes. 
This timing applies to all shot methods, i.e., the 1-shot, 2-shot, and 5-shot methods. 
It is evident from the adaptation results that adapting to another device or a newer version normally requires approximately one minute. 

The overall results indicate that training time costs for the 1-shot, 2-shot, and 5-shot methods are approximately 86, 96, and 106 minutes, respectively. 
Whereas, time costs for device and version adaptations corresponding to each shot method are approximately 6 minutes and 13 minutes, respectively. 
\textcolor{black}{
The lower time cost associated with the 1-shot method, when compared to the 2-shot and 5-shot methods, is due to efficient learning with a few examples. 
The 1-shot method is designed to learn from a single example corresponding to each class, requiring less computational time and resources. 
Therefore, the results demonstrate that the 1-shot method, which learns from a single example, is more time-efficient than the 2-shot and 5-shot methods that require learning from multiple examples. 
}

Observing the time cost difference between training and adaptation, one could argue that the training time is not significantly higher. 
Therefore, opting for training with a high-specification machine might be more feasible instead of adaptation. 
While the time difference might not seem substantial for a single device, this time cost can exponentially grow with the increased number and variety of devices and their multiple software versions. 
In the context of a healthcare IoT application, it is typical to have numerous interconnected medical devices of different types. 
Multiple vendors often provide the same type of device, and the software of each device continually evolves. 
Furthermore, healthcare IoT applications, developed in rapid-release environments, require frequent testing, especially during daily releases. 
Under these conditions, adapting a DT to another device or its newer version could save considerable time, thereby enhancing the efficiency of the testing process.

\begin{center}
    \begin{resboxblack}[colbacktitle=gray,left=3.5pt,right=3.5pt,top=3.5pt,bottom=3.5pt]{RQ4 Outcome}
    \textcolor{black}{
    The average training time to generate \iDts{} is approximately 17 minutes for the one-shot method, 18 minutes for the two-shot method, and 20 minutes for the five-shot method.
    When applying \approach{} to generate a DT, the device features must be considered during the selection of the few-shot method. 
    Devices with a large number of features (for instance, \Karie{}) combined with high-shot methods may lead to increased time costs. 
    For the adaptations across devices and versions, each few-shot method approximately takes just a minute, suggesting a lower time cost for creating \aDts{}. 
    These results suggest that \approach{} is cost-effective in addressing the \emph{evolution} challenge. 
    }
    \end{resboxblack}
\end{center}

\subsection{Threats to Validity}
To overcome \emph{external validity} threats, we evaluated \approach{} with five different types of real-world medical devices, which are widely used with varying features, making them well-representative case studies. Note that these devices are already operational in Oslo City.
For \emph{internal validity}, we designed our experiment carefully, e.g., using an industrial healthcare IoT application, API schema, and documentation provided by Oslo City. 
We discussed evaluation setup in various sessions with practitioners from Oslo City's health department and its industry partner who are providing the healthcare IoT application. 
For the hyperparameters selection, we followed the common practice of running multiple experimental trials~\cite{nam2023stunt}. 
Furthermore, we analyzed the experiment results using different metrics to minimize the chances of threats to \emph{construct validity}. 
We used two stopping criteria (i.e., the total number of iterations and model improvement) for our experiments. 
Moreover, we utilized metrics such as precision, recall, F1-score, and percentage similarity to analyze the results.  
These metrics are commonly employed to evaluate the quality of machine learning models.
We also reported descriptive statistics using mean and standard deviation. 
To reduce \emph{conclusion validity} threats, we executed experiments while consulting Oslo City's technical team. 
We also assessed results using statistical tests: Wilcoxon signed-rank and Cliff's Delta, following well-defined guidelines~\cite{arcuri2011practical}.

\section{Insights and Lessons Learned}\label{insights}
In this section, we discuss insights and lessons based on our experience \textcolor{black}{and the results of our experiments}.

\subsection{Training Data Generation} 
Since medical devices are assigned to patients, collecting real data for training leads to privacy issues. In particular, the General Data Protection Regulation (GDPR) \textcolor{black}{\st{regulation}} from the European Union prevents us from obtaining real data to train machine learning models. To this end, \approach{} relies on an automated data generation phase to collect data for training, which requires connecting medical devices (dedicated for testing) with a healthcare IoT application and sending requests for a specific time (e.g., eight hours each day in our experiments) to collect training data. 
Such data generation is needed whenever a new medical device is employed in the IoT application, as no historical data is available anyway. In the case of software upgrades of medical devices, it is only needed to execute the data generation for a small portion of time, e.g., 1/2th of the time required for a new device, to obtain sufficient data for fine-tuning the trained model based on our experience gained from the current experiments.   

Before generating data, there are two important factors to consider: (i) a device's operating mechanism and (ii) constraints on API requests specified by vendors. 
The device's actual operating mechanism determines the delay between two subsequent requests. % and it is essential for the device's safety. 
For example, a medicine dispensing operation takes 3-4 seconds to cut medicine from a roll and drop it in a pouch. 
Constraints on API requests help in deciding the execution limit of the data generation phase. If the execution limit is short, the data generation phase can be repeated at different time intervals.

\subsection{Configuring Meta-Learning}\label{sec:confml}
In our experiments, we observed that the meta-learning phase needs to be configured carefully. 
Since medical device datasets do not have large numbers or complex sets of features, one hidden layer in model architecture is sufficient. 
Adding more layers to model architecture slows down training and sometimes does not give high accuracy. 
The number of tasks to sample during training should be low (2/3-digit number, e.g., 10) as it adversely affects the performance and model accuracy. 
MAML is designed to work with fewer gradient steps and shots~\cite{finn2017model}; less number for adaptation steps and shots is a better pick. 
For adapting to a new device, we noticed that fine-tuning only requires 1-shot and 1/5th iterations used for training with 1/4th time steps. 
Moreover, when adjusting learning rates, we discovered that the learning rate for the Adam optimizer should be less than that for the MAML algorithm. Based on our experiments with a variety of medical devices, we observed that a simple model is sufficient. However, more complex models and adaptation steps might be needed for more complex medical devices.

\begin{figure}[htbp]
\centerline{\includegraphics[width=8.7cm, height=9.3cm, keepaspectratio]{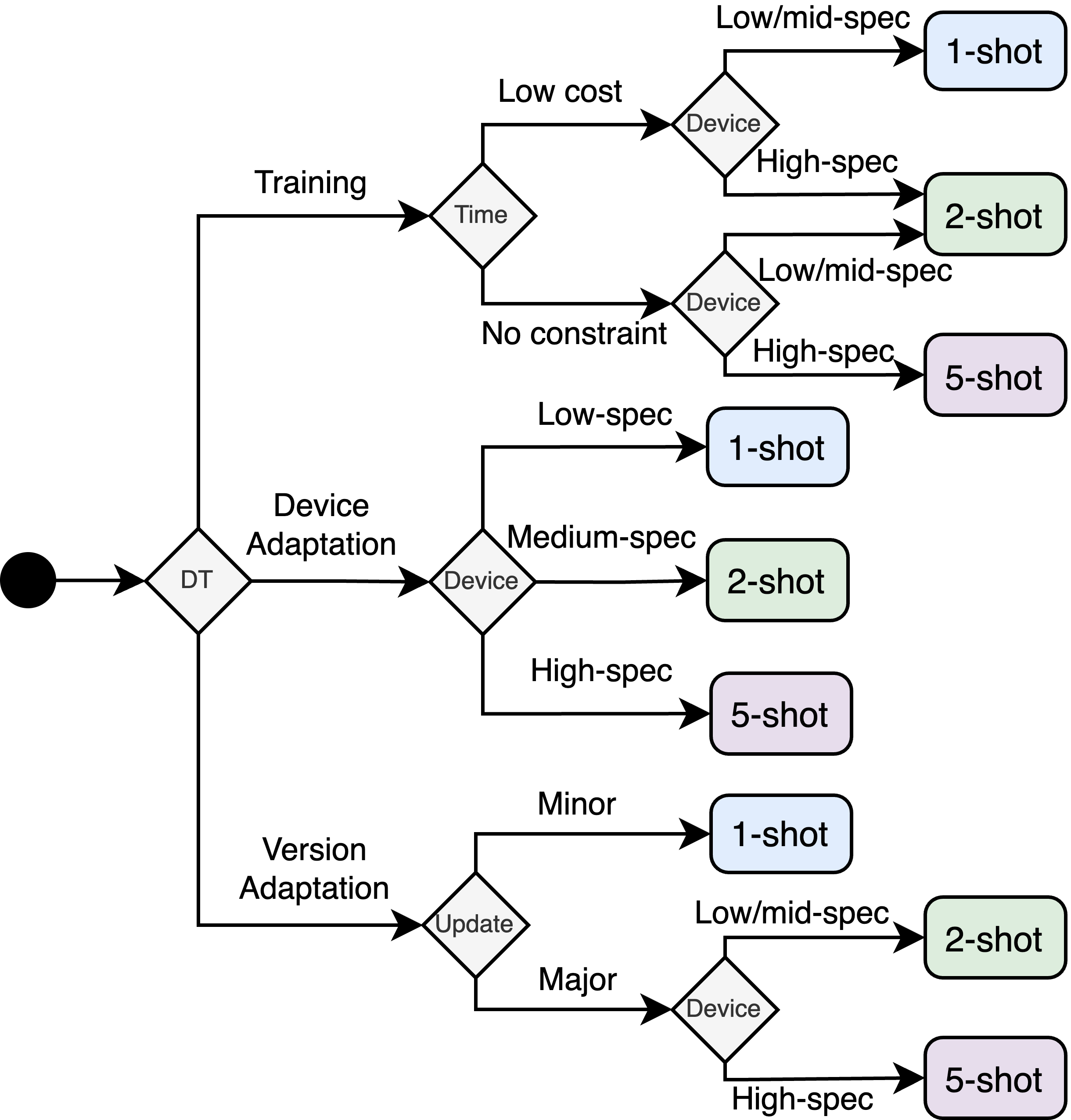}}
\caption{An illustration guiding the selection of a few-shot method based on device features, evolution, and time cost, as derived from experimental results. }
\label{fig:guide}
\end{figure}

\textcolor{black}{\subsection{Selecting Few-shot method}\label{sec:selguide}}
\textcolor{black}{
Based on insights derived from experimental results, \Cref{fig:guide} provides recommendations for practitioners to consider when selecting a specific few-shot method for applying \approach{}. 
To generate a DT from scratch with training, if the time constraints demand low cost and the device possesses a low to medium number of features, the 1-shot method can be used. 
However, for the device with a high number of features, the 2-shot method is recommended. 
If there is no time cost constraint, the 2-shot method is advised for devices with low to medium features, while the 5-shot method is recommended for a high-featured device. 
For DT adaptation across devices, the number of device features guides the selection of a few-shot method: 1-shot for low-featured devices, 2-shot for medium-featured devices, and 5-shot for high-featured devices. 
When adapting DTs across versions, the 1-shot method is advisable for minor updates. 
In case of major upgrades, the 2-shot method is recommended for devices with low to medium features, while the 5-shot method is suggested for high-featured devices. 
It is worth noting that for adaptations across devices and versions, experimental results have shown that the time differences among various few-shot methods are not significantly high. 
Consequently, device features and changes during evolution are important factors in selecting a specific few-shot method, further elaborated in \Cref{sec:generalization}. 
}

\subsection{Model Uncertainty}
From the analysis of experimental results, it was observed that DT models trained using different shot methods demonstrate a certain level of uncertainty. 
This is particularly apparent in the context of DT adaptations across various devices. 
In the case of device-to-device DT adaptations, the variations in the features of different devices influence the model's performance.
For example, \Karie{} supports five languages, while \Medido{} supports three languages that are also common to \Karie{}. 
If \Medido{}'s DT is adapted from \Karie{}, the variation in supported languages introduces an element of uncertainty to the model, potentially impacting the model's adaptation and prediction capabilities. 
Therefore, we observed variations in DT fidelity across different few-shot learning methods, particularly when operating multiple DTs.  
Given the potential impact of uncertainty, focusing on quantifying uncertainty for the DTs generated and adapted using meta-learning can be an intriguing future research direction. 
Uncertainty quantification could provide an additional metric to evaluate and improve the performance of models~\cite{catak2022uncertainty}, thereby enhancing the fidelity of DTs generated with meta-learning techniques.

\subsection{Communicating with Physical Devices}\label{sec:commpd}
A DT sends API requests to its corresponding medical device for communication. Note that each device vendor has API gateways to allow requests from authorized sources, which is the case in our context. 
For other contexts, such authorization information might need to be obtained to establish DTs' communication with their physical devices. 
Furthermore, \approach{} allows configuring a DT's communication with a physical device during testing. For instance, test engineers can decide when and how much DTs communicate with which physical devices based on specific testing purposes.   

\subsection{Testing Healthcare IoT Applications}\label{sec:testhiot}

To create a test setup for testing a healthcare IoT application (i.e., SUT) with \approach{} generated DTs, we advise the following steps for test engineers. 
First, decide the number and type of medical devices required for testing. 
Second, execute \approach{} to generate the desired number of DTs for each type of medical device. 
For multiple DT versions corresponding to different device upgrades needed for testing, assign each DT version with a unique SN. 
Third, integrate DTs with healthcare IoT applications using the APIs generated for each DT. Also, configure the DTs' communication with physical devices. 
In the end, it is ready for testing the SUT. 

While experimenting with 1000 DTs, we observed that significant computation resources are required for loading and inferencing trained models. 
For testing at scale, we recommend allocating distributed resources locally or remotely on the cloud. 
One potential setup involves an individual machine for each device type DT and running multiple DT batches concurrently. 
Such a setup can be easily created on a cloud with virtual machines for different devices DTs. 
The main benefit is that it is a one-time effort and provides a remote interface to healthcare authorities and industry partners for testing healthcare IoT applications remotely.

\subsection{Devices Adaptation and Generalization}\label{sec:generalization}
DT adaptation to a new medical device must match the category of the device; a DT can be adapted to another device from the same category with common features. For example, if a DT is generated for a particular medicine dispenser, it can only adapt to another type of dispenser but not a pulse oximeter. Adapting to a device of a different category with different features and behaviors is unnatural. 
In a rare case of DT adaptation to a downgraded version of the same type of devices, a DT can be fine-tuned with \approach{}. 
However, test engineers should analyze the DT behavior corresponding to the old device to ensure that their behaviors are matched. 
For example, if an older device has 10 features and a newer device has 15 features (10 old and 5 new), adapting from the new to the old retains 5 new features, which could lead to high fidelity.

\textcolor{black}{
Our results demonstrated \approach{}'s practicality in generating and adapting DTs to evolving medical devices. 
In practice, we observed that medical devices can also malfunction during testing, potentially due to several environmental factors~\cite{sartaj2024digital,sartaj2024uncertainty}. 
}
Since medical devices are part of the test infrastructure, the DT fidelity level achieved by \approach{} is sufficient to support automated testing. 
While operating 1000 DTs in different batches, we analyzed that \approach{} is scalable in generating and running many DTs without significantly affecting fidelity. This is a notable advantage for testing healthcare IoT applications with thousands of medical devices. 
\textcolor{black}{
Based on the results of five real-world medical devices, \approach{} can be used for other IoT-based healthcare applications with REST APIs and involving different types of devices, such as mobile security alarm devices for medical emergencies. 
Considering that, \approach{} has broader applicability to multiple diverse types of devices used in various IoT-based healthcare applications. 
}

\textcolor{black}{\subsection{Key Contributions and Outcome of our Study}\label{sec:findingslist}}
\textcolor{black}{
Our work contributes to the scientific body of knowledge by introducing a meta-learning-based approach (\approach{}) for generating adaptable DTs of evolving medical devices and operating these DTs within test infrastructure to facilitate testing of healthcare IoT applications. 
Moreover, our work contribution to the engineering body of knowledge is twofold. 
First, we provided an open-source implementation of \approach{} that can be customized and extended to apply in a practical context. 
Second, through experiments with five real-world medical devices, we demonstrated the applicability of \approach{}, and based on these results, we provided valuable insights and guidelines for practitioners. 
To summarize, the key outcomes of our experiments are as follows. 
\begin{itemize}
    \item \approach{} can effectively generate and adapt DTs with a low-shot method, such as 1-shot. 
    For the DT adaptations across devices or versions, a 2-shot or 5-shot method might be necessary depending on the device features and the extent of the version upgrade. 
    \item DTs generated using \approach{} with fewer-shot (1/2-shot) methods exhibit high fidelity with their physical counterparts, signifying their potential to serve as substitutes for physical devices during testing. 
    \item When applied to large-scale testing, \approach{} is scalable in generating and operating numerous DTs that represent a variety of medical devices, as well as preserving a high fidelity level. 
    \item For rapidly evolving medical devices, adapting a DT to a new variant takes one minute on average. This suggests that DTs generated with \approach{} can adapt to a different device or version without requiring high computational resources. 
\end{itemize}
}

\section{Conclusion}\label{conclusion}
This work is attributed to solving key challenges of integrating medical devices into a test infrastructure for automated and rigorous testing of healthcare IoT applications. 
In this aspect, we presented \approach{}, a meta-learning-based approach for building and adapting DTs of medical devices. 
We presented an empirical evaluation of \approach{} using five real-world medical devices comprising three commonly used medicine dispensers (i.e, \Karie{}, \Medido{}, and \Pilly{}) and two measurement devices (i.e., \BPMeter{} and \POximeter{}), which have been integrated with a healthcare IoT application from Oslo City. 
Our evaluation assessed \approach{}'s performance in generating DTs, adapting DTs for different medical devices and various evolving versions, and DTs fidelity in terms of their behavioral similarities with their corresponding physical medical devices. 
We also evaluated \approach{} in operating 1000 DTs in different batches, i.e., 100, 200, 400, 600, 800, and 1000 representing physical medical devices of each type. 
In addition, we analyzed the time cost associated with generating and adapting DTs using various few-shot methods.

\textcolor{black}{
The experimental findings indicated that \approach{} can generate and adapt DTs with high fidelity using a low-shot method like 1/-shot. 
It was discovered that the performance of a few shot method depends on device complexity, specifically the number of features, and the extent of device upgrades, whether major or minor.
For instance, to generate or adapt DTs for a feature-rich device like \Karie{}, a high-shot method such as a 5-shot method may be necessary. 
Conversely, for simpler devices like \POximeter{}, a 1-shot or 2-shot method can be sufficient. 
Similarly, if a device undergoes a major upgrade, a high-shot method may be required, while a 1-shot method could suffice for minor updates. 
When operating multiple DTs in different batches, it was observed that DTs generated or adapted using the 1-shot method exhibit stable performance, while maintaining high fidelity. 
This indicates \approach{}'s scalability in generating and operating multiple DTs simultaneously. 
Additionally, the time cost analysis demonstrated that training time for DT generation increases as the number of shots and device features increases, while the time cost for DT adaptations generally remains consistent. 
}

\textcolor{black}{
A potential future research direction could involve quantifying uncertainty in DT models trained with various few-shot meta-learning methods. 
In terms of practical application, future works could focus on two directions: 1) extending the usage of \approach{} to a wider variety of medical devices for generating and adapting DTs, and 2) evaluating testing effectiveness by employing \approach{}-generated DTs in place of physical devices.  
}

%%
%% Acknowledgments 
\section*{Acknowledgments}
This research work is a part of the WTT4Oslo project (No. 309175) funded by the Research Council of Norway. All the experiments reported in this paper are conducted in a laboratory setting of Simula Research Laboratory; therefore, they do not by any means reflect the quality of services Oslo City provides to its citizens. Moreover, these experiments do not reflect the quality of services various vendors provide to Oslo City.

\bibliographystyle{unsrt}
\bibliography{refs}

\begin{thebibliography}{10}

\bibitem{balasubramanian2021scalable}
Venki Balasubramanian and Alireza Jolfaei.
\newblock A scalable framework for healthcare monitoring application using the internet of medical things.
\newblock {\em Software: Practice and Experience}, 51(12):2457--2468, 2021.

\bibitem{sartaj2023testing}
Hassan Sartaj, Shaukat Ali, Tao Yue, and Kjetil Moberg.
\newblock {Testing Real-World Healthcare IoT Application: Experiences and Lessons Learned}.
\newblock In {\em Proceedings of the 31st ACM Joint European Software Engineering Conference and Symposium on the Foundations of Software Engineering}, ESEC/FSE 2023, page 2044–2049, New York, NY, USA, 2023. Association for Computing Machinery.

\bibitem{wtsproject}
Improving~Quality of~Healthcare Welfare~Technology.
\newblock \url{https://prosjektbanken.forskningsradet.no/project/FORISS/309175}, 2014.
\newblock [Online; accessed 15-May-2024].

\bibitem{tao2019digital}
Fei Tao, Meng Zhang, and Andrew Yeh~Chris Nee.
\newblock {\em Digital twin driven smart manufacturing}.
\newblock Academic Press, 2019.

\bibitem{sartaj2023hita}
Hassan Sartaj, Shaukat Ali, Tao Yue, and Julie~Marie Gjøby.
\newblock {HITA: An Architecture for System-level Testing of Healthcare IoT Applications}.
\newblock In {\em European Conference on Software Architecture}, pages 451--468, Cham, 2024. Springer.

\bibitem{li2018learning}
Da~Li, Yongxin Yang, Yi-Zhe Song, and Timothy Hospedales.
\newblock Learning to generalize: Meta-learning for domain generalization.
\newblock In {\em Proceedings of the AAAI conference on artificial intelligence}, number~1, pages 3490--3497. AAAI, 2018.

\bibitem{hospedales2021meta}
Timothy Hospedales, Antreas Antoniou, Paul Micaelli, and Amos Storkey.
\newblock Meta-learning in neural networks: A survey.
\newblock {\em IEEE transactions on pattern analysis and machine intelligence}, 44(9):5149--5169, 2021.

\bibitem{nam2023stunt}
Jaehyun Nam, Jihoon Tack, Kyungmin Lee, Hankook Lee, and Jinwoo Shin.
\newblock {STUNT}: Few-shot tabular learning with self-generated tasks from unlabeled tables.
\newblock In {\em International Conference on Learning Representations}, pages 1--19, 2023.

\bibitem{yue2021understanding}
Tao Yue, Paolo Arcaini, and Shaukat Ali.
\newblock Understanding digital twins for cyber-physical systems: a conceptual model.
\newblock In {\em Leveraging Applications of Formal Methods, Verification and Validation: Tools and Trends: 9th International Symposium on Leveraging Applications of Formal Methods, ISoLA 2020, Rhodes, Greece, October 20--30, 2020, Proceedings, Part IV}, pages 54--71, Cham, 2021. Springer.

\bibitem{rivera2020engineering}
Luis~F Rivera, Hausi~A M{\"u}ller, Norha~M Villegas, Gabriel Tamura, and Miguel Jim{\'e}nez.
\newblock On the engineering of {IoT}-intensive digital twin software systems.
\newblock In {\em Proceedings of the IEEE/ACM 42nd International Conference on Software Engineering Workshops}, pages 631--638. ACM, 2020.

\bibitem{kirchhof2020model}
J{\"o}rg~Christian Kirchhof, Judith Michael, Bernhard Rumpe, Simon Varga, and Andreas Wortmann.
\newblock Model-driven digital twin construction: synthesizing the integration of cyber-physical systems with their information systems.
\newblock In {\em Proceedings of the 23rd ACM/IEEE International Conference on Model Driven Engineering Languages and Systems}, pages 90--101. ACM, 2020.

\bibitem{sleuters2019digital}
Jack Sleuters, Yonghui Li, Jacques Verriet, Marina Velikova, and Richard Doornbos.
\newblock A digital twin method for automated behavior analysis of large-scale distributed {IoT} systems.
\newblock In {\em 2019 14th Annual Conference System of Systems Engineering (SoSE)}, pages 7--12, New York, NY, USA, 2019. IEEE.

\bibitem{christofi2022novel}
Nikolena Christofi and Xavier Pucel.
\newblock A novel methodology to construct digital twin models for spacecraft operations using fault and behaviour trees.
\newblock In {\em Proceedings of the 25th International Conference on Model Driven Engineering Languages and Systems: Companion Proceedings}, pages 473--480, New York, NY, USA, 2022. Association for Computing Machinery.

\bibitem{dalibor2020towards}
Manuela Dalibor, Judith Michael, Bernhard Rumpe, Simon Varga, and Andreas Wortmann.
\newblock Towards a model-driven architecture for interactive digital twin cockpits.
\newblock In {\em Conceptual Modeling: 39th International Conference, ER 2020, Vienna, Austria, November 3--6, 2020, Proceedings}, pages 377--387, Cham, 2020. Springer.

\bibitem{dalibor2022generating}
Manuela Dalibor, Malte Heithoff, Judith Michael, Lukas Netz, J{\'e}r{\^o}me Pfeiffer, Bernhard Rumpe, Simon Varga, and Andreas Wortmann.
\newblock Generating customized low-code development platforms for digital twins.
\newblock {\em Journal of Computer Languages}, 70:101117, 2022.

\bibitem{corradini2023dtmn}
Flavio Corradini, Arianna Fedeli, Fabrizio Fornari, Andrea Polini, and Barbara Re.
\newblock {DTMN} a modelling notation for digital twins.
\newblock In {\em Enterprise Design, Operations, and Computing. EDOC 2022 Workshops: IDAMS, SoEA4EE, TEAR, EDOC Forum, Demonstrations Track and Doctoral Consortium, Bozen-Bolzano, Italy, October 4--7, 2022, Revised Selected Papers}, pages 63--78, Cham, 2023. Springer.

\bibitem{robles2023opentwins}
Julia Robles, Cristian Mart{\'\i}n, and Manuel D{\'\i}az.
\newblock {OpenTwins}: An open-source framework for the development of next-gen compositional digital twins.
\newblock {\em Computers in Industry}, 152:104007, 2023.

\bibitem{picone2021wldt}
Marco Picone, Marco Mamei, and Franco Zambonelli.
\newblock {WLDT: A general purpose library to build IoT digital twins}.
\newblock {\em SoftwareX}, 13:100661, 2021.

\bibitem{sciullo2024relativistic}
Luca Sciullo, Alberto De~Marchi, Angelo Trotta, Federico Montori, Luciano Bononi, and Marco Di~Felice.
\newblock Relativistic digital twin: Bringing the {IoT} to the future.
\newblock {\em Future Generation Computer Systems}, 153:521--536, 2024.

\bibitem{xu2024pretrain}
Qinghua Xu, Tao Yue, Shaukat Ali, and Maite Arratibel.
\newblock Pretrain, prompt, and transfer: Evolving digital twins for time-to-event analysis in cyber-physical systems.
\newblock {\em IEEE Transactions on Software Engineering}, 50(6):1464--1477, 2024.

\bibitem{zhou2024toward}
Huiying Zhou, Longqiang Wang, Gaoyang Pang, Huimin Shen, Baicun Wang, Haiteng Wu, and Geng Yang.
\newblock Toward human motion digital twin: A motion capture system for human-centric applications.
\newblock {\em IEEE Transactions on Automation Science and Engineering}, pages 1--12, 2024.

\bibitem{shoukat2024smart}
Muhammad~Usman Shoukat, Lirong Yan, Jiawen Zhang, Yu~Cheng, Muhammad~Umair Raza, and Ashfaq Niaz.
\newblock Smart home for enhanced healthcare: exploring human machine interface oriented digital twin model.
\newblock {\em Multimedia Tools and Applications}, 83(11):31297--31315, 2024.

\bibitem{pirbhulal2024cognitive}
Sandeep Pirbhulal, Sabarathinam Chockalingam, Habtamu Abie, and Nathan Lau.
\newblock Cognitive digital twins for improving security in {IT-OT} enabled healthcare applications.
\newblock In {\em International Conference on Human-Computer Interaction}, pages 153--163. Springer, 2024.

\bibitem{bersani2022engineering}
Marcello~M Bersani, Chiara Braghin, Angelo Gargantini, Raffaela Mirandola, Elvinia Riccobene, and Patrizia Scandurra.
\newblock Engineering of trust analysis-driven digital twins for a medical device.
\newblock In {\em European Conference on Software Architecture}, pages 467--482. Springer, 2022.

\bibitem{elayan2021digital}
Haya Elayan, Moayad Aloqaily, and Mohsen Guizani.
\newblock Digital twin for intelligent context-aware {IoT} healthcare systems.
\newblock {\em IEEE Internet of Things Journal}, 8(23):16749--16757, 2021.

\bibitem{sartaj2024modelbased}
Hassan Sartaj, Shaukat Ali, Tao Yue, and Kjetil Moberg.
\newblock Model-based digital twins of medicine dispensers for healthcare {IoT} applications.
\newblock {\em Software: Practice and Experience}, 54(6):1172--1192, 2024.

\bibitem{kaul2023role}
Rohit Kaul, Chinedu Ossai, Abdur Rahim~Mohammad Forkan, Prem~Prakash Jayaraman, John Zelcer, Stephen Vaughan, and Nilmini Wickramasinghe.
\newblock The role of {AI} for developing digital twins in healthcare: The case of cancer care.
\newblock {\em Wiley Interdisciplinary Reviews: Data Mining and Knowledge Discovery}, 13(1):e1480, 2023.

\bibitem{david2021inference}
Istvan David, Jessie Galasso, and Eugene Syriani.
\newblock Inference of simulation models in digital twins by reinforcement learning.
\newblock In {\em 2021 ACM/IEEE International Conference on Model Driven Engineering Languages and Systems Companion (MODELS-C)}, pages 221--224, New York, NY, USA, 2021. IEEE.

\bibitem{kirchhof2021understanding}
J{\"o}rg~Christian Kirchhof, Lukas Malcher, and Bernhard Rumpe.
\newblock Understanding and improving model-driven {IoT} systems through accompanying digital twins.
\newblock In {\em Proceedings of the 20th ACM SIGPLAN International Conference on Generative Programming: Concepts and Experiences}, pages 197--209. ACM, 2021.

\bibitem{nguyen2022digital}
Luong Nguyen, Mariana Segovia, Wissam Mallouli, Edgardo Montes~de Oca, and Ana~R Cavalli.
\newblock Digital twin for {IoT} environments: A testing and simulation tool.
\newblock In {\em Quality of Information and Communications Technology: 15th International Conference, QUATIC 2022, Talavera de la Reina, Spain, September 12--14, 2022, Proceedings}, pages 205--219, Cham, 2022. Springer.

\bibitem{lehner2021aml4dt}
Daniel Lehner, Sabine Sint, Michael Vierhauser, Wolfgang Narzt, and Manuel Wimmer.
\newblock {AML4DT}: a model-driven framework for developing and maintaining digital twins with {automationML}.
\newblock In {\em 2021 26th IEEE International Conference on Emerging Technologies and Factory Automation (ETFA)}, pages 1--8, New York, NY, USA, 2021. IEEE.

\bibitem{croatti2020integration}
Angelo Croatti, Matteo Gabellini, Sara Montagna, and Alessandro Ricci.
\newblock On the integration of agents and digital twins in healthcare.
\newblock {\em Journal of Medical Systems}, 44:1--8, 2020.

\bibitem{somers2023digital}
Richard~J Somers, James~A Douthwaite, David~J Wagg, Neil Walkinshaw, and Robert~M Hierons.
\newblock Digital-twin-based testing for cyber--physical systems: A systematic literature review.
\newblock {\em Information and Software Technology}, 156:107145, 2023.

\bibitem{damjanovic2019open}
Violeta Damjanovic-Behrendt and Wernher Behrendt.
\newblock An open source approach to the design and implementation of digital twins for smart manufacturing.
\newblock {\em International Journal of Computer Integrated Manufacturing}, 32(4-5):366--384, 2019.

\bibitem{dobaj2022towards}
J{\"u}rgen Dobaj, Andreas Riel, Thomas Krug, Matthias Seidl, Georg Macher, and Markus Egretzberger.
\newblock Towards digital twin-enabled devops for cps providing architecture-based service adaptation \& verification at runtime.
\newblock In {\em Proceedings of the 17th Symposium on Software Engineering for Adaptive and Self-Managing Systems}, pages 132--143, New York, NY, USA, 2022. Association for Computing Machinery.

\bibitem{wang2019digital}
Jinjiang Wang, Lunkuan Ye, Robert~X Gao, Chen Li, and Laibin Zhang.
\newblock Digital twin for rotating machinery fault diagnosis in smart manufacturing.
\newblock {\em International Journal of Production Research}, 57(12):3920--3934, 2019.

\bibitem{xia2021digital}
Kaishu Xia, Christopher Sacco, Max Kirkpatrick, Clint Saidy, Lam Nguyen, Anil Kircaliali, and Ramy Harik.
\newblock A digital twin to train deep reinforcement learning agent for smart manufacturing plants: Environment, interfaces and intelligence.
\newblock {\em Journal of Manufacturing Systems}, 58:210--230, 2021.

\bibitem{iottwinmaker}
{AWS IoT TwinMaker}.
\newblock \url{https://docs.aws.amazon.com/iot-twinmaker/}, 2014.
\newblock [Online; accessed 21-May-2024].

\bibitem{iothub}
{Azure IoT Hub}.
\newblock \url{https://azure.microsoft.com/en-us/products/iot-hub/}, 2014.
\newblock [Online; accessed 21-May-2024].

\bibitem{azuredt}
Azure~Digital Twins.
\newblock \url{https://learn.microsoft.com/en-us/azure/digital-twins/}, 2014.
\newblock [Online; accessed 21-May-2024].

\bibitem{vorto}
{Eclipse Vorto}.
\newblock \url{https://www.eclipse.org/vorto/}, 2014.
\newblock [Online; accessed 21-May-2024].

\bibitem{hono}
{Eclipse Hono}.
\newblock \url{https://www.eclipse.org/hono/}, 2014.
\newblock [Online; accessed 21-May-2024].

\bibitem{ditto}
{Eclipse Ditto}.
\newblock \url{https://projects.eclipse.org/projects/iot.ditto}, 2014.
\newblock [Online; accessed 21-May-2024].

\bibitem{eclipseterms}
Eclipse~Terms of~Use.
\newblock \url{https://www.eclipse.org/legal/termsofuse.php}, 2014.
\newblock [Online; accessed 21-May-2024].

\bibitem{dalibor2022cross}
Manuela Dalibor, Nico Jansen, Bernhard Rumpe, David Schmalzing, Louis Wachtmeister, Manuel Wimmer, and Andreas Wortmann.
\newblock A cross-domain systematic mapping study on software engineering for digital twins.
\newblock {\em Journal of Systems and Software}, 193:111361, 2022.

\bibitem{lehner2022digital}
Daniel Lehner, J{\'e}r{\^o}me Pfeiffer, Erik-Felix Tinsel, Matthias~Milan Strljic, Sabine Sint, Michael Vierhauser, Andreas Wortmann, and Manuel Wimmer.
\newblock Digital twin platforms: requirements, capabilities, and future prospects.
\newblock {\em IEEE Software}, 39(02):53--61, 2022.

\bibitem{li2022domain}
Jia Li, Shiva Nejati, Mehrdad Sabetzadeh, and Michael McCallen.
\newblock A domain-specific language for simulation-based testing of {IoT} edge-to-cloud solutions.
\newblock In {\em Proceedings of the 25th International Conference on Model Driven Engineering Languages and Systems}, pages 367--378. ACM, 2022.

\bibitem{gutierrez2019evolutionary}
Lorena Guti{\'e}rrez-Madro{\~n}al, Antonio Garc{\'\i}a-Dom{\'\i}nguez, and Inmaculada Medina-Bulo.
\newblock Evolutionary mutation testing for {IoT} with recorded and generated events.
\newblock {\em Software: Practice and Experience}, 49(4):640--672, 2019.

\bibitem{gupta2021hierarchical}
Deepti Gupta, Olumide Kayode, Smriti Bhatt, Maanak Gupta, and Ali~Saman Tosun.
\newblock Hierarchical federated learning based anomaly detection using digital twins for smart healthcare.
\newblock In {\em 2021 IEEE 7th International Conference on Collaboration and Internet Computing (CIC)}, pages 16--25, New York, NY, USA, 2021. IEEE.

\bibitem{de2023digital}
Alessandra De~Benedictis, Francesco Flammini, Nicola Mazzocca, Alessandra Somma, and Francesco Vitale.
\newblock Digital twins for anomaly detection in the industrial internet of things: Conceptual architecture and proof-of-concept.
\newblock {\em IEEE Transactions on Industrial Informatics}, 19(12):11553--11563, 2023.

\bibitem{korzun2017internet}
DG~Korzun.
\newblock Internet of things meets mobile health systems in smart spaces: An overview.
\newblock {\em Internet of Things and Big Data Technologies for Next Generation Healthcare}, pages 111--129, 2017.

\bibitem{demetriou2023internet}
Demetra Demetriou, Kgomotso Mathabe, Georgios Lolas, and Zodwa Dlamini.
\newblock Internet of things in society 5.0 and the democratization of healthcare.
\newblock In {\em Society 5.0 and Next Generation Healthcare: Patient-Focused and Technology-Assisted Precision Therapies}, pages 111--130. Springer, 2023.

\bibitem{kelly2020internet}
Jaimon~T Kelly, Katrina~L Campbell, Enying Gong, and Paul Scuffham.
\newblock The internet of things: Impact and implications for health care delivery.
\newblock {\em Journal of Medical Internet Research}, 22(11):e20135, 2020.

\bibitem{rayan2022internet}
Rehab~A Rayan, Christos Tsagkaris, Andreas~S Papazoglou, and Dimitrios~V Moysidis.
\newblock The internet of medical things for monitoring health.
\newblock In {\em Internet of Things}, pages 213--228. CRC Press, 2022.

\bibitem{alsaig2023dependable}
Alaa Alsaig and Vangalur Alagar.
\newblock Dependable service-oriented design of healthcare iot.
\newblock In {\em 2023 IEEE International Conference on Smart Internet of Things (SmartIoT)}, pages 257--264, New York, NY, USA, 2023. IEEE.

\bibitem{thrun1998learning}
Sebastian Thrun and Lorien Pratt.
\newblock Learning to learn: Introduction and overview.
\newblock In {\em Learning to learn}, pages 3--17. Springer, Cham, 1998.

\bibitem{schmidhuber1987evolutionary}
Jürgen Schmidhuber.
\newblock {\em Evolutionary principles in self-referential learning, or on learning how to learn: The meta-meta-... hook}.
\newblock PhD thesis, Technische Universität München, 1987.

\bibitem{finn2017model}
Chelsea Finn, Pieter Abbeel, and Sergey Levine.
\newblock Model-agnostic meta-learning for fast adaptation of deep networks.
\newblock In {\em Proceedings of the 34th International Conference on Machine Learning}, volume~70 of {\em Proceedings of Machine Learning Research}, pages 1126--1135. PMLR, 06--11 Aug 2017.

\bibitem{nashid2023retrieval}
Noor Nashid, Mifta Sintaha, and Ali Mesbah.
\newblock Retrieval-based prompt selection for code-related few-shot learning.
\newblock In {\em 2023 IEEE/ACM 45th International Conference on Software Engineering (ICSE)}, pages 2450--2462, New York, NY, USA, 2023. IEEE.

\bibitem{wtsoslo}
National Welfare~Technology Program.
\newblock \url{https://www.helsedirektoratet.no/tema/velferdsteknologi}, 2014.
\newblock [Online; accessed 11-May-2024].

\bibitem{oslocity}
Norwegian health authority.
\newblock \url{https://www.oslo.kommune.no/etater-foretak-og-ombud/helseetaten/}, 2014.
\newblock [Online; accessed 11-May-2024].

\bibitem{arcuri2019restful}
Andrea Arcuri.
\newblock {RESTful API} automated test case generation with {EvoMaster}.
\newblock {\em ACM Transactions on Software Engineering and Methodology (TOSEM)}, 28(1):1--37, 2019.

\bibitem{isaku2023cost}
Erblin Isaku, Hassan Sartaj, Christoph Laaber, Tao Yue, Shaukat Ali, Thomas Schwitalla, and Jan~F Nyg{\aa}rd.
\newblock {Cost Reduction on Testing Evolving Cancer Registry System}.
\newblock In {\em 2023 IEEE International Conference on Software Maintenance and Evolution (ICSME)}, pages 508--518, New York, NY, USA, 2023. IEEE.

\bibitem{sartaj2024restapi}
Hassan Sartaj, Shaukat Ali, and Julie~Marie Gjøby.
\newblock {REST API Testing in DevOps: A Study on an Evolving Healthcare IoT Application}, 2024.

\bibitem{kotsiantis2006data}
Sotiris~B Kotsiantis, Dimitris Kanellopoulos, and Panagiotis~E Pintelas.
\newblock Data preprocessing for supervised leaning.
\newblock {\em International journal of computer science}, 1(2):111--117, 2006.

\bibitem{guyon2003introduction}
Isabelle Guyon and Andr{\'e} Elisseeff.
\newblock An introduction to variable and feature selection.
\newblock {\em Journal of machine learning research}, 3(Mar):1157--1182, 2003.

\bibitem{rfc9110}
Roy~T. Fielding, Mark Nottingham, and Julian Reschke.
\newblock {HTTP Semantics}.
\newblock RFC 9110, jun 2022.

\bibitem{daqi2005classification}
Gao Daqi and Ji~Yan.
\newblock Classification methodologies of multilayer perceptrons with sigmoid activation functions.
\newblock {\em Pattern Recognition}, 38(10):1469--1482, 2005.

\bibitem{akiba2019optuna}
Takuya Akiba, Shotaro Sano, Toshihiko Yanase, Takeru Ohta, and Masanori Koyama.
\newblock Optuna: A next-generation hyperparameter optimization framework.
\newblock In {\em Proceedings of the 25th ACM SIGKDD international conference on knowledge discovery \& data mining}, pages 2623--2631, New York, NY, USA, 2019. Association for Computing Machinery.

\bibitem{arnold2020learn2learn}
Sébastien M.~R. Arnold, Praateek Mahajan, Debajyoti Datta, Ian Bunner, and Konstantinos~Saitas Zarkias.
\newblock learn2learn: A library for meta-learning research, 2020.

\bibitem{scikit-learn}
F.~Pedregosa, G.~Varoquaux, A.~Gramfort, V.~Michel, B.~Thirion, O.~Grisel, M.~Blondel, P.~Prettenhofer, R.~Weiss, V.~Dubourg, J.~Vanderplas, A.~Passos, D.~Cournapeau, M.~Brucher, M.~Perrot, and E.~Duchesnay.
\newblock Scikit-learn: Machine learning in {P}ython.
\newblock {\em Journal of Machine Learning Research}, 12:2825--2830, 2011.

\bibitem{flask}
Flask.
\newblock \url{https://flask.palletsprojects.com/en/2.2.x/}, 2014.
\newblock [Online; accessed 02-March-2024].

\bibitem{Sartaj_MeDeT_2023}
Hassan Sartaj.
\newblock {MeDeT: Medical Devices Digital Twins Generation with Meta-learning}, Oct 2023.

\bibitem{karie}
{Karie Medicine Dispenser}.
\newblock \url{https://kariehealth.com/}, 2014.
\newblock [Online; accessed 21-May-2024].

\bibitem{medido}
{Automatic Medicine Dispenser Medido}.
\newblock \url{https://medido.com/en/}, 2014.
\newblock [Online; accessed 21-May-2024].

\bibitem{pilly}
{Pilly SMS}.
\newblock \url{https://responssenteret.no/responsskolen/brukere/manualer-videoer/Pilly.php}, 2014.
\newblock [Online; accessed 21-May-2024].

\bibitem{goodfellow2016deep}
Ian Goodfellow, Yoshua Bengio, and Aaron Courville.
\newblock {\em Deep Learning}.
\newblock MIT Press, 2016.

\bibitem{bishop2006pattern}
Christopher~M Bishop and Nasser~M Nasrabadi.
\newblock {\em Pattern recognition and machine learning}, volume~4.
\newblock Springer, New York, NY, USA, 2006.

\bibitem{arcuri2011practical}
Andrea Arcuri and Lionel Briand.
\newblock A practical guide for using statistical tests to assess randomized algorithms in software engineering.
\newblock In {\em Proceedings of the 33rd international conference on software engineering}, pages 1--10, New York, NY, USA, 2011. Association for Computing Machinery.

\bibitem{catak2022uncertainty}
Ferhat~Ozgur Catak, Tao Yue, and Shaukat Ali.
\newblock Uncertainty-aware prediction validator in deep learning models for cyber-physical system data.
\newblock {\em ACM Transactions on Software Engineering and Methodology (TOSEM)}, 31(4):1--31, 2022.

\bibitem{sartaj2024digital}
Hassan Sartaj, Shaukat Ali, and Julie~Marie Gjøby.
\newblock Digital twins environment simulation for testing healthcare {IoT} applications.
\newblock In {\em Proceedings of the 48th Annual Computers, Software, and Applications Conference}, COMPSAC 2024, page 900–901, New York, NY, USA, 2024. IEEE.

\bibitem{sartaj2024uncertainty}
Hassan Sartaj, Shaukat Ali, and Julie~Marie Gjøby.
\newblock Uncertainty-aware environment simulation of medical devices digital twins.
\newblock {\em Software and Systems Modeling}, pages 1--27, 2024.

\end{thebibliography}

\end{document}